\documentclass[smallextended]{svjour3}       % onecolumn (second format)
\RequirePackage{fix-cm}
\smartqed  % flush right qed marks, e.g. at end of proof
\usepackage{graphicx}

\usepackage{enumitem}
\usepackage{changepage} 
\usepackage[sort&compress, authoryear, comma]{natbib}
\usepackage{subfigure}
\usepackage{svg} % for svg import
\usepackage[many]{tcolorbox} %Used for summary boxes
\usepackage[utf8]{inputenc} %UTF 8 support
\usepackage[T1]{fontenc}
\usepackage{soul} %Allows text highlighting, so that we can use it for review
\usepackage[bookmarksdepth=4]{hyperref} %Makes crossrefs clickable
\usepackage[misc,geometry]{ifsym}  %Letter graphics for corresponding author
\usepackage{float} %Enables 'H' float
\usepackage{ifthen} %ifthenelse command (used by text commenting feature)
\usepackage{amssymb} %symbols (used by text commenting feature)
\usepackage{booktabs}
\usepackage{makecell} 
\usepackage{endnotes} 
\usepackage{moresize}
\usepackage[subtle]{savetrees}
\usepackage[title]{appendix}
\usepackage{rotating}
\usepackage{pifont}
\usepackage{ragged2e} %/justify
\usepackage{amsmath}
\usepackage[export]{adjustbox}
\usepackage{multirow} 
\usepackage{tikz}
\usepackage{float}
\usepackage{color, colortbl}
\definecolor{Gray}{gray}{0.9}
\usepackage[compatibility=false, singlelinecheck=false]{caption}

\usepackage{tabularx}

\usepackage{listings, xcolor}

\definecolor{verylightgray}{rgb}{0.99,0.99,0.99}

\lstdefinelanguage{Solidity}{
	keywords=[1]{anonymous, assembly, assert, balance, break, callcode, case, catch, class, constant, continue, constructor, contract, debugger, default, delete, do, else, emit, event, experimental, export, external, false, finally, for, function, if, implements, import, in, indexed, instanceof, interface, internal, is, length, library, log0, log1, log2, log3, log4, memory, modifier, new, payable, pragma, private, protected, public, pure, push, require, return, returns, revert, selfdestruct, send, solidity, storage, struct, suicide, super, switch, then, this, throw, transfer, true, try, typeof, using, value, view, while, with, addmod, ecrecover, keccak256, mulmod, ripemd160, sha256, sha3}, % generic keywords including crypto operations
	keywordstyle=[1]\color{black}\bfseries,
	keywords=[2]{address, bool, def, byte, bytes, bytes1, bytes2, bytes3, bytes4, bytes5, bytes6, bytes7, bytes8, bytes9, bytes10, bytes11, bytes12, bytes13, bytes14, bytes15, bytes16, bytes17, bytes18, bytes19, bytes20, bytes21, bytes22, bytes23, bytes24, bytes25, bytes26, bytes27, bytes28, bytes29, bytes30, bytes31, bytes32, enum, int, int8, int16, int24, int32, int40, int48, int56, int64, int72, int80, int88, int96, int104, int112, int120, int128, int136, int144, int152, int160, int168, int176, int184, int192, int200, int208, int216, int224, int232, int240, int248, int256, mapping, string, uint, uint8, uint16, uint24, uint32, uint40, uint48, uint56, uint64, uint72, uint80, uint88, uint96, uint104, uint112, uint120, uint128, uint136, uint144, uint152, uint160, uint168, uint176, uint184, uint192, uint200, uint208, uint216, uint224, uint232, uint240, uint248, uint256, var, void, ether, finney, szabo, wei, days, hours, minutes, seconds, weeks, years},	% types; money and time units
	keywordstyle=[2]\color{black}\bfseries,
    keywords=[3]{block, blockhash, coinbase, difficulty, gaslimit, number, timestamp, msg, delegate, sender, sig, value, now, tx, gasprice, origin},	% environment variables
	keywordstyle=[3]\color{violet}\bfseries,
	identifierstyle=\color{black},
	sensitive=false,
	comment=[l]{\#},
	morecomment=[s]{/*}{*/},
	commentstyle=\color{blue}\fontfamily{pcr}\selectfont\ssmall,
	stringstyle=\color{red}\fontfamily{pcr}\selectfont\ssmall,
	morestring=[b]',
	morestring=[b]",
 	keywordstyle=[4]\color{red}\bfseries,
    keywords=[4]{funct, gas, args},	% 
}

\lstdefinestyle{sidebyside}{
	language=Solidity,
	backgroundcolor=\color{verylightgray},
	extendedchars=true,
	basicstyle=\fontfamily{pcr}\selectfont\tiny,
	showstringspaces=false,
	showspaces=false,
	numbers=left,
	numberstyle=\tiny,
	numbersep=5pt,
	tabsize=2,
	breaklines=true,
	showtabs=false,
	captionpos=b,
	xleftmargin=3.0ex,
}

\definecolor{highlightcolor}{rgb}{1.0, 0.9, 0.9} % Light red for highlighting

\lstdefinestyle{similarinterfacecheck}{
    language=Solidity,
    backgroundcolor=\color{verylightgray},
    basicstyle=\fontfamily{pcr}\selectfont\small,
    showstringspaces=false,
    showspaces=false,
    numbers=left,
    numberstyle=\small,
    numbersep=5pt,
    tabsize=2,
    breaklines=true,
    showtabs=false,
    captionpos=b,
    xleftmargin=3.0ex,
    moredelim=**[is][\colorbox{highlightcolor}]{@}{@}, % Delimiters for highlighted text
}	% copy the file from this repo

\newcommand{\savefootnote}[2]{\footnote{\label{#1}#2}}

\graphicspath{{Images/}}

\newcommand{\countobservations}{
    \def \countobservations{1}
}
\newcounter{observation}
\newenvironment{observation}[1]
{   \refstepcounter{observation}
    \noindent \textbf{Observation~\theobservation) 
    \textit{#1}}
}

\newcommand{\countapplications}{
    \def \countapplications{1}
}
\newcounter{application}
\newboolean{showcomments}
\setboolean{showcomments}{true}

\ifthenelse{\boolean{showcomments}}
{\newcommand{\nbc}[3]{
 {\colorbox{#3}{\bfseries\sffamily\scriptsize\textcolor{white}{#1}}}
 {\textcolor{#3}{\sf\small$\blacktriangleright$\textit{#2}$\blacktriangleleft$}}
 }
}
{\newcommand{\nbc}[3]{}
 }

\newtcolorbox{mybox}[2][]{
top=0.15in,left=4pt,right=4pt,bottom=4pt,
fonttitle=\bfseries,
colbacktitle=gray,
colback=gray!5,
colframe=gray!40!black,
enhanced,
attach boxed title to top left={xshift=1.5em,yshift=-\tcboxedtitleheight/2},
boxed title style={size=small},
drop shadow={black!50!white},
title=#2,#1}

\newcommand{\countimplications}{
    \def \countimplications{1}
}
\newcounter{implication}
\countimplications

\begin{document}

\title{UPC Sentinel: An Accurate Approach for Detecting Upgradeability Proxy Contracts in Ethereum%\thanks{Grants or other notes
}
% \subtitle{Do you have a subtitle?\\ If so, write it here}


\author{Amir~M.~Ebrahimi \and Bram~Adams \and Gustavo~A.~Oliva \and Ahmed~E.~Hassan
}


\institute{Amir~M.~Ebrahimi, Gustavo~A.~Oliva, Ahmed~E.~Hassan \at Software Analysis and Intelligence Lab (SAIL), School of Computing \\
Queen’s University, Kingston, Ontario, Canada \\
\email{\{amir.ebrahimi,gustavo,ahmed\}@cs.queensu.ca}           %  \\
           \and
           Bram~Adams \at Lab on Maintenance, Construction and Intelligence of Software (MCIS), School of Computing \\
Queen’s University, Kingston, Ontario, Canada \\
\email{bram@cs.queensu.ca}
}

\date{Received: date / Accepted: date}

\maketitle
    \begin{abstract}
    Software applications that run on a blockchain platform are known as DApps. DApps are built using smart contracts, which are immutable after deployment. Just like any real-world software system, DApps need to receive new features and bug fixes over time in order to remain useful and secure. However, Ethereum lacks native solutions for post-deployment smart contract maintenance, requiring developers to devise their own methods. A popular method is known as the upgradeability proxy contract (UPC), which involves implementing the proxy design pattern (as defined by the Gang of Four). In this method, client calls first hit a proxy contract, which then delegates calls to a certain implementation contract. Most importantly, the proxy contract can be reconfigured during runtime to delegate calls to another implementation contract, effectively enabling application upgrades. For researchers, the accurate detection of UPCs is a strong requirement in the understanding of how exactly real-world DApps are maintained over time. For practitioners, the accurate detection of UPCs is crucial for providing application behavior transparency and enabling auditing. In this paper, we introduce UPC Sentinel, a novel three-layer algorithm that utilizes both static and dynamic analysis of smart contract bytecode to accurately detect active UPCs. We evaluated UPC Sentinel using two distinct ground truth datasets. In the first dataset, our method demonstrated a near-perfect accuracy of 99\%. The evaluation on the second dataset further established our method's efficacy, showing a perfect precision rate of 100\% and a near-perfect recall of 99.3\%, outperforming the state of the art. Finally, we discuss the potential value of UPC Sentinel in advancing future research efforts.
\end{abstract}

    \section{Introduction}
\label{sec:introduction}

Ethereum, since its inception, has revolutionized the blockchain landscape by introducing smart contracts technology. A smart contract is a self-executing software, with terms of agreement directly written into code lines~\citep{Tikhomirov18, Oliva21, Chen20, Ebrahimi23}. Beyond mere value transfer, which is the primary function of traditional blockchain platforms like Bitcoin, Ethereum's smart contracts pave the way for developing full-fledged decentralized applications (DApps). DApps are important because they offer a level of autonomy, security, and transparency that traditional centralized apps cannot. By running on a blockchain platform, they minimize the risks of data breaches and downtime, resist censorship, and give users enhanced control over their data and transactions, fostering trust and collaboration in digital interactions. DApps currently support several business domains, including decentralized finance (DeFi)~\citep{CorporateFinanceInstitute, FortuneBusinessInsights_2023} and supply chain management~\citep{Deloitte_2023}.

In blockchain platforms like Ethereum, the principle of immutability ensures transparency, trust, and security by guaranteeing that transactions cannot be rolled back and the bytecode of a deployed contract cannot be changed. However, immutability also comes at a great cost. As Decentralized Applications (DApps) become more complex, the need for periodic maintenance, such as bug fixes and feature improvements, becomes paramount~\citep{Chen20-2, Chen20}. As such, in a space where technology evolves rapidly and vulnerabilities can be discovered post-deployment, the immutability of smart contracts can hinder timely updates, potentially compromising functionality or security. Thus, developers are often caught in a delicate balance between the assurance of immutability and the practical need for application maintenance and evolution.

Despite being the most prominent infrastructure for hosting and executing smart contracts, Ethereum does \textit{not} provide any native mechanism to address the need for DApp maintenance and evolution. To bridge this gap, practitioners have designed and implemented various methods for upgrading (i.e., replacing) smart contracts. The \textit{proxy pattern} method stands out due to its flexibility in enabling the maintenance of DApps~\citep{smartcontract_upgrades, Salehi22, Ebrahimi23}. Inspired by the proxy design pattern~\citep{GoF}, a proxy smart contract enables upgradeability by separating the contract logic from its state. A \textit{proxy contract} delegates calls to an \textit{implementation contract} that contains the actual logic code, while the proxy itself maintains the state. If the logic needs to be updated, (i) a new implementation contract is deployed, and (ii) the proxy is set to redirect calls to this new contract instead of the old one, allowing the original implementation contract to be upgraded without losing its state or changing its blockchain address.

Given the increasing popularity of the proxy pattern~\citep{Ebrahimi23}, there exists a pressing need for tools that can detect and trace upgradeability proxy contracts (UPCs)~\citep{eip1967}. As the blockchain ecosystem grows in complexity and scale, ensuring the transparency, security, and reliability of these proxy contracts becomes of prime importance. For end-users and stakeholders, such tools provide increased transparency by alerting them about the existence of underlying proxy mechanisms and changes in contract behavior. Furthermore, regulators and auditors require these tools to monitor the blockchain for compliance, ensuring that upgrade mechanisms are not misused. Finally, these tools are also critical for researchers who aim to understand how exactly real-world DApps are maintained over time~\citep{eip1967, Ebrahimi23, Salehi22, Bodell23}.

% it is notable that, as of Sep. 2022, only 23.5\% of contracts on Etherscan have their source code published (a.k.a., verified contracts). 
% 
Several studies have proposed methods for detecting UPCs, with the majority relying on source code analysis~\citep{Bodell23, qasse2023, Feist19}. However, the applicability of these source-level UPC detectors is limited, as the majority of smart contracts are closed-source~\citep{Junzhou2020,Huang2021,Wenkai2024,Ting2019,Chen21_Maintenance, Jiachi_DefectChecker_2022}. In contrast, methods focusing on bytecode analysis are naturally not hindered by the absence of source code. However, to the best of our knowledge, only one study specifically addresses the detection of UPCs at the bytecode level~\citep{Salehi22}. Most importantly, while several different standards and patterns have been introduced for UPCs~\citep{Meisami23, ERCs}, only three specific patterns are covered by~\citet{Salehi22}: Regular proxies~\citep{palladino_2021}, Beacon~\citep{beacon_proxy}, and Universal Upgradeability Proxy Standard~\citep{eip_1822}. More broadly, there is a lack of empirical studies evaluating the effectiveness of bytecode-based methods for detecting UPCs and comparing them to source-code-based approaches.

Upgradeability proxy detection methods also apply a range of static and dynamic analysis techniques. Methods that rely on static analysis can suffer from a lack of information, especially when roles and responsibilities are spread across several independently deployed contracts. In such cases, only a dynamic analysis technique can extract the right dependency, expanding the static analyzer search space. Methods based exclusively on dynamic analysis can also be insufficient, as the indications of upgradeability may not necessarily manifest at the behavioral level (e.g. if an upgrade has never taken place). Therefore, proper orchestration between the static and dynamic techniques is required to effectively detect UPCs. Among existing methods, only a few combine these approaches, but current techniques face challenges in orchestration. These include limitations such as analyzing only one dependency (e.g., implementation contract) at a time, applying these techniques to a narrow range of patterns, or focusing solely on immediate dependencies, which restricts their ability to detect upgradeability mechanisms in transitive dependencies (i.e., dependencies of dependencies). Such a deeper, multi-level dynamic dependency analysis is currently overlooked by literature.

To address the limitations of current methods, we propose UPC Sentinel, a three-layer algorithm that leverages both static and dynamic analysis techniques on smart contract bytecode to detect UPCs. UPC Sentinel is grounded in a comprehensive survey of the most common upgradeability proxy design patterns~\citep{Meisami23}. Specifically, for a given input contract, the algorithm first employs dynamic analysis to determine whether the contract is an \texttt{active} proxy (i.e., a proxy whose functionality has been utilized at least once since its creation)~\citep{Ebrahimi23}. If the contract is identified as an active proxy, a combination of static and dynamic analyses is applied to evaluate whether it qualifies as a UPC. Finally, UPC Sentinel performs a fine-grained classification, identifying the specific upgradeability pattern the UPC adheres to. We carried out an empirical evaluation using two independent ground truth datasets to gauge the performance of UPC Sentinel from various perspectives, contrasting it with findings from prior research. 

\smallskip \noindent \textbf{Key contributions:} (i) introducing UPC Sentinel, an effective and accurate algorithm for detecting upgradeability proxy contracts (UPC) at the bytecode level (ii) centered on three derived bytecode-level reference designs for upgradeability, grounded in the seminal list of upgradeability proxy patterns and their characteristics curated by~\citet{Meisami23}; (iii) the ability to identify the specific upgradeability proxy pattern to which a detected UPC conforms at the bytecode level; (iv) compiling a new ground truth dataset with over 3,177 UPCs; (v) identifying and fixing labeling inconsistencies from a dataset released by~\citet{Bodell23} that includes over 994 contracts; (vi) conducting a thorough empirical evaluation using the aforementioned ground truths including more than 4,000 data points.

\smallskip \noindent \textbf{Paper structure:} Section~\ref{sec:background} discusses the key terminologies and concepts used throughout the paper. Section \ref{sec:method} outlines the design of UPC Sentinel, while Section \ref{sec:data-collection} details our data sources, the characteristics of the employed ground truth datasets, and the data quality checks we performed. Section~\ref{sec:eval} elaborates on our empirical evaluations. Section~\ref{sec:discussion} discusses the potential value of UPC Sentinel, and examines the time cost associated with its use. Section \ref{sec:related-work} provides an in-depth explanation of related studies and presents a thorough comparative analysis between our method and previous studies. Section~\ref{sec:threats} highlights potential threats to the validity of our study and proposes future directions for addressing the limitations of UPC Sentinel. Lastly, Section~\ref{sec:conclusion} concludes our study.
    \section{Background} 
\label{sec:background}

\subsection{The Anatomy of Proxy Contracts} \label{subsec:back-proxy-contract}

The proxy contract design pattern involves two main components: the \textit{proxy contract} and the \textit{implementation contract} (Figure \ref{fig:back-proxy-pattern-design}):

\begin{figure}[!htbp]
 \centering
 \includegraphics[scale=0.75]{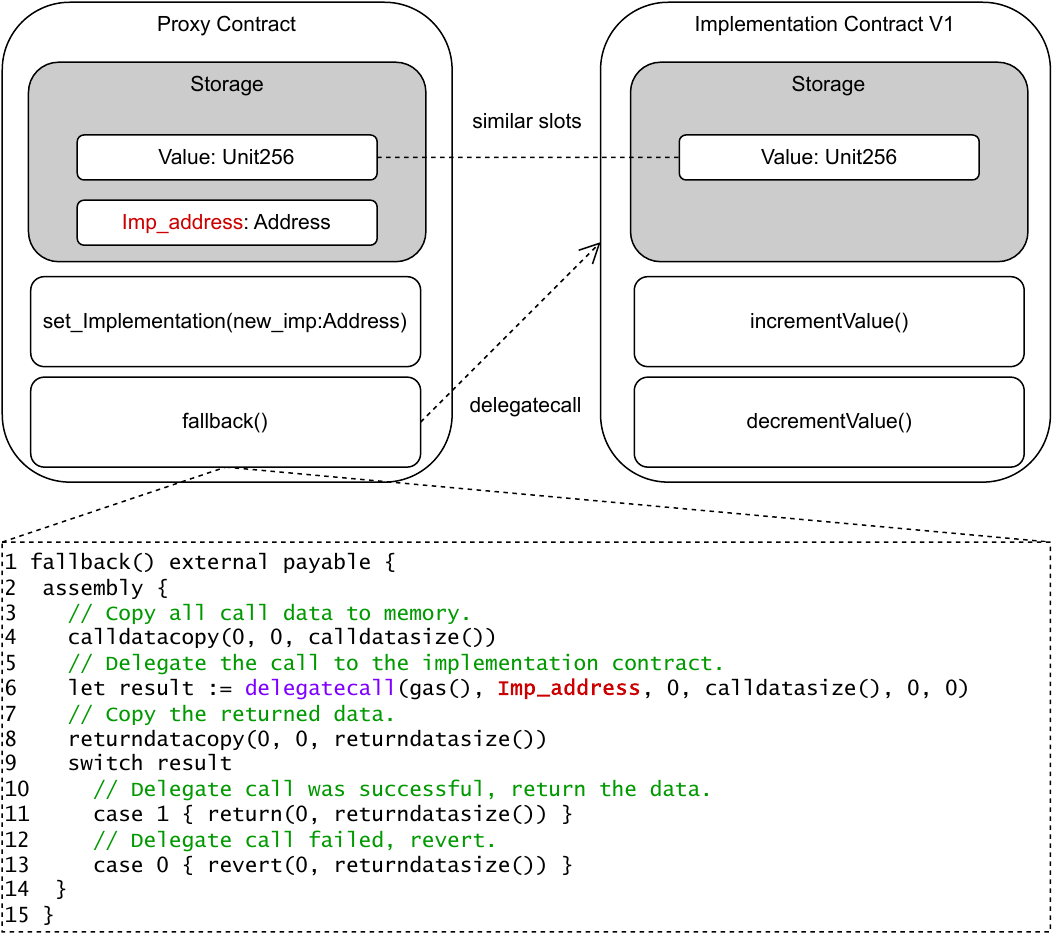}
 \caption{The basic proxy contract design.}
  \label{fig:back-proxy-pattern-design}
\end{figure}

\smallskip\noindent\textbf{Proxy contract}. This contract keeps the storage (i.e., state) and delegates calls to the implementation contract~\citep{openzeppelin_2020_proxy_pattern}. At the storage level, the proxy typically has a state variable that keeps track of the current implementation contract address. We briefly call this variable the \texttt{implementation variable}. Additionally, the proxy can define other state variables. At the delegation level, a proxy is typically implemented inside a \texttt{fallback} function where the calls are caught and subsequently delegated to the implementation contract using a low-level operation called \texttt{delegatecall}. The delegatecall uses the input \texttt{calldata}\footnote{Calldata is the data payload included in a smart contract call, containing the function selector and any encoded parameters necessary for executing the function on the Ethereum Virtual Machine (EVM).} to extract the first 4 bytes, which specify the function selector of the function it delegates the call to.

When a delegatecall is made from the proxy contract to the implementation contract, the code from the implementation contract is executed as if it were part of the proxy contract itself. Thus, if the implementation contract modifies a storage variable, it is actually modifying the storage of the proxy contract. In Figure \ref{fig:back-proxy-pattern-design}, when a call is delegated to either ``decrementValue'' or ``incrementValue'' functions, it results in decreasing or increasing the ``value'' variable stored in the proxy storage. This is because the Solidity compiler determines the layout of storage variables by the order in which they are declared in the contract. Since both the proxy and the implementation contract declare ``value'' as the first storage variable, it resides at storage slot 0 in both contracts. When delegatecall is executed, the code from the implementation contract references storage slot 0, which due to the delegation, points to the proxy contract's storage slot 0 where its own ``value'' variable is stored. Hence, any modification to ``value'' by the implementation contract through the delegatecall operation directly updates the ``value'' in the proxy contract's storage.

\smallskip\noindent\textbf{Implementation contract (a.k.a., logic contract)}. The implementation contract contains the business logic that will be executed. The implementation contract does not store any long-term data. Instead, any changes to data variables are reflected in the proxy's storage.

\subsection{Fallback Function}\label{subsec:back-fallback}
A \texttt{fallback} function is a special type of function in Ethereum smart contracts written in Solidity~\citep{soliditylang}. It does not have a name nor parameters. This allows the contract to react to unexpected or arbitrary calls. Figure \ref{fig:back-proxy-pattern-design} shows the logic of a fallback function in a typical proxy contract. Implementing the proxy functionality within the fallback functions allows the proxy to effectively exposes all the the public and external functions that the implementation contract exposes, even though it does not define any of them explicitly. The fallback function acts like a "one to many" mechanism, capable of handling all function calls and \textit{seamlessly} delegating them to an appropriate implementation contract's function. On the other hand, if one is certain about the specific set of functions they want the proxy support for, they can implement only those in the proxy and forward calls only for them. While this approach is less flexible and does not fully leverage the potential of proxy contracts, it can be more explicit about what is allowed.

    \section{UPC Sentinel Approach}
\label{sec:method}

\subsection{Technical Requirements}\label{subsec:problem-domain}

The primary objective of UPC Sentinel is to accurately identify UPCs within the Ethereum blockchain, which is the most common method for upgrading smart contracts on the Ethereum network~\citep{Salehi22}. Hence, other upgradeability methods (see Appendix \ref{subsec:landscape-of-upgradeability-methods}), such as \texttt{self-destruct}, \texttt{create2} mechanism, and strategy pattern are outside the scope of this paper. 

UPC Sentinel takes a smart contract address or a list of addresses as input. First, we aim to distinguish if the input contract is a proxy contract with high recall and precision (\textbf{requirement R1}). This step ensures comprehensive and accurate identification of proxy contracts. Subsequently, if the contract in question is a proxy contract, UPC Sentinel must accurately determine whether the proxy is specifically designed for upgradeability or if it serves other purposes (\textbf{requirement R2}).

Furthermore, the majority of deployed smart contracts lack source code~\citep{Junzhou2020,Huang2021,Wenkai2024,Ting2019,Chen21_Maintenance, Jiachi_DefectChecker_2022}, which limits the applicability of UPC detectors that rely on this artifact. Therefore, UPC Sentinel shall effectively detect UPCs based on the contract's bytecode (\textbf{requirement R3}). This approach ensures that the identification of UPCs can occur even when their source code is not available.

% To shed light on the ratio of smart contracts that have their source code published as of Sept. 2022, we did a study on 50,845,833 deployed smart contracts (see Section \ref{sec:data-collection}), using Etherscan's REST API. Our collection involved a two-step process for each contract address. Initially, we verified whether the source code was accessible and specifically written in Solidity, the most prominent programming language for smart contracts\footnote{\url{https://etherscan.io/dashboards/contract-statistics}}. If these conditions were met, we proceeded to retrieve the source code. Otherwise, we excluded that contract from our analysis. We obtained 11,251,864 (22.1\%) verified contracts through Etherscan API.

% Furthermore, although some contracts may not be directly verified in Etherscan, they can have a cloned version (i.e., a contract with an identical bytecode) that was already verified. Therefore, such contracts can be also considered as verified, yet indirectly. To account for these cases, we consider any contract that has identical bytecode to those 11,251,864 verified contracts as verified too, yielding 11,973,102 (23.5\%) verified contracts. This means that 76.5\% of deployed contracts as of Sep. 2022 have \textit{never} been verified. 

Moreover, various implementation strategies (e.g., Transparent Upgradeability Proxy, Beacon Upgradeability Proxy, etc.) exist for designing UPCs (see Section \ref{subsec:upgradeability-proxy-patterns}). While these patterns are based on the foundational concepts of the proxy design pattern, each possesses distinct characteristics. For instance, variations might occur in how and where the implementation contract address is stored or how and where the upgrade functions are implemented. Therefore, UPC Sentinel must not only achieve a high detection rate regardless of the specific upgradeability design pattern but also identify the specific type of upgradeability pattern a detected UPC represents (\textbf{requirement R4}).

\subsection{Three-layered Design of UPC Sentinel}\label{subsec:solution-domain}
\begin{figure}[!t]
\centering
  \includegraphics[width=0.95\columnwidth]{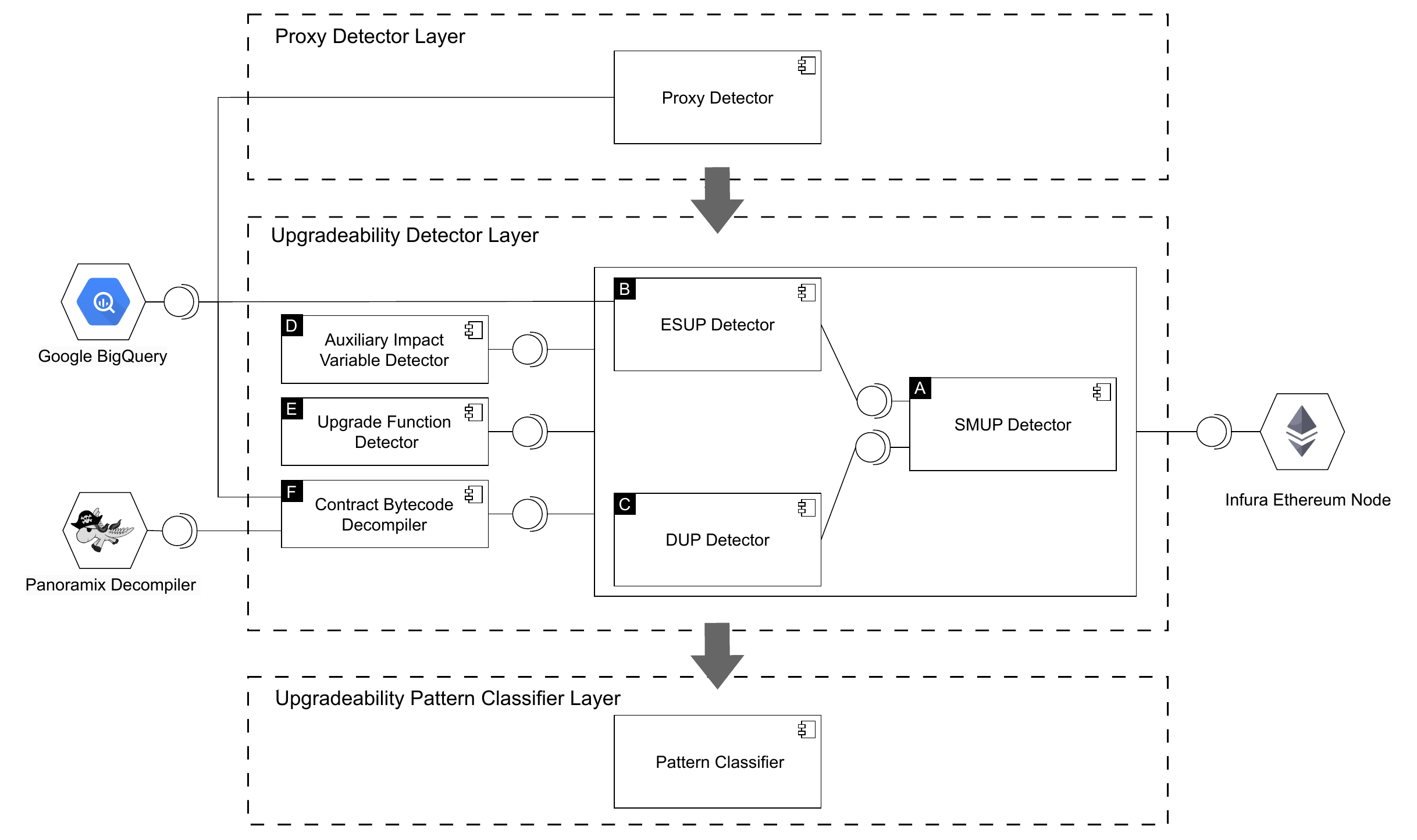}
  \caption{UPC Sentinel architecture.}
  \label{fig:upc-sentinel-architecture}
\end{figure}

We now introduce the UPC Sentinel algorithm that detects UPCs through a two-layered approach. The input of UPC sentinel is a list of smart contract addresses. Figure \ref{fig:upc-sentinel-architecture} depicts the architecture of UPC Sentinel, which consists of three distinct layers: i) the proxy contract detector, ii) the upgradeability detector, and iii) the upgradeability pattern classifier. The proxy contract detector ascertains if a given contract operates as a proxy, fulfilling R1. The upgradeability detector examines whether an identified proxy contract is specifically tailored for upgradeability purposes, fulfilling R2. Finally, the upgradeability pattern classifier provides a detailed classification, identifying the specific upgradeability proxy pattern the detected UPC adheres to. For every input contract address, UPC Sentinel outputs a binary label indicating whether the contract is an upgradeability proxy contract (UPC) or not. If classified as a UPC, it also identifies the type of upgradeability pattern the UPC represents, fulfilling R4. In the following Sections \ref{subsec:proxy-detector-layer}, \ref{subsec:upgradeability-detector-layer}, and \ref{subsec:upgradeability-pattern-classifier-layer}, we provide a detailed discussion of each layer.

\subsection{Proxy Detector Layer}\label{subsec:proxy-detector-layer}

We employed the dynamic method that we proposed in our previous study to detect proxy contracts~\citep{Ebrahimi23}. Specifically, we enumerated two \textit{inherent and conceptual properties} of a proxy contract based on existing guidelines and standards~\citep{GoF, Agarwal_2023, Wasnik_2024, EIP-2535, Fravoll_2018b, OpenZeppelin_2017b}. Take contract $C$ as an example. One can argue that $C$ is a proxy contract if it satisfies both inherent properties of proxy contracts:

\begin{itemize}[label=\textbullet, itemsep = 3pt, topsep = 3pt]
    \item \textbf{Property \#1.} $C$ must \underline{expose} - to its users - at least one public or external interface/function, $F_C$, within which scope the call is delegated  (through \texttt{delegatecall} statement) to an appropriate implementation contract’s function, $F_I$. $F_C$ is typically the $C$'s fallback function \citep{Ebrahimi23, Bodell23, Salehi22, qasse2023, OpenZeppelin_2017b}.

    \item \textbf{Property \#2.} Assuming Property \#1 is satisfied, the function selectors/signatures of $F_C$ and $F_I$ must match exactly. This requirement ensures that a proxy contract preserves the interface of the object it represents \citep{GoF, Agarwal_2023, Wasnik_2024}. When the delegatecall uses the first four bytes of the calldata (i.e., $F_C$'s function selector), it essentially guarantees that $F_C$ and $F_I$ correspond to the same function, enabling seamless delegation to the implementation contract~\citep{EIP-2535, Fravoll_2018b, Ebrahimi23}. If this condition is not satisfied, $C$ is classified as an adapter contract rather than a proxy contract \citep{GoF}.
\end{itemize}

\begin{figure}[!t]
\centering
  \includegraphics[scale=0.75]{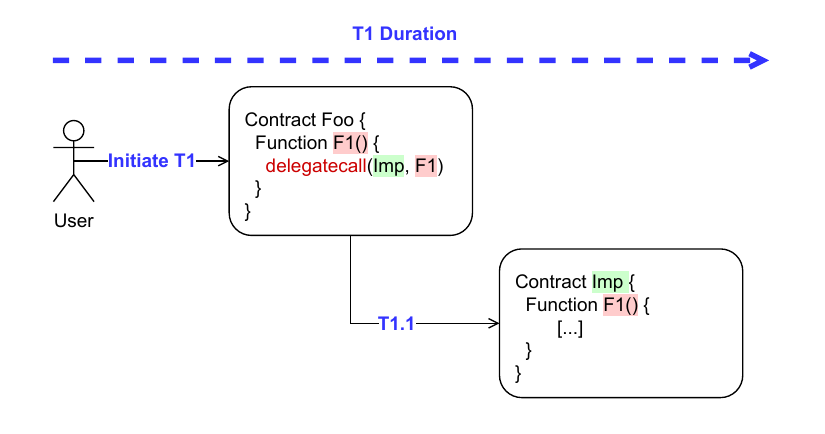}
  \caption{An illustrative example demonstrating how a proxy is identified based on its two inherent properties.
}
  \label{fig:proxy-detector-example}
\end{figure}

In other words, both the proxy and implementation contracts must have at least one function with a matching selector, and the proxy contract must employ the delegatecall to forward calls within this function to the corresponding function in the implementation contract.  
In our previous study on detecting proxy contracts~\citep{Ebrahimi23}, we employed these two properties as our filtering criteria, and devised a method that contrasts the runtime behavior (i.e., transactions) of smart contracts with these two criteria to identify proxy contracts. To clarify how our proxy detector functions at the transaction level, consider the example shown in Figure \ref{fig:proxy-detector-example}. If a user initiates a transaction session T1 that calls the function F1 of the Foo contract, and during T1, if the Foo’s F1 function uses a delegatecall statement (i.e., through the internal transaction T1.1) to relay the call to the Imp contract’s F1 function, which has an identical signature to the Foo’s F1 function, then Foo is indeed a proxy contract. For any given smart contract address, this layer outputs its proxy status (i.e., whether it is a proxy or not) and all the implementation contracts to which the proxy contract has ever delegated calls. 

There are two primary reasons for our choice of this method. First, this method is \textit{not} based on source code analysis, which satisfies our R3 requirement. Second, we demonstrated that this method achieves perfect precision and recall in detecting \texttt{active} proxy contracts and detects more proxy contracts than a previous approach by~\citep{Salehi22}. We defined an active proxy contract as one that has had at least one transaction following its creation, in which the proxy delegates calls to its implementation contract~\citep{Ebrahimi23}. Hence, the adoption of this method introduces a specific focus to our methodology, narrowing our scope to detecting only active UPCs (see Section \ref{sec:threats} for a discussion about the potential impacts of this design choice).

\subsection{Upgradeability Detector Layer} \label{subsec:upgradeability-detector-layer}

After identifying a proxy, the next important step is to determine if it is an \emph{upgradeable} proxy contract (UPC). We propose an algorithm that incorporates both static and dynamic analyses to determine the type of a given proxy contract. Particularly, we applied a static analysis on the decompiled contract's bytecode, which satisfies R3. In the following sections, we first describe how we derived three upgradeability reference designs (URDs) that guided the development of our algorithms at the bytecode level. Next, we briefly introduce each of the URDs along with a general detection strategy for each. Finally, we delve into the implementation details of the six UPC Sentinel's upgradeability detector components.

\subsubsection{Deriving Upgradeability Reference Designs (URDs)}\label{subsec-deriving-urds}
Given our objective of detecting UPCs at the bytecode level, we set out to analyze the bytecode-level mechanisms onto which established upgradeability proxy patterns are mapped, then to aggregate these mechanisms into higher-level abstractions (i.e., upgradeability reference designs). Ultimately, by focusing on the identification of those abstractions, our goal is to be able to detect any UPCs adhering to those abstractions irrespective of their specific implementation choices (e.g., their storage layout, the inheritance hierarchy, type of variables, etc.) and availability of source code. 

The input of this study are eight established standards and well-proven upgradeability proxy patterns that are commonly endorsed by the Ethereum community. In particular, \citet{Meisami23} compiled a curated list of proxy patterns, which served as our starting point. From this list, we selected the eight patterns that are specifically known for upgradeability. A comprehensive technical discussion of these patterns is provided in Appendix~\ref{subsec:upgradeability-proxy-patterns}.

% At its core, an upgradeability proxy contract extends a standard proxy contract by incorporating a mechanism (e.g., an upgrade or setter function) that facilitates the dynamic replacement of the implementation contract address to which it delegates calls. This principle serves as the foundation of the upgradeability mechanism in all existing upgradeability proxy patterns~\citep{OpenZeppelin_2018a_unstructured, eip1967, Palladino20, palladino_2021, Mudge18, beacon_proxy, eip_1822, OpenZeppelin_inherited_git, OpenZeppelin_eternal_git, Diligence_2017}. This implies that to 

To identify a given contract as an upgradeability proxy contract, it must satisfy two key conditions at an abstract level\citep{Johnson_2016, AL2023Review, OpenZeppelin_2018b_proxy_upgrade_pattern, OpenZeppelin_2018b_writing_upgradeable_contracts, Ethereum.org_2023}: i) the contract must function as a proxy, and ii) it must include an upgrade mechanism or function that allows the proxy’s implementation address to be modified. While the first condition is addressed by the proxy detector layer (Section \ref{subsec:proxy-detector-layer}), the second condition encapsulates the core essence of upgradeability. Consequently, identifying the upgradeability status of a proxy contract can be formulated as a search for two critical syntactic features: (i) the location where the state variable storing the proxy’s implementation contract address is declared, and (ii) the location where the upgrade function enabling the modification of this state variable is defined. We briefly call the former and latter features as the ``implementation variable location'' and the ``upgrade function location'', respectively.

Given the critical role of these two features in determining the upgradeability status of a proxy contract, we aim to analyze their evolution across various upgradeability proxy patterns and derive a generalized guided approach to detecting upgradeability at the bytecode level. Notably, \citet{Bodell23} conducted an extensive study on a wide range of syntactic features specific to upgradeability proxy patterns at the source code level including the mentioned features. While their findings are foundational and inform our analysis, the transformations that occur during compilation to bytecode—such as inheritance linearization, storage layout standardization, and the unification of read/write operations—necessitate a deeper investigation into how the features' values are encoded and altered at the bytecode level.

To analyze each upgradeability pattern, the first and third authors independently studied its online documentation, then analyzed it using deductive coding based on the two features and the initial (source code level) values from \citeauthor{Bodell23}’s study). Afterwards, they met to discuss any disagreements (zero cases) in their classification. Table \ref{tab:comp-imp-upg-location} represents the values, similarities, and dissimilarities of the two features across all eight patterns for both our study (bytecode level) and the study by \citet{Bodell23} (source code level). In particular, in three (i.e., Inherited Storage Upgradeability Proxy Pattern, Eternal Storage Upgradeability Proxy Pattern, and ERC-2535 Diamonds Upgradeability Proxy Pattern) out of eight upgradeability patterns, we observe different values for the features at the bytecode level compared to the source code level findings. However, in the remaining five patterns our findings concur with prior study's findings \citep{Bodell23}.

\begin{table}[!t]
\caption{Comparison of the implementation variable location and upgrade function location feature values across patterns, analyzed at the source code (top) and bytecode (bottom) levels. Red-highlighted values indicate differences between the bytecode and source code analyses. Red-highlighted values indicate differences between the bytecode and source code analyses \citep{Bodell23}.}
\label{tab:comp-imp-upg-location}
\resizebox{\textwidth}{!}{%
\begin{tabular}{lllcllllll}
\toprule
\textbf{Study} & \textbf{Feature} & \multicolumn{8}{c}{\textbf{Upgradeability Proxy Patterns}} \\ \cline{3-10} 
 &  & 
\multicolumn{1}{c}{\rotatebox{90}{\textbf{\color[HTML]{FF0000} Inherited Storage Upgradeability Proxy}}} & 
\multicolumn{1}{c}{\rotatebox{90}{\textbf{\color[HTML]{FF0000} Eternal Storage Upgradeability Proxy}}} & 
\multicolumn{1}{c}{\rotatebox{90}{\textbf{ERC-1967 Upgradeability Proxy (Unstructured Store)}}} & 
\multicolumn{1}{c}{\rotatebox{90}{\textbf{Transparent Upgradeability Proxy}}} & 
\multicolumn{1}{c}{\rotatebox{90}{\textbf{\color[HTML]{FF0000} ERC-2535 Diamonds Upgradeability Proxy}}} & 
\multicolumn{1}{c}{\rotatebox{90}{\textbf{ERC-1822 Universal Upgradeable Proxy}}} & 
\multicolumn{1}{c}{\rotatebox{90}{\textbf{Beacon Upgradeability Proxy}}} & 
\multicolumn{1}{c}{\rotatebox{90}{\textbf{Registry Upgradeability Proxy}}} \\ \hline
 & \begin{tabular}[c]{@{}l@{}}Implementation \\ variable location\end{tabular} & 
{\color[HTML]{FF0000} Inherited} & {\color[HTML]{FF0000} -} & Proxy & Proxy & 
{\color[HTML]{FE0000} Proxy} & Proxy \& Implementation & External & External \\ \cline{2-10} 
\multirow{-2}{*}{\begin{tabular}[c]{@{}l@{}}\citet{Bodell23}\\(Source code level)\end{tabular}} & \begin{tabular}[c]{@{}l@{}}Upgrade\\ function location\end{tabular} & 
\multicolumn{1}{c}{{\color[HTML]{FF0000} -}} & {\color[HTML]{FF0000} -} & Proxy & Proxy & 
Implementation & Implementation & External & External \\ \hline
 & \begin{tabular}[c]{@{}l@{}}Implementation\\ variable location\end{tabular} & 
{\color[HTML]{FF0000} Proxy} & \multicolumn{1}{l}{{\color[HTML]{FF0000} Proxy}} & Proxy & Proxy & 
{\color[HTML]{FE0000} Proxy \& Implementation} & Proxy \& Implementation & External & External \\ \cline{2-10} 
\multirow{-2}{*}{\begin{tabular}[c]{@{}l@{}}Our Study\\(Bytecode level)\end{tabular}} & \begin{tabular}[c]{@{}l@{}}Upgrade\\ function location\end{tabular} & 
{\color[HTML]{FF0000} Proxy} & \multicolumn{1}{l}{{\color[HTML]{FF0000} Proxy}} & Proxy & Proxy & 
Implementation & Implementation & External & External \\ \hline
\end{tabular}
}
\end{table}

In particular, for the first feature (i.e., the implementation variable location), according to the documentation of both the Inherited Storage Upgradeability Proxy Pattern \citep{OpenZeppelin_inherited_git} and the Eternal Storage Upgradeability Proxy Pattern \citep{OpenZeppelin_eternal_git}, the proxy contracts inherit from another contract where the upgradeability-related state variables are declared. At the source code level, this means that the implementation variable location resides in the inherited contract for both patterns. \citep{Bodell23} also noted this for the former pattern. However, when considering bytecode, Solidity uses the C3 linearization~method to flatten the inheritance tree into a linear order~\citep{Wikipedia_solidity}. This process integrates the storage layout of parent contracts into the proxy's bytecode. As a result, for both the Inherited Storage and Eternal Storage Upgradeability Proxy Patterns, the implementation variable is ultimately embedded within the proxy's bytecode.

In the case of the ERC-2535 Diamonds Upgradeability Proxy Pattern \citep{Mudge_2020_diamonds, EIP-2535}, the Diamond Proxy maintains a mapping (i.e., the implementation variable) of function selectors to their corresponding implementation (a.k.a., facet) contract addresses. The upgrade process is orchestrated by a key implementation contract, the DiamondCutFacet. To ensure that the upgrade function can effectively modify the Diamond Proxy’s mapping state variable, the Diamond Proxy and DiamondCutFacet must adhere to a consistent storage layout. The storage consistency is achieved by defining the mapping state variable in a shared library utilized by both the Diamond Proxy and DiamondCutFacet. Consequently, both the proxy and its implementation contract that manages the upgrade mechanism have access to the implementation variable at the same storage slot. The upgrade itself must be initiated by the Diamond Proxy through the delegatecall mechanism, ensuring that changes to the mapping occur in the Diamond Proxy's context. At the bytecode level, since Solidity compiles the library code alongside the dependent contracts (in this case, the Diamond Proxy and DiamondCutFacet), the implementation variable is present in both the proxy and its implementation bytecodes.

After the deductive coding process~\citep{Hsieh2005}, we further abstracted the UPC patterns into three distinct groups, referred to as Upgradeability Reference Designs (URDs), based on the bytecode-level values of two key features. This abstraction provides a higher-level understanding that transcends the specific source code implementation details of individual patterns. By focusing on core features and bytecode-level concepts, these reference designs enable a robust and generalizable detection method capable of identifying UPCs whose upgradeability mechanisms align with one of the three designs, regardless of implementation specifics. Thus, it offers a resilient methodology that requires less effort to adapt to small variations occurring over time, reducing the need for case-by-case reanalysis.

The first group, Self-Managed Upgradeability Proxy (SMUP), includes the Inherited Storage Upgradeability Proxy, Eternal Storage Upgradeability Proxy, ERC-1967 Upgradeability Proxy (Unstructured Storage) and Transparent Upgradeability Proxy. In these designs, both the implementation variable and the upgrade function are embedded within the proxy bytecode. The second group, Delegated Upgradeability Proxy (DUP), consists of the ERC-1822 Universal Upgradeable Proxy and ERC-2535 Diamonds Upgradeability Proxy. Here, the implementation variable is stored in both the proxy and the implementation bytecode, while the upgrade function is located in the implementation bytecode. The final group, Externally Supervised Upgradeability Proxy (ESUP), encompasses two patterns including Beacon Upgradeability Proxy and Registry Upgradeability Proxy with both the implementation variable and the upgrade function are managed externally, residing in an external contract bytecode. Notably, this final reference design was also identified and explained in \citeauthor{Bodell23}’s study, where it is referred to as the Registry Patterns. In the following sections, we provide a detailed explanation of each reference design.

\begin{figure}[!t]
\centering
  \includegraphics[width=0.95\columnwidth]{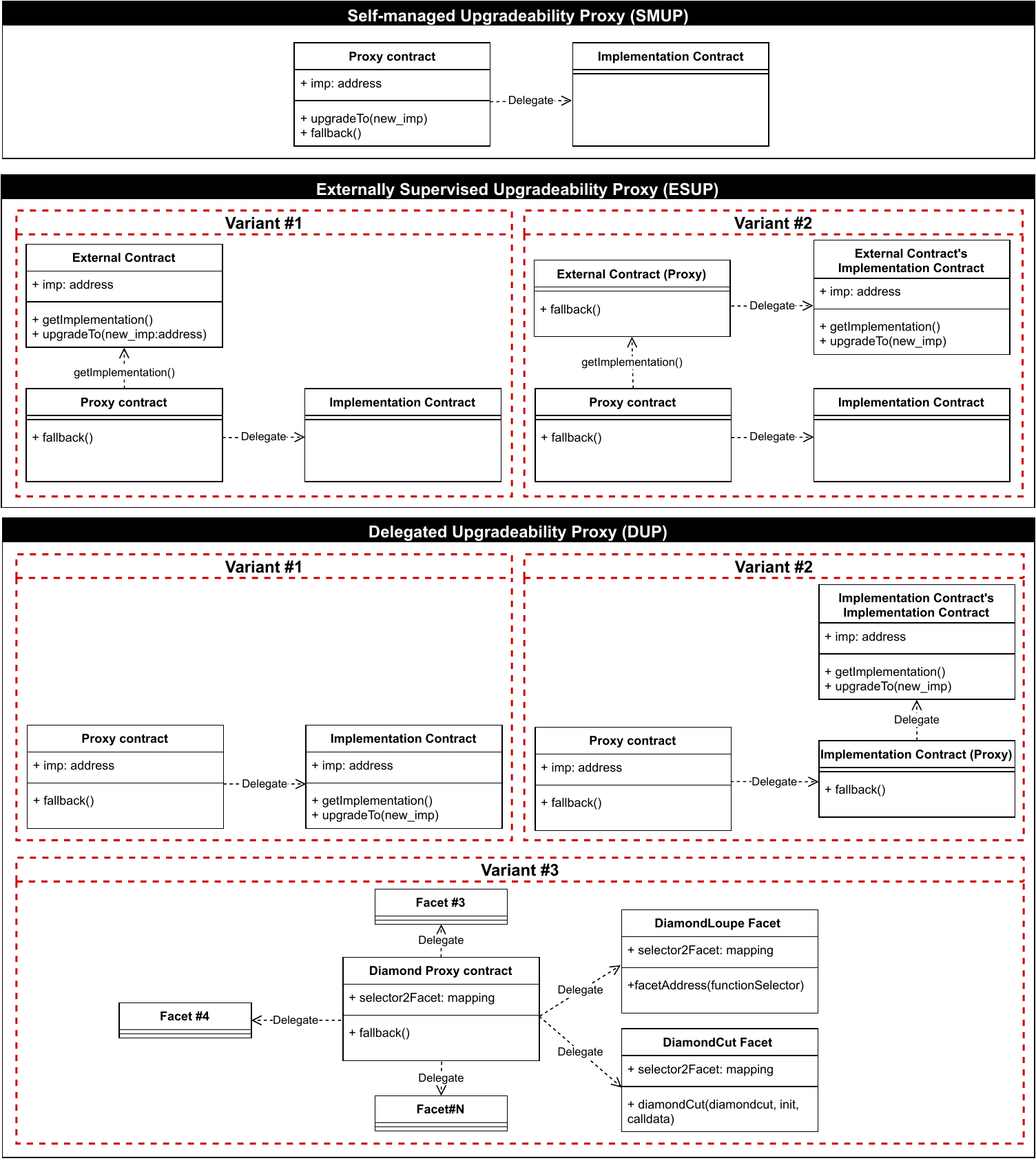}
  \caption{The structure of the three URDs and their variants. Note that all the used identifiers (e.g., function names or variable names) are examples, and in practice, one can use any arbitrary identifiers.}
  \label{fig:upgradeability-reference-designs}
\end{figure}

\paragraph{URD \#1: Self-managed Upgradeability Proxy (SMUP).} The first and most common reference design is when the proxy contract itself implements the upgradeability mechanism. Figure \ref{fig:upgradeability-reference-designs} shows a simple UML class diagram of this URD. In this case, the proxy contract implements at least one external or public function through which the old implementation contract address-stored in the proxy's storage-can be replaced with the new one. For instance, the Transparent Upgradeability Proxy, Inherent Upgradeability Proxy, Eternal Upgradeability Proxy, ERC-1967 Upgradeability Proxy (Unstructured Storage) are examples of this class. This type can be detected by running a static analyzer on the proxy contract's bytecode.

\paragraph{URD \#2: Externally Supervised Upgradeability Proxy (ESUP).}\label{urd-esup-def} The second reference design is when both the implementation variable and the upgradeability mechanism are located in an external contract that the proxy contract relies on. Figure \ref{fig:upgradeability-reference-designs} shows a simple UML class diagram of this URD. In this case, the proxy typically makes an inquiry to the external contract to obtain the latest implementation contract address before delegating calls to it. Beacon Upgradeability Proxy or Registry Upgradeability Proxy are two common examples of this class. To detect this type, a dynamic method is needed to first detect the external contract address. Once the external contract is identified, a static analyzer can assess its bytecode to validate whether it has implemented the upgradeability mechanism or not (variant \#1). In a more complex scenario, the external contract itself can be a proxy and the upgradeability mechanism is implemented within the external contract's implementation contracts (variant \#2). 

\paragraph{URD \#3: Delegated Upgradeability Proxy (DUP).} The third reference design is when the upgradeability mechanism is implemented by the proxy's implementation contracts. Figure \ref{fig:upgradeability-reference-designs} shows a simple UML class diagram of this URD. In this case, the proxy dedicates a specific slot in its storage that stores the implementation contract address. We briefly call this slot the \texttt{implementation slot}. The implementation contract also dedicates an exactly similar implementation slot in its own storage. Then, during the upgrade, the proxy delegates the upgrade operation to its implementation where the upgrade function is actually defined. Since the execution occurs in the context of the proxy during the delegatecall, upon updating the implementation slot within the upgrade function it indeed replaces the corresponding implementation slot defined in the proxy contract (see Section \ref{subsec:back-proxy-contract} for an example). ERC-1822 Universal Upgradeable Proxy, and ERC-2535 Diamonds Upgradeability Proxy are key patterns under this class. To detect this type, a reliable method is required to identify \textit{all} implementation contracts of the proxy. Subsequently, a static analyzer must individually examine their bytecodes to verify: (i) whether the implementation has allocated an identical implementation slot and (ii) whether it incorporates the upgradeability mechanism (variant \#1). Similar to the ESUP reference design, a more complex scenario is likely if the implementation contract is a proxy itself. In this variant, one needs to assess the proxy's implementation's implementation contracts (a.k.a., secondary dependencies) and repeat a similar detection procedure for indication of upgradeability (variant \#2). 

Regarding the ERC-2535 Diamonds Upgradeability Proxy, the Diamond proxy maintains a mapping between function selectors and the specific facet contract where they are implemented. Per the standard~\citep{Mudge_2020_diamonds, EIP-2535}, DiamondCutFacet is a facet that executes the upgrade function, commonly named “diamondCut” function, because the Diamond proxy communicates with its facets via delegatecall. However, if the Diamond has never transacted with DiamondCutFacet to conduct upgrades, it is not possible to detect this facet via regular transaction analysis. To address this, we propose a different dynamic method. In particular, DiamondLoupeFacet is another fundamental facet of this pattern that acts as a search engine for functionalities that Diamond's facets offer. Specifically, when one invokes the ``facetAddress'' function from the DiamondLoupeFacet with an input function selector, this facet yields the facet contract address where the specified function resides. Therefore, as we can determine the ``diamondCut'' function selector beforehand, its facet address (i.e., DiamondCutFacet) can be queried through the DiamondLoupeFacet. If DiamondLoupeFacet yields the DiamondCutFacet's address, then the upgrade function is identified, and the proxy is upgradeable. We classify this as the third variant of the DUP reference design due to the incorporation of this additional behavioral detection method (variant \#3).

\subsubsection{Major Components of UPC Sentinel}\label{subsec:major-components-of-upc-sentinel}
In this section, we provide a detailed explanation of the six components of the upgradeability detector layer depicted in Figure \ref{fig:upc-sentinel-architecture}.

\begin{figure}[t]
\centering
  \includegraphics[scale=0.4]{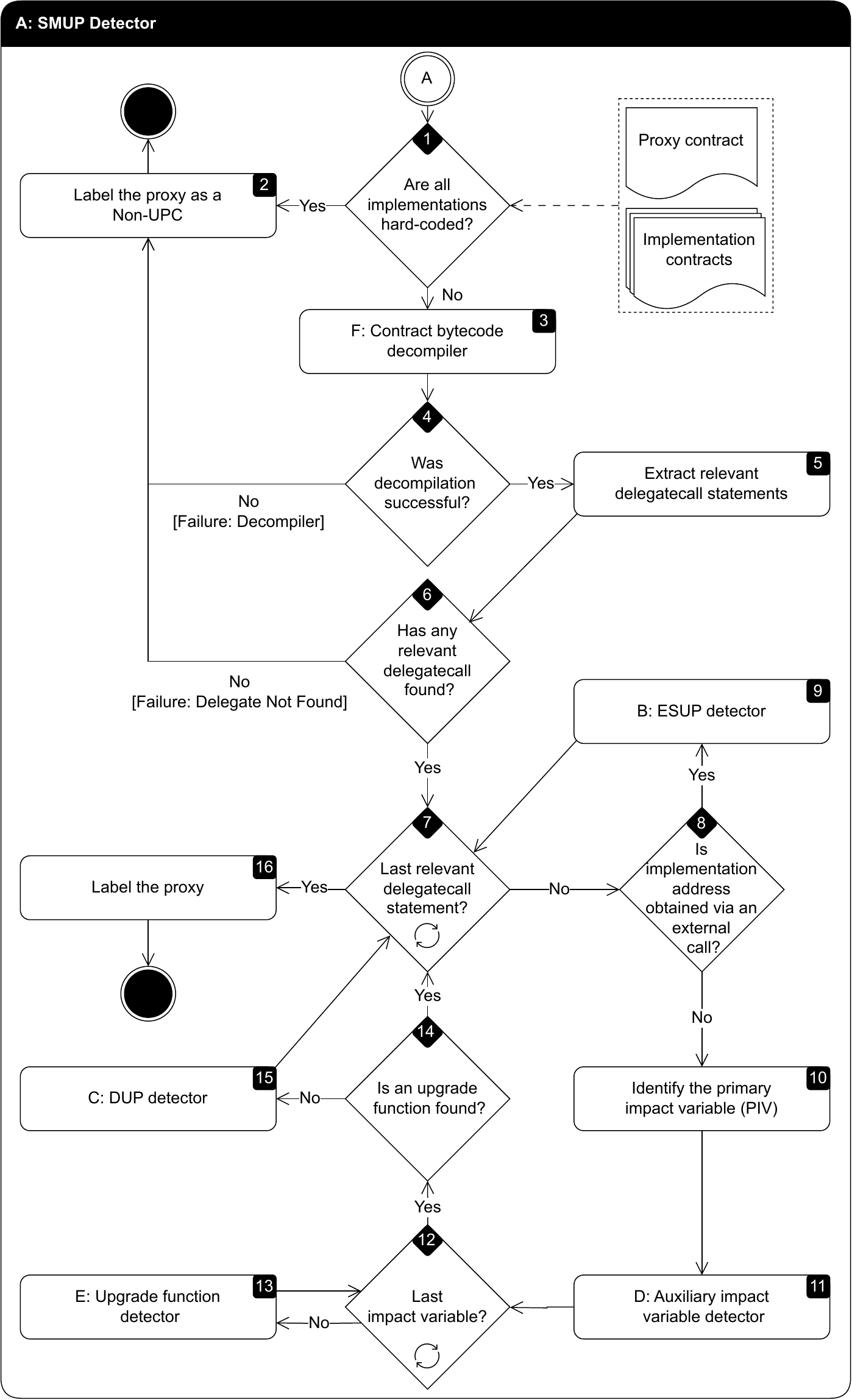}
  \caption{The detailed design of the ``A: SMUP Detector'' module.}
  \label{fig:A-smup-detector}
\end{figure}

\paragraph{A: SMUP Detector.}This module analyzes the proxy contract itself to find signs of upgradeability. Figure \ref{fig:A-smup-detector} depicts our algorithm. The module takes as inputs a proxy contract and all its implementation contracts detected by the proxy detector layer. To begin with, we examine all the implementation contracts to determine if their addresses are hard-coded within the proxy's bytecode. This is because if an implementation contract address is hard-coded in the proxy, it cannot be upgraded. For each implementation contract that is not hard-coded, we record its address for further investigation (A1). If all the implementations are hard-coded, the proxy is labeled as Non-UPC (a.k.a., forwarder). Otherwise, we proceed to the next step (A2). Next, we pass the proxy contract address to the ``F: Contract Bytecode Decompiler'' module, which retrieves the bytecode of the proxy and subsequently decompiles it using Panoramix decompiler~\citep{Palkeo_2019} (A3). After this, we check if Panoramix generated a decompiled source file. If not, we halt the process with the “Failure: Decompiler” flag, and label the proxy contract as Non-UPC. Otherwise, we proceed to the next step (A4). 

In the next step (A5), we identify where proxy functionality is implemented at the decompiled bytecode level, specifically by extracting all \texttt{relevant delegatecall} statements. As outlined in Section \ref{subsec:proxy-detector-layer}, proxy contracts possess two fundamental properties: Property \#1 and Property \#2. Given that bytecode is a machine-interpretable representation of source code, these properties should be discernible in the decompiled bytecode of any proxy contract. Therefore, every proxy contract must include at least one function that uses a delegatecall to process the calldata at the decompiled bytecode level. To validate this, we conducted an automated analysis of the decompiled bytecode from a statistically significant random sample of 16,603 active proxy contracts drawn as of September 2022 (with a 99\% confidence level and a 1\% confidence interval). We identified bytecode patterns corresponding to Properties \#1 and \#2 in 16,000 samples (99.98\%), with only 3 exceptions (0.02\%) corresponding to rare edge cases.

\begin{figure}[!t]
\centering
  \includegraphics[scale=0.4]{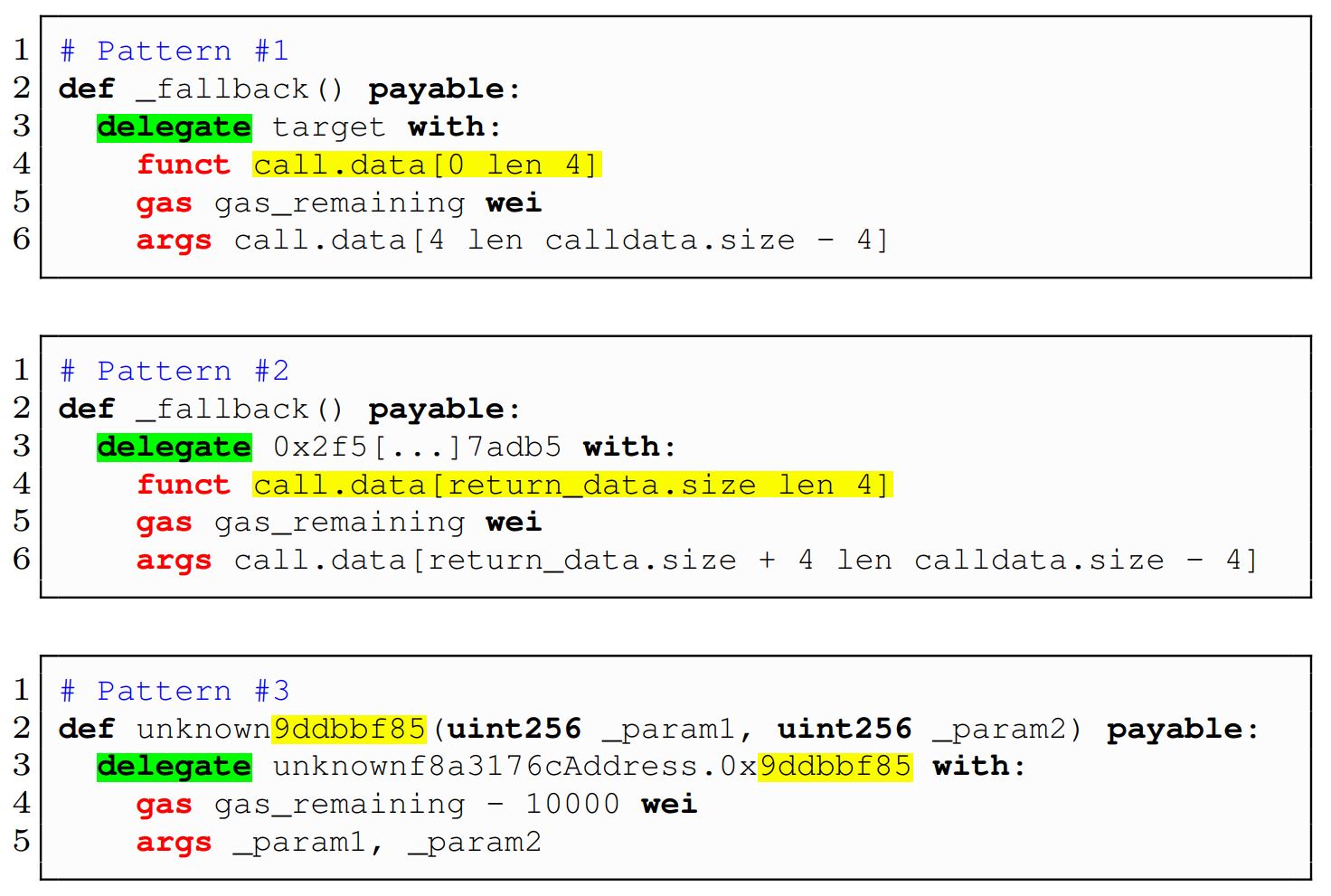}
  \caption{Three patterns that show the two conceptual properties of proxy contracts at the decompiled bytecode level.}
  \label{fig:pattern-concepual-property}
\end{figure}

Figure~\ref{fig:pattern-concepual-property} illustrates the bytecode patterns we identified for Property \#1 and Property \#2. For Property \#1 (highlighted in green), a proxy function must include a delegatecall statement within a function. For Property \#2 (highlighted in yellow), the selector of the implementation contract's function to which calls are delegated must either:
\begin{itemize}[label=\textbullet, itemsep = 3pt, topsep = 3pt]
    \item be \texttt{call.data[0 len 4]} or \texttt{call.data[return\textunderscore data.size len 4]}, indicating the use of the first four bytes of the calldata to determine the function to which calls must be delegated.
    \item alternatively, if both the proxy and implementation function selectors are identical and hard-coded, this suggests a specialized proxy function beyond the typical fallback implementation.
\end{itemize}

Using the extracted patterns from the above study, we next scan the proxy's decompiled bytecode to identify all relevant delegatecall statements, allowing us to extract the specific variable that stores the implementation contract address (A5).

Having analyzed the decompiled bytecode of a proxy contract, If we cannot locate any relevant delegatecall (i.e., proxy functionality), We halt the entire detection process process with “Failure: Delegate Not Found” flag, and label the proxy contract as Non-UPC (A6). Otherwise, we then loop over the set of all relevant delegatecall statements to examine them one by one through steps A8 to A15. The control flow of the loop is governed by a conditional check at A7, labeled ``Last relevant delegatecall statement?''. This condition ensures that the loop terminates once the last relevant delegatecall statement is processed, thereby preventing an infinite loop. Once all relevant delegatecall statements are evaluated, we proceed to step A16 where we categorize the proxy as UPC and Non-UPC based on the information collected in a global memory (A7). Specifically, our algorithm uses a global memory (akin to a memory scratchpad) to store information that is relevant to all three detectors. This global memory is shared across the A: SMUP, B: ESUP, and C: DUP detector which can update it through the ``E: Upgrade Function Detector'' module at steps A13, B13, and C9, respectively. More specifically, for each relevant delegatecall, a detector records the signature of an upgrade function that it identifies, its line number, the contract address where the upgrade functions are found in the global memory. 

For each relevant delegatecall, we first check how the implementation contract address is obtained. There are two possibilities. First, the implementation address can be obtained by an external call to another contract that stores this address (e.g., variant \#1 and variant \#2 of the ESUP design in Figure \ref{fig:upgradeability-reference-designs}). Second, this address can be stored inside the proxy storage (e.g., SMUP design in Figure \ref{fig:upgradeability-reference-designs}). In this case, a corresponding variable definition can be found inside the proxy storage, which is termed the \texttt{implementation variable}. In the former case, we proceed to A9, whereas in the latter case, we proceed to steps A10-A15 to examine if the implementation variable can be updated inside the proxy or its implementation contracts (A8). 

In the case where the implementation address is retrieved by an external call, we execute the ``B: ESUP Detector'' module to analyze the external dependencies of the proxy for an indication of upgradeability. After running this module for the current relevant delegatecall, we return to step A7 to examine the next relevant delegatecall.

In the case where the implementation variable is stored in the proxy, we parse the delegatecall statement to extract it. (e.g., see the ``imp'' variable in SMUP reference design in Figure \ref{fig:upgradeability-reference-designs}). This variable holds the address of the implementation contract. Therefore, if this variable is reassigned within a public or external function (e.g., see the ``upgradeTo'' function), it results in a change of the implementation contract address, thereby indicating an upgrade. We refer to the implementation variable as the \texttt{primary impact variable}. Thereafter, we extract the storage slot of the primary impact variable (a.k.a., the implementation slot) (A10). 

It is not always the case that the primary impact variable is directly replaced by a function's parameter at runtime. For example, developers may initially store the new implementation address in a pending implementation variable (e.g., as a proposal). Upon approval, such as through governance or voting mechanisms, the pending implementation variable replaces the primary impact variable. Consequently, the primary impact variable can also be substituted by other variables indirectly. These variables, which have the potential to trigger upgrades, are referred to as \texttt{auxiliary impact variables}. To extract such variables, the primary impact variable and its associated storage slot are passed to the ``D: Auxiliary Impact Variable Detector'' module. Both the primary and auxiliary variables are termed \textit{impact variables} because any modifications to these variables can result in an upgrade of the implementation contract (A11). We then process each impact variable (A12). For a given impact variable, we evaluate all the assignments in the proxy contract to identify if its value is replaced in a function. This process is carried out by the ``E: Upgrade Function Detector'' module. If we find an upgrade function, we record its function signature and proceed to the next impact variable to find all possible upgrade functions (A13). 

Having processed all impact variables, we then check if we found at least one upgrade function that replaces the value of one of these impact variables (A14). If yes, we proceed to A7 to evaluate the next delegatecall statement. Otherwise, it can be the case that the impact variables are replaced by one of the proxy's implementation contracts. In the latter case, we proceed to A15 where we analyze if the proxy complies with the DUP reference design by assessing the proxy's implementation contracts through the ``C: DUP Detector'' module. Subsequently, we proceed to A7 to evaluate the next relevant delegatecall statement. 

After processing all the relevant delegatecall statements, we check the global memory. If there is at least one relevant delegatecall statement for which an upgrade function is found through either of the detectors, we label the proxy as a UPC. Otherwise, it is labeled as a Non-UPC (A16).

\begin{figure}[!t]
\centering
  \includegraphics[scale=0.4]{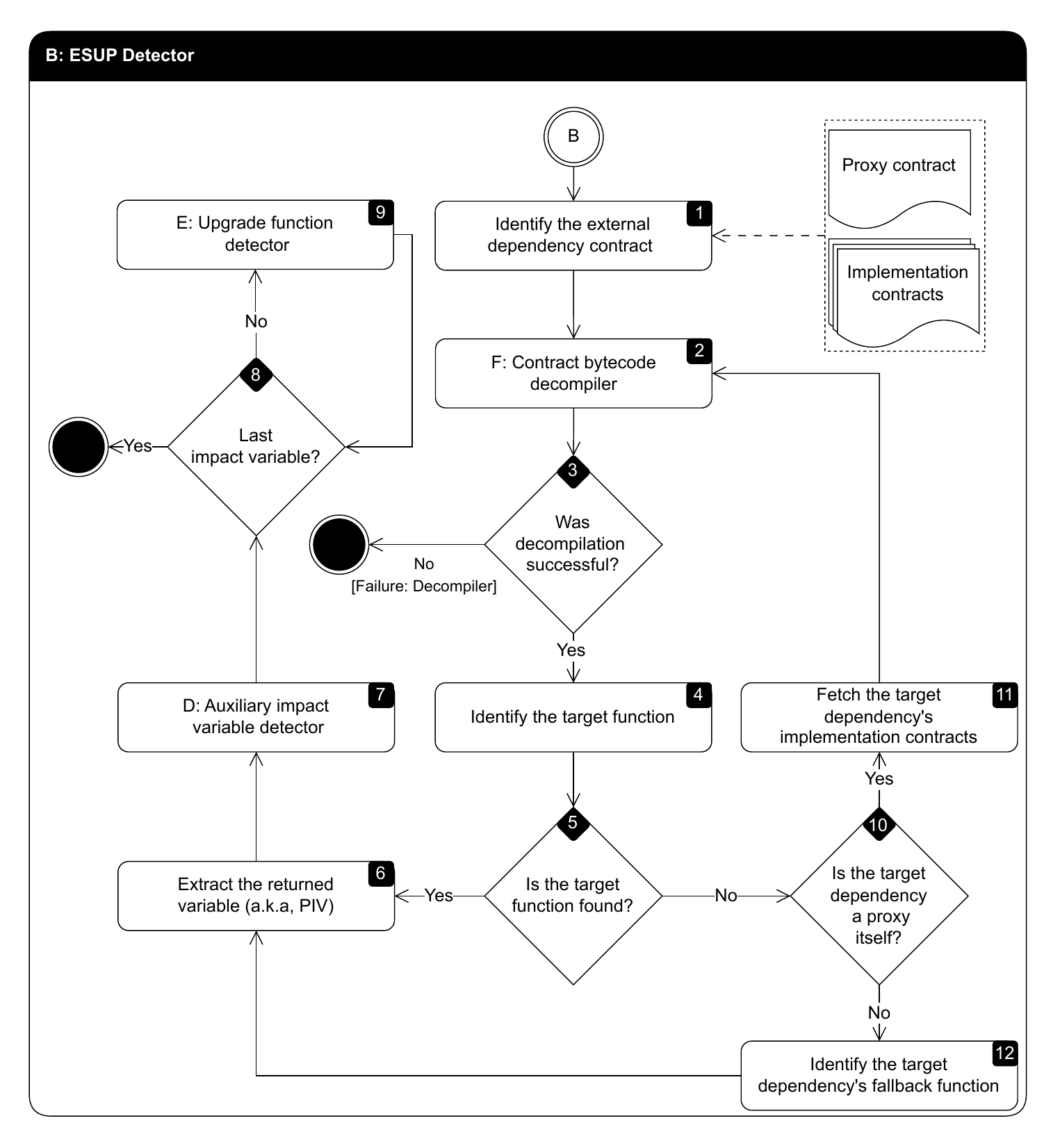}
  \caption{The detailed design of the ``ESUP Detector'' module.}
  \label{fig:B-esup-detector}
\end{figure}

\paragraph{B: ESUP Detector.}This module detects UPCs that are based on the Externally Supervised Upgradeability Proxy (ESUP) reference design (see variant \#1 and \#2 of the ESUP reference design in Figure \ref{fig:upgradeability-reference-designs}). In particular, if the address of the implementation contract is obtained by a call to another external contract (i.e., target dependency) prior to the delegatecall operation, then the target dependency can implement the upgradeability functions. Thus, the main objective of this module is to analyze the proxy's dependencies to identify indications of upgradeability. The module takes as inputs the proxy contract and its implementation contracts. Figure \ref{fig:B-esup-detector} depicts our algorithm.

The first step is to detect the target dependency, as a proxy may have several dependencies. the target dependency is a contracts that store the address of the implementation contract address, As mentioned earlier, prior to the delegatecall operation, the proxy makes an external call (e.g., ``getImplementation'' function call) to the target dependency contract to obtain the implementation contract address. Therefore, we conduct a dynamic analysis to identify the target dependency contract. More specifically, we analyze the proxy delegatecall transactions and retrieve the target trace that exactly precedes the delegatecall trace, and its output contains the implementation contract address to which the delegatecall trace is delegating. The recipient of the target trace is the target dependency contract that returns the implementation contract address (B1). We collect and decompile the target dependency contract bytecode (B2). If Panoramix cannot generate a decompiled source file, we halt the entire detection process with the “Failure: Decompiler” flag and return to step A7 of ``A: SMUP Detector'' and continue with the next relevant delegatecall statement (B3).

Once the target dependency's bytecode is decompiled, we search for a target function (e.g., the ``getImplementation'' function) that returns the implementation contract address, because the value returned by the target function indicates the implementation variable inside the target dependency contract. To find this function, we use the target trace's calldata. In particular, the first four bytes of the calldata show the function selector of the target function called from the target dependency contract. Then, we extract all the functions inside the target dependency's decompiled bytecode and subsequently compute their function selectors (B4).

We match the selector of the target function against all function selectors found in the target dependency contract to identify if the target function is implemented within the contract. If we find a match, we proceed to step B6. Otherwise, step B11 (B5). We extract the variable returned by the target function (a.k.a, the primary impact variable) (B6). We extract the auxiliary impact variables (B7). We then process every impact variable in step B9. Having processed all the impact variables, we return to step A7 of ``A: SMUP Detector'' (B8). For a given impact variable, we evaluate all assignments within the target dependency contract to identify if an upgrade takes place. If so, we store the signature of the function where the upgrade took place and proceed to the next impact variable (B9).

If we cannot find a match for the target function within the target dependency contract, there are two possible reasons for this. First, the target dependency contract itself can be a proxy contract that redirects the calls to its implementation contract where we can find the target function (i.e., the variant \#2 of the ESUP design in Figure \ref{fig:upgradeability-reference-designs}). Otherwise, if the target dependency is not a proxy, the target function is the fallback function of the target dependency because the fallback is the default function and gets triggered automatically when a function call cannot be resolved within the target dependency contract (B10). As mentioned in the second variant of the ESUP reference design, if the target dependency is a proxy contract, we identify its implementation contracts. Then, we recursively repeat the process - from B2 - but this time for the target dependency's implementation contracts (B11). However, if the target dependency is not a proxy, we extract its fallback function and treat it as the target function, and then continue from step B6 (B12).

\begin{figure}[!t]
\centering
  \includegraphics[scale=0.4]{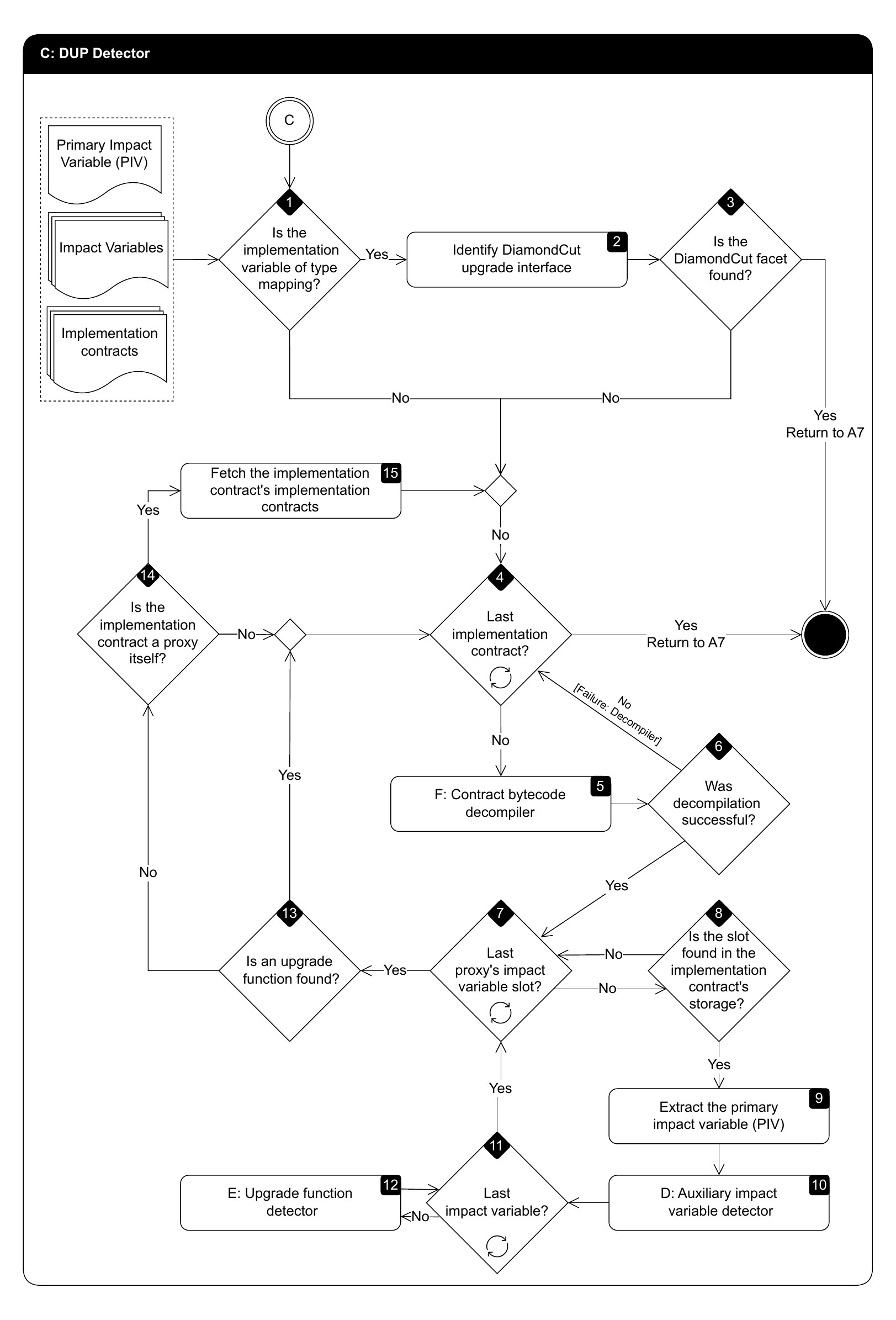}
  \caption{The detailed design of the ``C: DUP Detector'' module. Note that the small diamonds without a decision criteria are actually UML flowchart merge nodes.}
  \label{fig:C-dup-detector}
\end{figure}

\paragraph{C: DUP Detector.}This module detects UPCs that are based on the Delegated Upgradeability Proxy (DUP) reference design. In particular, if we cannot find an upgrade function for the set of impact variables found in the proxy contract (see A12), we need to analyze the proxy's implementation contracts to identify if an implementation contract implement a function that can update one of the impact variables. The input of this module are the proxy's implementation contracts, the primary impact variable, and the set of all impact variables and their storage slot fetched during the A10 and A11 steps. Figure \ref{fig:C-dup-detector} depicts our algorithm.

More specifically, we first examine the data type of (i.e., the primary impact variable found in the proxy contract). If it is not a mapping, we move to step C4. Otherwise, the proxy can be a Diamond proxy, so we advance to step C2 (i.e., the variant \#3 of the DUP reference in Figure \ref{fig:upgradeability-reference-designs}). In this step, we examined three renowned implementations~\citep{Mudgen} (i.e., diamond-1-hardhat, diamond-2-hardhat, and diamond-3-hardhat) of the Diamond proxy pattern and extracted the ``diamondCut'' function signatures, resulting in one unique function signature\footnote{diamondCut((address,uint8,bytes4[])[],address,bytes)} for this function. Using an Infura Ethereum node, we call the DiamondLoupe's ``facetAddress'' function with each diamondCut function selector (C2) as input. If the call yields a facet address other than 0x0, the proxy follows the Diamond Upgradeability Proxy pattern, and we then revert to the ``A: SMUP Detector'' process. If not, we move to step C4 (C3). In step C4, we analyze the proxy's implementation contracts one by one.

After processing all implementations, we return to the ``A: SMUP Detector'' process (C4). For each proxy's implementation contract, we gather and decompile its bytecode (C5). If Panoramix cannot generate a decompiled source file, we halt the process for the current implementation contract with the “Failure: Decompiler” flag, and continue with the next implementation contract (C6). Otherwise, we match every input impact variable's storage slot against the implementation contract storage. Specifically, for every proxy's impact variable storage slot (C7), if we identify a similar slot in the implementation contract's storage (C8), we record its variable name and its storage slot (a.k.a., the primary impact variable) (C9). If not, we continue with the next impact variable's storage slot. Steps C10, C11, and C12 mirror A11, A12, and A13, respectively. After analyzing all the impact variables, If an upgrade function is identified, we record it and continue with the next implementation contract. Otherwise, we transition to C14 (C13). 

The reason why we could not find an upgrade function within the implementation contract could be that the implementation itself is a proxy and indeed the upgrade function is implemented in one of the implementation's implementation contracts (i.e., the second variant of this reference design). Otherwise, we proceed with the next proxy's implementation contract (C14). If the former is true, we first fetch the implementation's implementation contracts (A15) and recursively repeat the process for each from A4. 

\begin{figure}[!t]
\centering
  \includegraphics[scale=0.4]{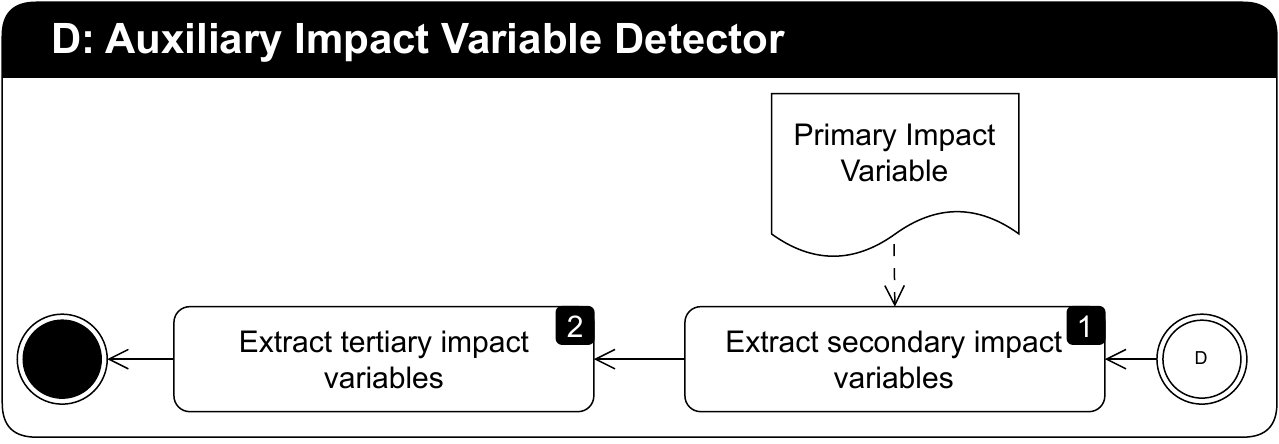}
  \caption{The detailed design of the ``D: Auxiliary Impact Variable Detector'' module.}
  \label{fig:D-auxilary-impact-variable-detector}
\end{figure}

\paragraph{D: Auxiliary Impact Variable Detector.} Once we detect the primary impact variable, we need to identify other auxiliary variables that affect the primary implementation variable. Figure \ref{fig:D-auxilary-impact-variable-detector} depicts our algorithm. More specifically, we categorized these variables into two categories, i.e., the secondary and tertiary impact variables. More specifically, we call an impact variable a secondary one if the variable storage slot is similar to the primary impact variable storage slot. For such variables, we extract the variable name and their storage slot. To identify secondary impact variables, we evaluate all the storage variables using the above condition (D1).

We call an impact variable a tertiary one if it is assigned to either a primary or secondary impact variable and it is found in the contract storage. For such variables, we extract the variable name and their storage slot. To identify these variables, we evaluate all the assignment expressions using the above two conditions (D2).

\begin{figure}[!t]
\centering
  \includegraphics[scale=0.4]{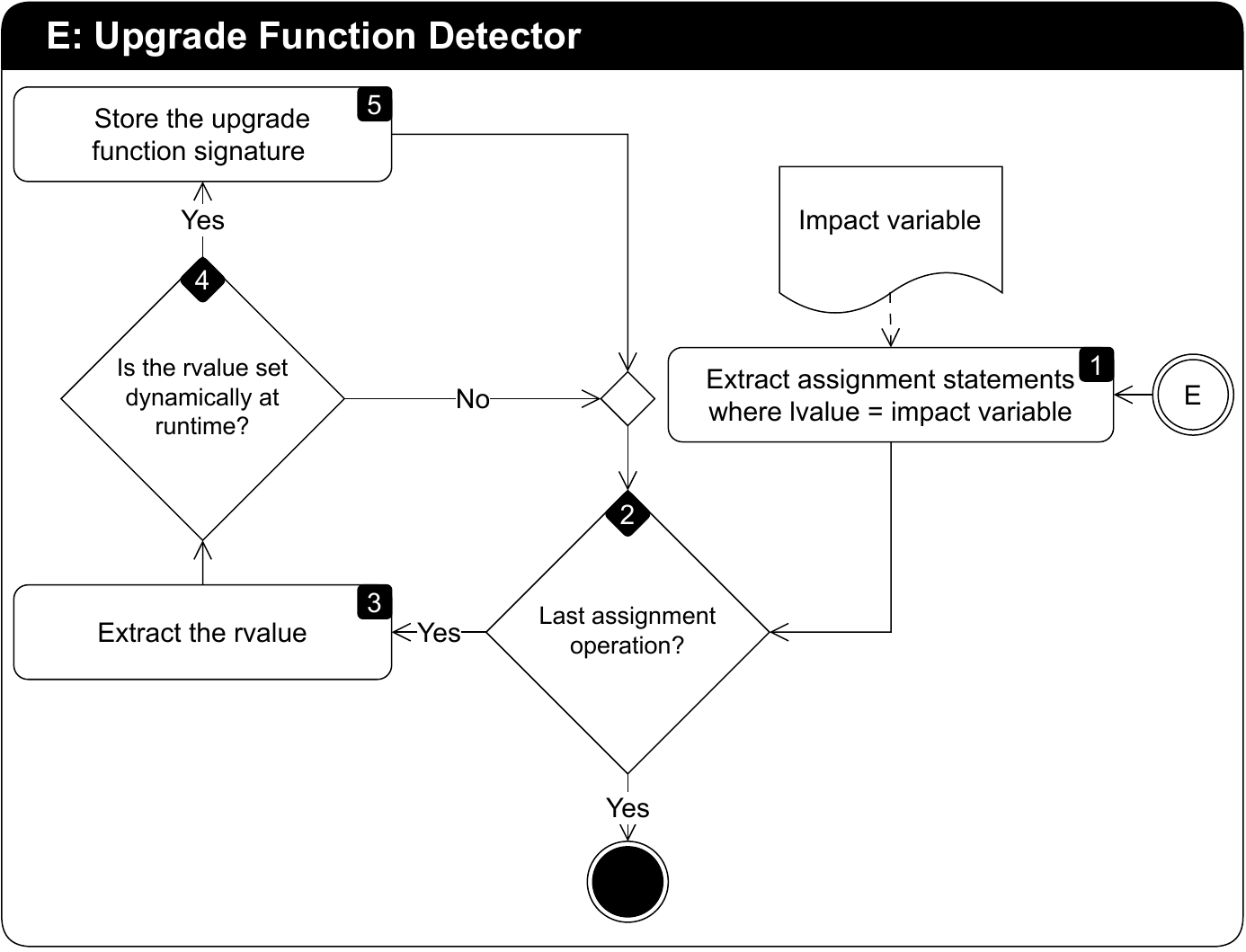}
  \caption{The detailed design of the ``E: Upgrade Function Detectors'' module.}
  \label{fig:E:upgrade-function-detector}
\end{figure}

\paragraph{E: Upgrade Function Detector.}The goal of this module is to detect the functions in which the value of an impact variable is replaced by a new value. The inputs of the module are an impact variable and a decompiled contract's source. Figure \ref{fig:E:upgrade-function-detector} depicts our algorithm. More specifically, we first retrieve all the assignments where the left operand (i.e., lvalue) is equal to the impact variable (E1). Then, we process every assignment (E2). For a given assignment, we extract its right operand (i.e., the rvalue) (E3), after which we check the rvalue source. If the rvalue is either a parameter of the function where the assignment occurs or obtained dynamically through a call to another contract, we detect an upgrade function. Otherwise, we proceed with the next assignment (E4). We store the signature of the upgrade function, its corresponding line of code, and the contract address where the upgrade function is found in the global memory. We then proceed with the next assignment to find all the possible upgrade functions (E5).

\begin{figure}[!t]
\centering
  \includegraphics[scale=0.4]{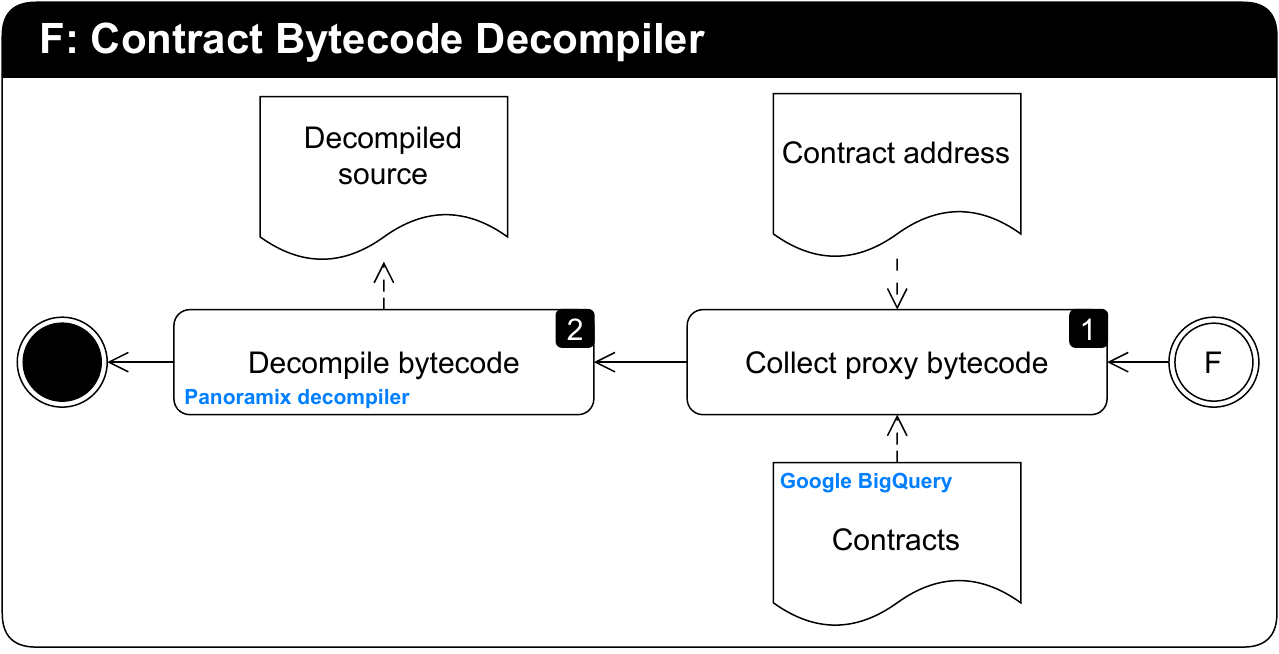}
  \caption{The detailed design of the ``F: Contract Bytecode Decompiler'' module.}
  \label{fig:F-contract-bytecode-detector}
\end{figure}

\paragraph{F: Contract Bytecode Decompiler.} The goal of this module is to collect and decompile a contract's bytecode (Figure \ref{fig:F-contract-bytecode-detector}). The input of this module is a contract address. From that address, we collect the bytecode of the given address by querying the contracts table from the BigQuery Ethereum public dataset and fetching the bytecode of the contract (F1). Next, we use the Panoramix decompiler (Appendix \ref{apx:panoramix-decompiler}) to decompile the bytecode (F2). 

\subsection{Upgradeability Pattern Classifier Layer}\label{subsec:upgradeability-pattern-classifier-layer}

Our upgradeability detector layer employs both static and dynamic techniques to determine whether a given contract is an upgradeability proxy contract (UPC) or not (Non-UPC). While this layer first identifies UPCs at a high level by determining the upgradeability reference design they adhere to (URDs), UPC Sentinel also further classifies a UPC by detecting the specific upgradeability proxy pattern used for its implementation (e.g., the Transparent Upgradeable Proxy Pattern, ERC-1822 Universal Upgradeable Proxy Pattern, or Beacon Upgradeability Proxy Patterns), since such fine-grained classification can provide added value in several key areas. 

First, understanding the specific upgradeability pattern enhances transparency, allowing users and stakeholders to verify how upgrades are handled and what control mechanisms are in place, ultimately fostering trust. Additionally, identifying the pattern enables automated analysis tools to apply targeted validation logic, improving the efficiency and accuracy of contract assessments. Lastly, recognizing the upgradeability pattern aids in compliance verification, by ensuring that UPC implementations align with recognized industry standards, thus supporting best practices and regulatory adherence.

As such, this additional Pattern Classifier Layer provides a fine-grained classification of the detected UPCs by identifying the specific type of upgradeability proxy pattern they are an instance of, at the decompiled bytecode level. In prior work,~\cite{Bodell23} characterized the syntactic features of different types of upgradeability proxy patterns at the source code level. While these syntactic features allow to identify how different patterns manifest at the source code level, the lower-level nature of bytecode and the linearization effects introduced during the compilation process necessitate a re-examination to determine whether and how specific patterns manifest at the decompiled bytecode level. For this, we also leveraged the specifications of upgradeability proxy patterns to identify key distinguishing features that enable us to detect these patterns at the decompiled bytecode level.

One primary linearization effect of the compilation process is that compilers often produce a flattened output, where the hierarchical parent-child relationships between contracts, typically defined through inheritance, are no longer preserved.  Instead, these relationships are replaced by a monolithic representation in which the inherited storage layout and logic are directly merged into the derived contract. As a result, the bytecode reflects a unified structure that does not explicitly indicate the original inheritance hierarchy, making it impossible for decompilers to accurately reconstruct the contracts’ original source-level architectural design or dependencies. This, for instance, has a direct effect on the detection of pattern types (i.e., Inherited, Eternal or EIP-1822 UUPS Upgradeability Proxy Patterns), all of which use inheritance at their core. This, for instance, has a direct effect on the detection of pattern types (i.e., Inherited Storage Upgradeability Proxy~\citep{OpenZeppelin_inherited_git}, Eternal Storage Upgradeability Proxy~\citep{OpenZeppelin_eternal_git}. or ERC-1822 Universal Upgradeable Proxy Patterns~\citep{eip_1822}), all of which use inheritance at their core. 

Furthermore, some storage related information can be lost at the decompiled bytecode level. Specifically, bytecodes do not directly embody the contract's storage layout, and decompilers like Panoramix have to infer the storage layout by interpreting \texttt{SSTORE} (i.e., write) and \texttt{SLOAD} (i.e., read) bytecode instructions. Therefore, if a variable—specifically those whose visibility is of type \texttt{internal}—has never been used in any computations (i.e., a read or write operation) across the code, the decompiler cannot detect its existence based on the bytecode instructions; thus, it cannot infer that variable in the storage layout of the decompiled contract. Yet, accurate storage layout at the decompiled bytecode level is critical for identification of certain patterns like Inherited and Eternal Upgradeability Proxy contracts, where the consistency of the storage layout between proxy and implementation contracts is a primary factor for correct identification. 

\begin{figure}[!t]
\centering
  \includegraphics[width=1.0\columnwidth]{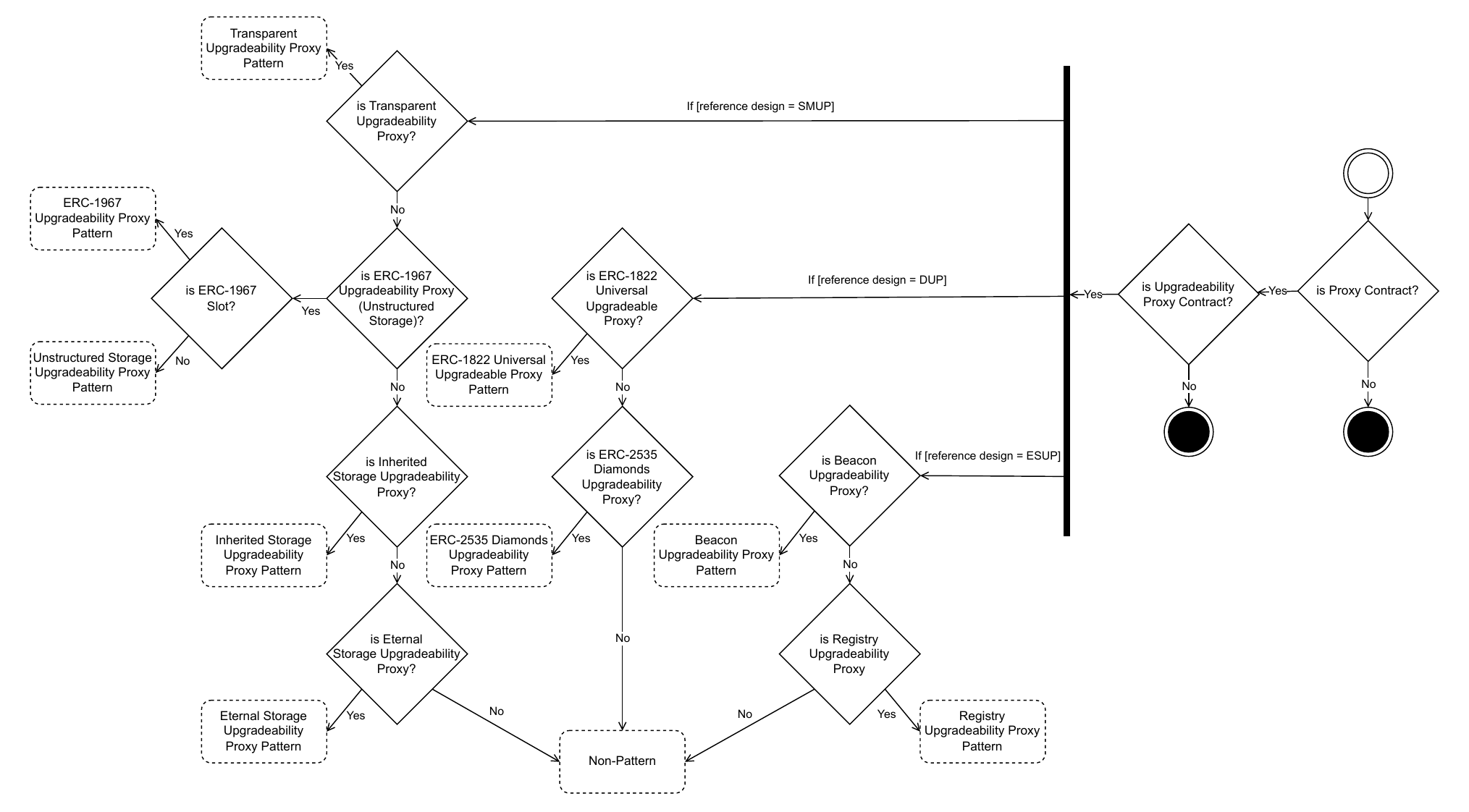}
  \caption{A systematic top-down approach for identifying the type of upgradeability proxy pattern that a UPC adheres to.}
  \label{fig:upgradeability-proxy-pattern-process}
\end{figure}

Figure~\ref{fig:upgradeability-proxy-pattern-process} outlines the decision-making process used to determine the specific upgradeability pattern a given UPC is an instance of. When classifying the upgradeability pattern of a given UPC, we follow a systematic approach. First, the upgradeability detector layer determines the upgradeability status and the reference design of the UPC. Next, we identify the most specific upgradeability pattern that the UPC conforms to within its reference design. We include a Non-pattern category to classify any UPC that does not adhere to recognized patterns or for cases where its type cannot be determined.

More specifically, based on our three reference upgradeability designs (see Section~\ref{subsec-deriving-urds}), a UPC can only be classified under one reference design at a time (i.e., either SMUP, DUP, or ESUP). This is because each reference design is based on a distinct upgradeability mechanism (see Section~\ref{subsec-deriving-urds}). Similarly, at the upgradeability pattern level, a proxy cannot be categorized under upgradeability patterns from multiple reference designs. For example, a proxy based on the ERC-1822 Universal Upgradeable Proxy Pattern cannot simultaneously be classified under the Inherited Storage Upgradeability Proxy Pattern. This is due to the fundamental difference in the underlying upgradeability mechanisms of patterns across different reference designs: in the former pattern, the implementation contract is responsible for handling the upgrade logic, whereas in the latter pattern, the proxy itself directly implements the upgrade functions.

Furthermore, of the four patterns in the SMUP upgradeability reference design, three patterns—the Inherited Storage Upgradeability Proxy, Eternal Storage Upgradeability Proxy, and ERC-1967 Upgradeability Proxy (Unstructured)—introduce fundamental principles for designing the storage layout of both proxy and its implementation contracts to minimize the risk of storage collisions~\citep{openzeppelin_2020_proxy_pattern, Bodell23}. Storage collisions occur when the proxy and its implementation contracts use the same storage slots for different purposes, potentially corrupting critical data such as the admin or implementation contract address. Importantly, storage collisions can occur in any proxy contract regardless of the specific upgradeability pattern employed.

Although the three aforementioned upgradeability patterns were initially created for designing SMUP-based UPCs, their \textit{storage layout principles} (i.e., Inherited Storage, Eternal Storage, and ERC-1967 (Unstructured) Storage) are widely adopted across other patterns to prevent storage collisions. For instance, either of the three storage layout principles can be used to design the storage layout of the Transparent Upgradeability Proxy Pattern (the fourth pattern in the SMUP reference design). In this case, we classify the UPC under the most specific pattern which is the Transparent Upgradeability Proxy Pattern because this pattern specifies an extra unique condition (i.e., a mechanism to control who can call the proxy functions) that differentiates it from the other three patterns. Similarly, ERC-1967 (Unstructured) Storage is frequently used for designing storage layouts in other proxy patterns, such as the ERC-1822 Universal Upgradeability Proxy Pattern or the Beacon Upgradeability Proxy Pattern. Hence, adopting these storage layout principles across other upgradeability patterns does not create a new reference design or upgradeability pattern; it simply applies proven strategies to avoid collisions. 

In the following sections, we detail the specific methods used to detect each pattern illustrated in Figure~\ref{fig:upgradeability-proxy-pattern-process} within its corresponding upgradeability reference design.

\subsubsection{Detecting Patterns under the SMUP Reference Design}
For any detected UPC conforming to the SMUP reference design, we evaluate it against the following types in the specified order: Transparent Upgradeability Proxy Pattern, ERC-1967 Upgradeability Proxy Pattern (Unstructured Storage), Inherited Storage Upgradeability Proxy Pattern, or Eternal Storage Upgradeability Proxy Pattern. Additionally, for any UPC that does not conform to the previously known patterns, we categorize it under the Non-pattern category. This typically occurs when the proxy's storage layout deviates from recognized standards.

\paragraph{Transparent Upgradeability Proxy Pattern.} This pattern resolves the issue of function clashing between a proxy and its logic contract by routing calls based on the caller’s address. If the caller is the proxy’s admin, the proxy handles the call internally; otherwise, it delegates the call to the logic contract \citep{palladino_2021, Bodell23}. This is a key aspect that distinguishes this pattern compared to Inherited, Eternal and ERC-1967 Upgradeability Proxy (Unstructured Storage) Patterns. Listing \ref{listing:pattern-transparent-erc-1967} shows this pattern at decompiled bytecode level for a contract selected from wild. To detect this, UPC sentinel analyzes the decompiled bytecode of the detected proxy function from its definition (Line 4) up to where the delegatecall is called (Line 7) to assess whether there is a conditional check (Line 5) on the caller of the proxy function before delegation. If we find one, we flag the UPC as an instance of the Transparent Upgradeability Proxy Pattern. Otherwise, the contract does not conform to this pattern.

\noindent
\begin{minipage}[!t]{\linewidth}
  \begin{lstlisting}[frame=single,language=Solidity, caption= An example illustrating how the Transparent Upgradeability Proxy routes calls based on the caller's address as demonstrated in lines 5 and 6., label={listing:pattern-transparent-erc-1967}]
def storage:
  stor3608 is addr at storage 0x360894a13ba1a3210667c828492db98dca3e2076cc3735a920a3ca505d382bbc
[...]
def _fallback() payable: # default function
  if caller == addr(storB531.field_0):
      revert with 0x8c[...]00, 32, 66, 0x74[...]65, mem[230 len 30]
  delegate uint256(stor3608.field_0) with:
     funct call.data[0 len 4]
     gas gas_remaining wei
     args call.data[4 len calldata.size - 4]
  if not delegate.return_code:
     revert with ext_call.return_data[0 len return_data.size]
  return ext_call.return_data[0 len return_data.size]
[...]
\end{lstlisting}
\end{minipage}

\paragraph{ERC-1967 Upgradeability Proxy Pattern (Unstructured Storage).} In this pattern, the implementation address is stored in a dedicated, fixed storage slot within the proxy contract, unlike the Inherited or Eternal Upgradeability Proxy Patterns, where storage slots are allocated sequentially. This dedicated slot is derived using a hashing mechanism to minimize the risk of collisions with the state variables of the logic contract. UPC Sentinel identifies this pattern by examining the storage slot of the implementation variable (i.e., the variable that stores the implementation contract address). If the storage slot is represented as a long hex string (a 32-byte storage slot), the UPC is classified as an instance of this pattern. Building on this, we further refine the classification by dividing this pattern into two subcategories.

The concept of using a hashing mechanism is a shared foundation between Unstructured Storage~\citep{OpenZeppelin_2018a_unstructured} and ERC-1967 Storage~\citep{eip1967}. The latter, however, represents an official Ethereum proposal that specifies a standardized slot for the implementation contract address. Given the significance of ERC-1967, we further divide this category into two subcategories: (i) ERC-1967 Upgradeability Proxy Pattern, identified when the implementation variable uses the exact slot specified by the ERC-1967 standard, (ii) Unstructured Storage Upgradeability Proxy Pattern, identified when a hashed slot other than the one defined by the ERC-1967 standard is used. Listing \ref{listing:pattern-transparent-erc-1967} illustrates an example of ERC-1967 storage slot used to store the implementation variable (\texttt{stor3608}), which is explicitly declared on line 2 following the storage keyword.

\paragraph{Inherited Storage Upgradeability Proxy Pattern.} The pattern eliminates storage clashes by introducing a shared storage structure inherited by both the proxy and the implementation contracts. Specifically, a dedicated contract defines and maintains the state variables required for upgradeability (e.g., the admin address or implementation contract address). By ensuring that both the proxy and the implementation contract inherit this shared storage structure as the \underline{first} base contract, the storage layout is consistently aligned. In addition to the shared storage, the implementation contract can optionally define its own state variables. However, to avoid conflicts, these variables must be declared after inheriting from the shared storage \citep{OpenZeppelin_inherited_git}. This ensures that the new storage slots defined by the implementation contract do not overlap with those used by the proxy. 

Given this, to detect this type at the decompiled bytecode level where inheritance relationships are flattened, when we examine the storage layout of the proxy with its implementation contracts, we must observe that the proxy's storage layout is a consistent subset of the implementation contract’s storage layout such that all state variables of the proxy appear at identical storage slots in the implementation contract, while any additional variables defined by the implementation contract occupy subsequent, non-overlapping slots. If so, we flag the UPC as an instance of the Inherited Storage Upgradeability Proxy Pattern. Otherwise, the contract does not conform to this pattern. For instance, lines 2 to 4 of Listing \ref{listing:pattern-inherited-proxy-storage} and Listing \ref{listing:pattern-inherited-imp-storage} demonstrate how the three upgradeability-related state variables from the proxy are consistently preserved in the implementation contract’s bytecode, demonstrating the alignment of shared storage and the correct allocation of additional variables in non-overlapping slots.

\noindent
\begin{minipage}[t]{0.47\textwidth}
  \begin{lstlisting}[frame=single, style=sidebyside, 
                      caption=An example of proxy contract's storage layout in the Inherited Storage Upgradeability Proxy pattern., 
                      label={listing:pattern-inherited-proxy-storage}]
def storage:
  owner is addr at storage 0
  adminAddress is addr at storage 1
  implementationAddress is addr at storage 2




[...]
  \end{lstlisting}
\end{minipage}
\hfill
\begin{minipage}[t]{0.47\textwidth}
  \begin{lstlisting}[frame=single, style=sidebyside, 
                      caption= As example of the implementation contract's storage layout in the Inherited Storage Upgradeability Proxy pattern., 
                      label={listing:pattern-inherited-imp-storage}]
def storage:
  owner is addr at storage 0
  adminAddress is addr at storage 1
  implementationAddress is addr at storage 2
  stor3 is mapping of uint8 at storage 3
  stor4 is mapping of uint8 at storage 4
  stor5 is mapping of uint8 at storage 5
  [...]
[...]

  \end{lstlisting}
\end{minipage}

\paragraph{Eternal Storage Upgradeability Proxy Pattern.} The pattern provides a mechanism to ensure consistent and immutable storage structures by centralizing all state variables, including predefined internal mapping variables (e.g., for integers, addresses, and strings), within a generic EternalStorage contract. A proxy’s implementation contracts inherit from this centralized contract, ensuring uniform access to the state variables. The proxy contract, must also inherit from EternalStorage as the first base contract to ensure alignment of its storage layout with that of the implementation contracts. Following this inheritance, the proxy also inherits from another contract, which introduces specific state variables related to upgradeability, such as the admin address and the implementation contract address~\citep{OpenZeppelin_eternal_git}.

This pattern exhibits distinct characteristics~\citep{OpenZeppelin_eternal_git}. First, the proxy and its implementation inherits from EternalStorage as the first base contract; thus, sharing consistent storage slots for variables of type mapping. Second, the proxy then allocates additional slots for upgradeability-related variables which are not shared with the implementation contract. Third, the implementation contract preserves the immutability of the storage layout by avoiding the declaration of new state variables. As opposed to the Inherited Storage Upgradeability Proxy Pattern, here the implementation’s storage layout is a consistent subset of the proxy contract’s storage layout.

A key feature of this pattern is its separation of concerns. The implementation contracts are responsible for managing updates to the mappings within EternalStorage, while the proxy contract focuses on ensuring that these updates take effect by securely delegating calls. Importantly, the proxy does not typically access the mapping variables directly through operations like reading or writing; instead, it delegates these tasks to the implementation contracts. In Solidity, storage slots are typically allocated sequentially for declared state variables, ensuring a predictable layout for upgrades. Decompilers, however, reconstruct the storage layout of contracts by analyzing bytecode opcodes (e.g., \texttt{SSTORE} and \texttt{SLOAD} for storage operations). When variables with internal visibility (i.e., the standard visibility choice for the mapping variable in this pattern) are declared but remain unused in computations or logic, their associated storage operation opcodes do not appear in the bytecode. Without these opcodes, decompilers cannot infer these variables in the reconstructed source code, even though they occupy storage slots. For the Eternal Storage Upgradeability Proxy Pattern, this means that internal mapping variables that are not accessed in the proxy or implementation contract will not appear in the storage layout of these contracts in their decompiled bytecode.

Listings \ref{listing:pattern-eternal-proxy-storage} and \ref{listing:pattern-eternal-imp-storage} show an example of the storage layout for the proxy and implementation contract of a UPC of this type. In this example, both the proxy and its implementation contracts inherit from the EternalStorage contract, where six mappings of different types are declared. Consequently, the compiler allocates the first six storage slots to these mapping. However, in the proxy contract, the first recognized variable slot is slot 6, as slots 0 through 5 are assigned to the mappings but are not inferred in the contract's decompiled bytecode. This occurs because none of the six mappings are referenced in any computations within the proxy’s source code. Similarly, Listing \ref{listing:pattern-eternal-imp-storage} illustrates the implementation contract, where mappings are defined at slots 0, 2, and 4, but slots 1, 3, and 5 remain unrecognized. Therefore, it is more challenging to examine the consistency between the proxy’s and implementation’s storage layout, an important aspect for identification of this pattern.

\noindent
\begin{minipage}[!t]{0.47\textwidth}
  \begin{lstlisting}[frame=single, style=sidebyside, 
                      caption=As example of the proxy contract's storage layout in the Eternal Storage Upgradeability Proxy pattern., 
                      label={listing:pattern-eternal-proxy-storage}]
def storage:
  upgradeabilityOwner is addr at storage 6
  version is uint256 at storage 7
  implementationAddress is addr at storage 8

[...]
  \end{lstlisting}
\end{minipage}
\hfill
\begin{minipage}[!t]{0.47\textwidth}
  \begin{lstlisting}[frame=single, style=sidebyside, 
                      caption=As example of the implementation contract's storage layout in the Eternal Storage Upgradeability Proxy pattern., 
                      label={listing:pattern-eternal-imp-storage}]
def storage:
 deployedABlock is mapping of uint256 at storage 0
 unknown871c0760 is mapping of addr at storage 2
 stor4 is mapping of uint8 at storage 4

[...]

  \end{lstlisting}
\end{minipage}

To deal with this issue, we rely on the design principles of the storage layout in this pattern~\citep{OpenZeppelin_eternal_git}. Particularly, as mentioned earlier, after inheriting from the EternalStorage, the proxy must also inherit from another contract where the upgradeability-related state variables are declared~\citep{OpenZeppelin_eternal_git}. These variables though will be inferred under the proxy storage because they are used by the proxy (see lines 2 to 4), occupying slots immediately following the last slot allocated to EternalStorage mappings. For example, in Listing 3, the variable upgradeabilityOwner begins at slot 6, following the last mapping variable at slot 5 (which is not inferred). Therefore, we first scan the proxy storage layout, and look for the first variable whose slot is an integer. This slot, termed the indicator slot, serves as a key marker. If the indicator slot is greater than zero (e.g., slot 6 on line 2), it means that the previous slots are likely allocated to internal EternalStorage’s mapping variables, but just not inferred. Given that the implementation contract must also first inherit from the EternalStorage and use its mappings, if all of its inferred state variables whose slot is smaller than the indicator slot are of type mapping, then it means the implementation contracts are inheriting from the EternalStorage, and the indicator slot is indeed the starting slot of the first upgradeability state variable within the proxy after the slot allocated to the non-inferred mapping variables. Thus, we flag the UPC type as an instance of the Eternal Storage Upgradeability Proxy Pattern. Otherwise, the UPC does not conform to this pattern as well, and we classify it under the Non-pattern category.

\subsubsection{Detecting Patterns under the DUP Reference Design}
For any detected UPC that follows the DUP reference design, we further classify it into one of the following categories: ERC-1822 Universal Upgradeable Proxy Pattern, ERC-2535 Diamonds Upgradeability Proxy Pattern.

\paragraph{ERC-1822 Universal Upgradeable Proxy Pattern.} According to the ERC-1822 standard, the proxy stores the implementation contract address in a specific predefined fixed slot, calculated in an unstructured manner. Additionally, every implementation contract must inherit from a contract, namely the Proxiable contract \citep{Bodell23, eip_1822}, which defines two key functions. The first function, updateCodeAddress, is an internal function that allows for the safe update of the implementation contract address. The second function, proxiableUUID, is a public or external view function that returns the storage slot where the implementation address is stored. A key distinguishing aspect of this pattern is the compatibility check performed during an upgrade via the updateCodeAddress function. More specifically, before updating the implementation address, updateCodeAddress verifies whether the new implementation contract supports the Proxiable interface and uses the same storage slot for the implementation address. This compatibility check is implemented through a call to proxiableUUID on the new implementation contract address, ensuring that the upgrade adheres to the required storage structure and maintains consistency.

Detecting this pattern at the decompiled bytecode level entails different measures. At the decompiled bytecode level, inheritance relationships are flattened and merged into a single continuous sequence of operations within the bytecode, obscuring the original contract structure and hierarchy. Therefore, we cannot simply check that the implementation contract inherits from the Proxiable interface. Yet, according to the standard~\citep{eip_1822}, the public upgrade function within the implementation contract must call the Proxiable’s updateCodeAddress function as part of its logic. Since the updateCodeAddress is an internal function (i.e., its interface is not exposed by the implementation contract), its logic as explained above will be copied under the implementation contract’s public upgrade function upon decompilation. Therefore, we detect this type by analyzing if the public upgrade function includes the logic of the proxiableUUID compatibility check as explained above. Listing~\ref{listing:pattern-erc-1822} presents an excerpt of the upgrade function from the implementation contract associated with a proxy of this type.

\noindent
\begin{minipage}[!t]{\linewidth}
  \begin{lstlisting}[frame=single,language=Solidity, caption= An example excerpt of the upgrade function from the implementation contract associated with an ERC-1822 Universal Upgradeable Proxy contract., label={listing:pattern-erc-1822}]

def unknown912a9885(addr _param1) payable: 
  [...]
  if not caller:
    revert with [...]
  [...]
  static call _param1.0x52d1902d with:
    gas gas_remaining wei
  [...]
  if ext_call.return_data[0] != 0xc5f16f0fcc639fa48a6947836d9850f504798523bf8c9a3a87d5876cf622bcf7:
    revert with 0, 'Not compatible'
  unknownabd108baAddress = _param1
  [...]
  
\end{lstlisting}
\end{minipage}

As shown, the function accepts a new address as input (line 2). To verify compatibility, the upgrade function calls the proxiableUUID function—identified by the selector \texttt{0x52d1902d}—on the provided address (line 7). It then compares the returned storage slot with the predefined fixed storage slot for the implementation contract's address (line 10). If these two checks are present within the upgrade function of the implementation contract, we flag the UPC as an instance of the ERC-1822 Upgradeable Proxy Pattern. Otherwise, the contract does not conform to this pattern. Additionally, details such as the location of the upgrade function(s) in the decompiled bytecode and the storage slot for the implementation contract's address are precomputed by the UPC Sentinel’s upgradeability detector layer during the UPC detection phase and stored in the global memory scratchpad (see Section~\ref{subsec:major-components-of-upc-sentinel}) for detection of this pattern.

\paragraph{ERC-2535 Diamonds Upgradeability Proxy Pattern.} We classify a UPC as an instance of this type if it precisely aligns with one of the three official implementations of this pattern (i.e., diamond-1-hardhat, diamond-2-hardhat, and diamond-3-hardhat) (see steps C1 to C3 in Figure~\ref{fig:C-dup-detector})~\citep{Mudge18}. Otherwise, the UPC does not conform to this pattern as well, and we classify it under the Non-pattern category.

\subsubsection{Detecting Patterns under the ESUP Reference Design}
For any detected UPC that follows the ESUP reference design, we further classify it into one of the following categories: Beacon Upgradeability Proxy Pattern and Registry Upgradeability Proxy Pattern.

\paragraph{Beacon Upgradeability Proxy Pattern.} This pattern relies on a defined standard where the BeaconProxy interacts with its Beacon contract via a specific interface. This interface includes a function implemented by the Beacon contract to return the implementation contract address it stores. In our approach, UPC Sentinel’s ESUP detector component verifies the presence of this function in the Beacon contract. As a result, any UPC conforming to the ESUP reference design inherently satisfies this condition. Additionally, the standard mandates that the Beacon contract stores the address of a single implementation contract, which serves as the key distinguishing feature compared to the Registry Upgradeability Proxy Pattern~\citep{beacon_proxy}. To confirm this, we analyze the type of the implementation variable (i.e., the variable holding the implementation contract address) that Beacon returns to the BeaconProxy. If the variable is of a type that stores a single address, we classify the UPC as an instance of this pattern. Otherwise, the contract is not considered to conform to this pattern.

\paragraph{Registry Upgradeability Proxy Pattern.} This pattern is very similar to the Beacon Upgradeability Proxy Pattern in terms of mechanism unless it allows to keep track of the different versions of the implementation contract address~\citep{Diligence_2017, Bodell23}. In other words, the Registry contract can store multiple addresses as opposed to the Beacon which only stores a single address. Therefore, we flag the UPC type as an instance of this pattern if the implementation variable exhibits multiplicity (e.g., array or mapping of addresses/objects)~\citep{Bodell23, Diligence_2017}. This multiplicity means that the contract can maintain a record of past and present implementation addresses, offering a versioning mechanism. Otherwise, the UPC does not conform to this pattern, and we classify it under the Non-pattern category.
    \section{Available Contract Data Sources}
\label{sec:data-collection}

Our approach and empirical study require data about Ethereum smart contracts and their associated transactional activities. We collected this data using Google BigQuery, which is a fully-managed, cloud-native data warehouse offered by the Google Cloud Platform. Google uses Ethereum ETL to extract and keep the \texttt{crypto\textunderscore ethereum}\savefootnote{ethereum_dataset}{bigquery-public-data.crypto\textunderscore ethereum} dataset up to date~\citep{GoogleEthereumETL}. Ethereum ETL is a tool designed to facilitate the extraction, transformation, and loading (ETL) of data from the Ethereum blockchain into a more accessible format for analysis~\citep{EthereumETL}. We also used the Infura\footnote{\href{https://www.infura.io/}{https://www.infura.io/}} node to interact with Ethereum smart contracts and retrieve on-chain data from Ethereum blockchain. Infura is a service offering developers scalable and reliable access to the Ethereum network without the need to run their own node.

More specifically, we extracted data from the \texttt{crypto\textunderscore ethereum} dataset, which records information about transactions, contracts, blocks, events, tokens and token holders, focusing on the extracted data from the \texttt{contracts} and \texttt{traces} tables. While the former contains information regarding the deployed smart contracts (e.g., address, bytecode, creation timestamp, etc.), the latter records detailed information about smart contract transactions (e.g., transaction hash, input, output, etc.). 

We used a snapshot of the mentioned tables encompassing data from Aug-07-2015 to Sep-01-2022, which aligns with the commencement of our experiments. Notably, the \texttt{contracts} table has records of 50,845,833 unique smart contracts. Meanwhile, the \texttt{traces} table provides details on 1,695,517,186 distinct transactions and includes 5,503,071,306 traces up to the specified data collection date.

% encompassing data as of Sep. 2022, which is the moment we started our experiments. Notably, the \texttt{contracts} table has records of XX unique smart contracts. Meanwhile, the \texttt{traces} table provides details on XX distinct transactions and includes XX traces up to the specified data collection date.

% We used a snapshot of the mentioned tables encompassing data from \hl{Jan. 2018}, as this is the time when upgradeability proxy patterns were officially recognized and widely adopted, up to \hl{Sep. 2022}, which is the moment we started our experiments. Notably, the \texttt{contracts} table has records of XX unique smart contracts. Meanwhile, the \texttt{traces} table provides details on XX distinct transactions and includes XX traces up to the specified data collection date.

% To conduct our empirical evaluation, we used two different ground truth datasets. In RQ1, we first collected a ground truth dataset (GE) of 3,177 contracts that are flagged as UPC by Etherscan (see Section \ref{subsec:eval-on-ge}), spanning from Jan. 2018 to Sep. 2022. To further assess and benchmark UPC Sentinel performance against findings from prior studies, we utilized another dataset (GB1) released by~\citet{Bodell23} including 994 labeled contracts (825 UPCs and 169 Non-UPCs) in RQ2, RQ3, and RQ4. Based on a set of preprocessing steps, we iteratively refined GB1 to produce GB2 and GB3 (see Sections \ref{subsec:eval-on-gb-proxy} and \ref{subsec:eval-upc-on-gb}).

To conduct our empirical evaluation, we used two distinct ground truth datasets. First, we compiled a dataset (GE) consisting of 3,177 contracts flagged as UPCs by Etherscan (see Section \ref{subsec:eval-on-ge}), covering the period from January 2018 to September 2022. We used the GE ground truth in Research Question 1 (see Section~\ref{subsec:eval-on-ge}). To further assess and benchmark the performance of UPC Sentinel against findings from prior studies, we utilized a second dataset (GB), released by~\citet{Bodell23}, which includes 994 labeled contracts (825 UPCs and 169 Non-UPCs) and was employed in Research 2 (see Section~\ref{subsec:eval-on-gb-proxy}), 3 (see Section~\ref{subsec:eval-upc-on-gb}) and 4 (see Section~\ref{subsec:eval-on-different-patterns}).

In the following sections, we examine the characteristics of the two ground truth datasets (i.e., GE and GB) and conduct systematic sanity checks to assess and validate their quality. Specifically, we follow established guidelines for analyzing data quality~\citep{Bhatia2024}, defining clear criteria for dataset modifications to ensure that each adjustment is both necessary and well-justified. Additionally, we adhere to general empirical software engineering principles by thoroughly documenting each modification, including its rationale and supporting evidence, to promote full transparency~\citep{Ralph2021}. Lastly, we have made all results and findings accessible through our replication package, facilitating easy access, thorough peer review, and reproducibility for future research.

\subsection{GE Ground Truth}\label{subsec:ge-groundtruth}
We carefully curated the ground truth data of UPCs and assessed UPC Sentinel against these data. Specifically, we employed our proxy detector method, detailed in Section \ref{subsec:proxy-detector-layer}, to extract all proxy contracts with at least two implementation contracts (see Section \ref{sec:data-collection}). We then queried each proxy contract on Etherscan. If Etherscan reported at least two implementation contracts for a specific proxy, we classified it as a UPC. This classification is based on the premise that having two implementations typically indicates an upgrade, transitioning from an older to a newer implementation. 

\begin{figure}[!t]
\centering
  \includegraphics[scale=0.75]{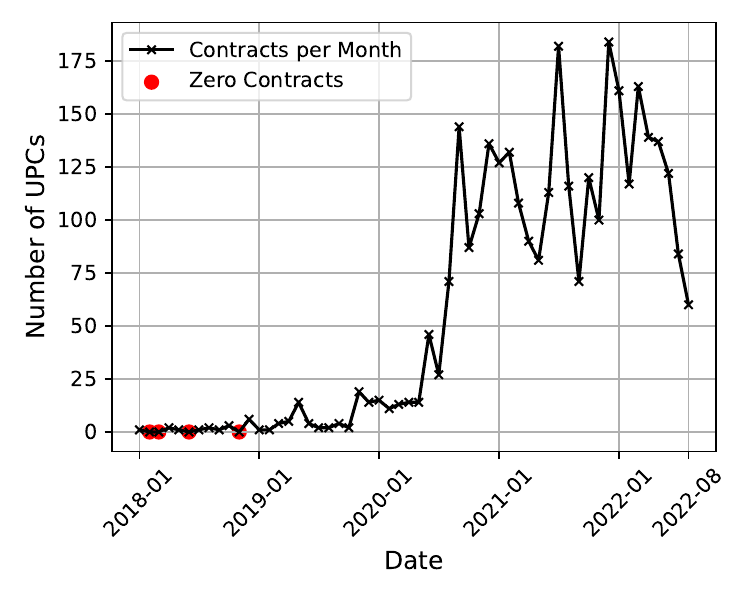}
  \caption{Monthly deployment of UPCs from our dataset of 3,177 UPCs collected from Etherscan.
}
  \label{fig:ge-upc-count-per-mont}
\end{figure}

Of the 31,481 distinct filtered proxy contracts, Etherscan reported 3,177 (10.1\%) as UPCs. We will refer to this ground truth as GE, which contains UPCs from Jan. 2018 (i.e., this is the time when upgradeability proxy patterns were officially recognized and widely adopted) to Sep. 2022. Figure \ref{fig:ge-upc-count-per-mont} illustrates the monthly distribution of collected UPCs. Notably, with the exception of four months marked by red dots, there has been at least one UPC recorded each month. The data reveals a steady escalation in the number of UPCs from 2018 to 2022. Initially, the growth rate was modest during the years 2018 to 2019. However, from 2020 onward, there was a noticeable acceleration in the frequency of UPCs, indicating an increasing trend in using UPCs.

\subsubsection{Mislabels}\label{subsec:mislabels-ge}
Mislabels refer to response variables that have been incorrectly labeled due to a flawed labeling strategy. This occurs when the assignment of labels does not align with the true underlying data or intended classification criteria, leading to systematic errors that can impact the accuracy and reliability of data-driven analyses and models~\citet{Bhatia2024}. Since the GE dataset may be influenced by the heuristics used in Etherscan's proxy verification service\footnote{\url{https://etherscan.io/proxyContractChecker}}, we took additional steps to ensure the accuracy of upgradeability labels by conducting a detailed review of the collected UPCs. However, given the labor-intensive process of manually validating the entire UPC dataset, we opted to verify a statistically representative random sample of contracts to maintain both accuracy and efficiency. The sample size (n = 94) was determined using Cochran's formula~\citep{Nanjundeswaraswamy21}, targeting a 95\% confidence level with a 10\% confidence interval. These standard parameters are widely accepted in empirical software engineering research to ensure the representativeness of the sample~\citep{Yang2022,Qiuyuan2021,Zhou2021,Zhou2020,Dayi2019b,Dayi2019a,Safwat2017,Safwat2018}.

For each contract, its associated metadata including its address, source code, implementation contract addresses, and decompiled bytecode was systematically reviewed. The first and third authors, each with significant experience in smart contract research (four years and six years, respectively), independently classified the 94 samples into UPC and Non-UPC categories. Leveraging established knowledge of upgradeability reference designs and proxy patterns, the authors (1) meticulously reviewing the source code of each contract, (2) identifying the proxy functionality, (3) locating and analyzing the implementation variable, and (4) evaluating whether an upgrade function was implemented to modify the value of the implementation variable. If both authors agreed that a sample contract was a UPC or Non-UPC, the contract was labeled accordingly. 

In cases where the authors disagreed (two cases), they engaged in discussions to reach a consensus. Disagreements in the final label arose from two scenarios: (i) inability to detect the proxy functionality where the proxy is implemented, and (ii) inability to detect an upgrade function, particularly if it is implemented outside the proxy contract, either in an external contract or an implementation contract. Given that in 92 out of 94 cases both authors agreed on the label, the observed proportionate agreement between the two authors was 97.98\%. After resolving these disagreements, our verification confirmed that all reviewed samples were indeed UPCs. This outcome reinforced our confidence in the accuracy of the collected ground truth dataset. For additional details about this analysis, please refer to our replication package\footnote{\href{https://github.com/SAILResearch/replication-24-amir-upc_sentinel/blob/88441d92c2d62e2ca96b8b6d3f290aadd0f8ac71/quality\%20checks\%20on\%20ground\%20truths/4.1.1\%20Mislabels\%20(Manual\%20Study\%20On\%2094\%20Instances\%20from\%20the\%20GE\%20Ground\%20Truth).xlsx}{GE Quality Check.xlsx}}. In the following we explain the disagreement cases. 

\smallskip\noindent\textbf{Case \#1:} Consider the case of ZeroEx\footnote{\href{https://etherscan.io/address/0xdef1c0ded9bec7f1a1670819833240f027b25eff\#code\#F1\#L1}{https://etherscan.io/address/0xdef1c0ded9bec7f1a1670819833240f027b25eff\#code\#F1\#L1}}. The proxy functionality of ZeroEx is implemented within its fallback function. The storage for ZeroEx is defined in a library called LibProxyStorage\footnote{\href{https://etherscan.io/address/0xdef1c0ded9bec7f1a1670819833240f027b25eff\#code\#F8\#L1}{https://etherscan.io/address/0xdef1c0ded9bec7f1a1670819833240f027b25eff\#code\#F8\#L1}}, which maintains a mapping between function selectors and the addresses of their corresponding implementation contracts. This structure gives developers greater control, enabling them to distribute roles and responsibilities (i.e., different services) across multiple implementation contracts, which can then be registered within the mapping for client access. When a ZeroEx client requests a service, ZeroEx first retrieves the appropriate implementation contract address from its mapping, then relays the call to that address. Our proxy detector identified 35 distinct implementation contracts that ZeroEx has delegated calls to over time. Below is a summary of the discussion between the two authors (i.e., the first and third authors of the manuscript) who conducted a manual analysis of this proxy.
\begin{itemize}[label=\textbullet, itemsep = 3pt, topsep = 3pt]
    \item Both authors confirmed that the ZeroEx proxy does not include a public or external function for the owner to perform upgrades, meaning it does not follow the SMUP reference design. They also agreed that the proxy contract retrieves the implementation address from its own storage, rather than from an external contract, meaning it does not follow the ESUP reference design. As a result, they began examining ZeroEx's implementation contracts to determine if the upgrade function is implemented there, assessing whether ZeroEx follows the DUP reference design.
     
    \item The third author identified the upgradeability functionality, while the first author did not, due to the complexity and high number (35) of ZeroEx implementation contracts. The third author noted that only one implementation contract, SimpleFunctionRegistry\footnote{\href{https://etherscan.io/address/0x843c1ae003a7217a122e9e2a7336377ee51b660f/advanced\#code\#F1\#L1}{https://etherscan.io/address/0x843c1ae003a7217a122e9e2a7336377ee51b660f/advanced\#code\#F1\#L1}} was responsible for release management operations, such as adding, removing, or rolling back a service in the proxy’s mapping, providing upgradeability for other implementation contracts within ZeroEx. He also observed that SimpleFunctionRegistry’s services are initialized during ZeroEx’s setup, thereby enabling release management from the outset. The first author, by contrast, was expecting each implementation contract to include upgradeability functions similar to those outlined in EIP-1822. Without this direct approach, he struggled to fully follow the third author’s reasoning and therefore initially flagged the contract as Non-UPC. 
    
    \item After discussing their differing interpretations, both authors agreed that ZeroEx is upgradeable. This highlights the need for a method that analyzes all implementation contracts linked to a proxy, particularly for UPCs following the DUP reference design. As shown in Step 4 of Figure 16, UPC Sentinel scans all implementation contracts associated with a proxy to identify these complex UPCs.
\end{itemize}

\smallskip\noindent\textbf{Case \#2:} Consider the case of BeaconProxy\footnote{\href{https://etherscan.io/address/0x87931e7ad81914e7898d07c68f145fc0a553d8fb\#code}{https://etherscan.io/address/0x87931e7ad81914e7898d07c68f145fc0a553d8fb\#code}}. The proxy functionality of BeaconProxy is handled within its fallback function, which first retrieves the address of the implementation contract and then relays the call to it. To obtain the implementation address, BeaconProxy calls the childImplementation getter function from an IBeacon contract, the address of which is stored in the proxy’s storage. Below is a summary of the discussion between the two authors (i.e., the first and third authors of the manuscript) who conducted a manual analysis of this proxy.

\begin{itemize}[label=\textbullet, itemsep = 3pt, topsep = 3pt]
    \item Both authors confirmed their understanding of the BeaconProxy and agreed that the implementation contract address is managed by an external contract, not the proxy itself, meaning BeaconProxy does not follow the SMUP or DUP reference designs. After examining BeaconProxy's transaction history and storage on Etherscan, they identified that the proxy calls an external contract, namely AdminUpgradeabilityProxy\footnote{\href{https://etherscan.io/address/0xbe86f647b167567525ccaafcd6f881f1ee558216\#code}{https://etherscan.io/address/0xbe86f647b167567525ccaafcd6f881f1ee558216\#code}} to fetch the implementation contract address in advance of delegatecall. 

    \item At this point, the first author was able to identify the upgradeability functionality in AdminUpgradeabilityProxy, whereas the third author was not. The first author initially searched AdminUpgradeabilityProxy’s source code to locate the childImplementation function but could not find it. Upon further investigation, he realized that AdminUpgradeabilityProxy is itself a proxy that delegates to another implementation contract, NFTXVaultFactoryUpgradeable\footnote{\href{https://etherscan.io/address/0xfa750439a60e385d9e5f3d25eb9db289b74b7062\#code}{https://etherscan.io/address/0xfa750439a60e385d9e5f3d25eb9db289b74b7062\#code}}. By examining NFTXVaultFactoryUpgradeable’s source code, he was able to locate both the childImplementation\footnote{\href{https://etherscan.io/address/0xfa750439a60e385d9e5f3d25eb9db289b74b7062\#readContract\#F2}{https://etherscan.io/address/0xfa750439a60e385d9e5f3d25eb9db289b74b7062\#readContract\#F2}} function and the upgradeChildTo\footnote{\href{https://etherscan.io/address/0xfa750439a60e385d9e5f3d25eb9db289b74b7062\#writeContract\#F17}{https://etherscan.io/address/0xfa750439a60e385d9e5f3d25eb9db289b74b7062\#writeContract\#F17}} function, which handles upgrades. In contrast, the third author, while following a similar procedure, overlooked the fact that AdminUpgradeabilityProxy is a proxy and stopped the analysis after failing to find childImplementation in its source code. 
    
    \item After discussion, both authors agreed that BeaconProxy is a UPC. This example illustrates the second variant of the ESUP reference design, where the external contract (the target dependency) is a proxy. In such cases, it is necessary to analyze the target dependency's implementation contracts to determine if it qualifies as a UPC. As shown in Step B10 of Figure~\ref{fig:B-esup-detector}, UPC Sentinel uses a recursive approach to examine the implementation contracts of the target dependency when it is identified as a proxy.

\end{itemize}

\subsection{GB Ground Truth}\label{subsec:gb-groundtruth}
As part of their replication package, ~\citet{Bodell23} released a ground truth dataset (GB) used in their experiments. Since this dataset forms the basis for our evaluations and comparisons in Sections \ref{subsec:eval-on-gb-proxy}, \ref{subsec:eval-upc-on-gb}, and \ref{subsec:eval-on-different-patterns}, we start by assessing the GB dataset using systematic guidelines for data quality analysis~\citep{Bhatia2024}. This ensures that our evaluations are conducted with precision, reliability, and fairness. According to~\citet{Bodell23}, the GB ground truth dataset comprises 994 proxy contracts, categorized into 826 UPCs and 168 Non-UPCs. In the following subsections, we provide a detailed discussion of each investigated quality criterion.

\subsubsection{Missing Values}\label{subsec:mising-values-gb}
Missing values refer to situations where certain data points are absent or not recorded in a dataset~\citep{Bhatia2024}. As part of their data collection,~\citet{Bodell23} manually added 25 contracts to their ground truth that notably lack associated Ethereum addresses, unlike the other entries. This absence of Ethereum addresses makes it impossible to verify whether these 25 contracts were sampled from live deployments or artificially created. Moreover, UPC Sentinel requires a valid Ethereum address to retrieve essential information such as a contract’s bytecode, transaction history, and complete set of implementation contracts. Consequently, we excluded these 25 instances from our analysis. After this step, the ground truth contains 969 instances. For transparency, we flagged these 25 cases in our replication package, which is available online\footnote{\href{https://github.com/SAILResearch/replication-24-amir-upc_sentinel/blob/7cb99786b6de89c725f7e3113be565316d544748/quality\%20checks\%20on\%20ground\%20truths/4.2.1\%20Mising\%20Values.xlsx}{Missing Values.xlsx}}.

\subsubsection{Schema Violations}\label{subsec:schema-violation-gb}
Schema violations occur when the data does not conform to the expected structure or constraints outlined in the schema~\citep{Bhatia2024}. \citet{Bodell23} reported that their ground truth dataset includes 994 labeled proxy contracts, implying that all labeled instances should predominantly align with the "proxy" category. However, our review uncovered eight instances labeled as "NOT EVEN A PROXY!". Additionally, these instances lacked predicted labels from USCHUNT. Due to these inconsistencies, we excluded these eight cases from our analysis. After this step, the ground truth contains 961 instances. For transparency, we have flagged these cases in our replication package, which is accessible online\footnote{\href{https://github.com/SAILResearch/replication-24-amir-upc_sentinel/blob/7cb99786b6de89c725f7e3113be565316d544748/quality\%20checks\%20on\%20ground\%20truths/4.2.2\%20Schema\%20Violations.xlsx}{Schema Violations.xlsx}}.

\subsubsection{Duplicates}\label{subsec:duplicate-gb}
A duplicate refers to repeated entries in a dataset where one or more data points are identical to others, potentially introducing bias~\citep{Bhatia2024}. As stated in Section~\ref{subsec:problem-domain} of the manuscript, the input of UPC Sentinel is just the address of the contract that one wishes to examine, as each contract address in Ethereum is unique and distinct, serving as an immutable identifier for a specific deployed smart contract. Yet, we identified 40 clusters of duplicate entries, each containing two entries, except for one cluster with three entries. 

To prevent duplicate bias in our findings, we retained only one instance per cluster and removed the duplicates. In 32 out of 40 clusters, the ground truth labels were consistent across both duplicates, hence we randomly selected one instance per cluster. In the remaining 8 clusters, we selected the instance whose ground truth label was reported as “y” (i.e., UPC) to avoid underestimating USCHUNT’s performance in the subsequent analyses. After this step, the ground truth contains 920 instances. For transparency, we flagged these 41 cases in our replication package, which is available online\footnote{\href{https://github.com/SAILResearch/replication-24-amir-upc_sentinel/blob/7cb99786b6de89c725f7e3113be565316d544748/quality\%20checks\%20on\%20ground\%20truths/4.2.3\%20Duplicates.xlsx}{Duplicates.xlsx}}.

\subsubsection{Mislabels}\label{subsec:mislables-gb}
Mislabels refer to response variables that have been incorrectly labeled due to a flawed labeling strategy ~\citep{Bhatia2024}. This occurs when the assignment of labels does not align with the true underlying data or intended classification criteria, leading to systematic errors that can impact the accuracy and reliability of data-driven analyses and models. Two measures for identifying and addressing mislabels are (i) Human Verification, where skilled labelers or third-party services review and correct errors; (ii) Programmatic Labeling, which employs algorithms or functions to systematically identify and relabel data points~\citep{Bhatia2024}.

In the following two subsections, we conduct two sanity checks to evaluate the accuracy of the ground truth in terms of correct proxy labeling and correct upgradeability proxy labeling, respectively. 

\paragraph{A) Sanity Checks of Correct Proxy Labeling.}\label{subsec:mislables-gb-proxy} 
In this process, we systematically verify the GB ground truth to ensure that the 920 included contracts are at a minimum legitimate proxy contracts. These checks combine multiple automated and effective methods to minimize errors, supplemented by manual verification to enhance confidence in the results. 

In Section~\ref{subsec:proxy-detector-layer}, we discussed the two inherent properties of proxy contracts, defined as Property \#1 and Property \#2, identified in prior studies. Subsequently, in~\ref{subsec:major-components-of-upc-sentinel}, we described how we mapped these properties to the decompiled bytecode level through an automated analysis of a statistically representative random sample of proxy contracts. As we observed these properties in 99.98\% of the sampled proxy contracts (with the missing 0.02\% corresponding to rare edge cases), we utilized these properties to systematically evaluate the ground truth contracts at the decompiled bytecode level. To ensure that none of the edge cases would be filtered out, this automated analysis was then followed by manual verification to enhance confidence in the results.  

Below, we provide a detailed explanation of these two verification steps. For each step, we report the observed proportion of agreement~\citep{McHugh2012} between the assigned labels and the ground truth labels. We will further discuss why Cohen’s Kappa was not applicable in this specific context in the threats to validity section (see Section \ref{subsec:threat-construct-validity}).

\begin{itemize}[label=\textbullet, itemsep = 3pt, topsep = 3pt]
    \item \textbf{A.1) Verifying Against Property \#1.} This check focuses on Property \#1, which is the presence of a delegatecall statement. Figure \ref{fig:A1-examine-contracts-against-the-first-property} illustrates the protocol we followed to systematically conduct this process. To mitigate potential risks related to decompilation issues, errors, or inaccuracies, we utilized two well-known online decompilers, Panoramix and Dedaub, to verify whether the ground truth contracts contained a delegatecall statement at the decompiled bytecode level (step 1). Upon analyzing the decompiled bytecode for the presence of the delegatecall statement, we observed that in 206 contracts, both decompilers unanimously did not generate a delegatecall statement at the decompiled bytecode level. Also, in 3 contracts, Dedaub generated a delegatecall statement that Panoramix did not. The delegatecalls that Dedaub generated are library calls, as Solidity uses delegatecall under the hood to handle library calls. However, library calls are not an indication of a proxy contract, and thus the output of Panoramix for these three cases remains accurate.
    
    \begin{figure}[!t]
    \centering
      \includegraphics[width=1.0\columnwidth]{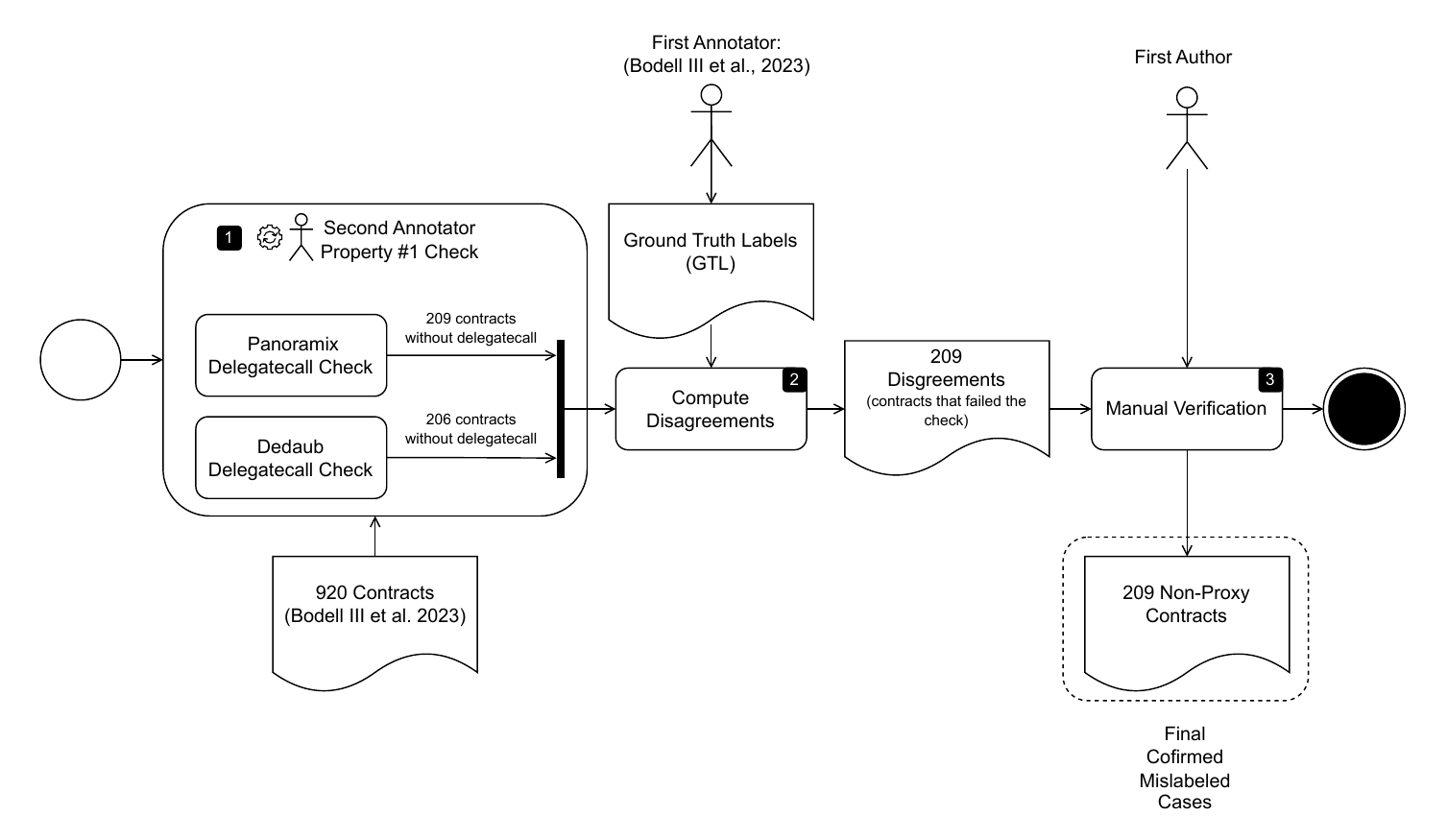}
      \caption{The protocol for examining the ground truth contracts against Property \#1 (i.e., the presence of delegatecall).}
      \label{fig:A1-examine-contracts-against-the-first-property}
    \end{figure}
    
    \hspace{2em}This observation suggested that 209 contracts do not meet Property \#1, indicating they physically cannot be proxy contracts (step 2). In 711 out of the 920 analyzed contracts (77.3\%), our automatic check (i.e., second annotator) aligned with the ground truth (i.e., first annotator), confirming that they satisfy Property \#1 and can potentially be proxy contracts. This unexpected finding prompted further investigation. To address the discrepancies between our automatically verified labels and the ground truth labels and to uncover the reasons behind this observation, the first author of this study—an expert with four years of experience in smart contract research—conducted a detailed analysis of the source code and UML class diagrams (both retrieved from Etherscan) for these 209 contracts (step 3). In all cases, the first author concurred with the findings of the two decompilers and identified a consistent explanation across all 209 instances. 
    
    \hspace{2em}This analysis shows that the verified source code of contracts from etherscan can be noisy, containing unreachable pieces of code, particularly unreachable proxy functionality, in all studied 209 cases. Since the proxy functionality is not reachable from the main contract, the solidity compiler does not generate any bytecode for it, which confirms why none of the employed decompilers generated a delegatecall statement in their reconstructed source code. In Section~\ref{subsec:lessons-learned}, we discuss the three patterns of unreachable proxy functions observed in the dataset, along with specific examples. Details of the remaining cases can be found in our replication package, available online\footnote{\href{https://github.com/SAILResearch/replication-24-amir-upc_sentinel/blob/7cb99786b6de89c725f7e3113be565316d544748/quality\%20checks\%20on\%20ground\%20truths/4.2.4.A1)\%20Verify\%20Against\%20Property\%20\%231.xlsx}{A.1) Verify Against Property \#1.xlsx}}.

    \item \textbf{A.2) Verifying Against Property \#2.} This check focuses on Property \#2, which is whether the proxy contract maintains the interface of the object it represents. Figure~\ref{fig:A1-examine-contracts-against-the-second-property} illustrates the protocol we followed to systematically conduct this verification process. Note that we no longer analyze the 209 non-proxy contracts from the first sanity check here, as they failed to satisfy Property \#1.

    \hspace{2em}As discussed in Section~\ref{subsec:major-components-of-upc-sentinel}, this property is satisfied when delegatecall uses the first four bytes of the calldata (e.g. \texttt{call.data[return\_data.size len 4]}, \texttt{call.data[0 len 4]} or, or when proxy and implementation function selectors are identical and hard-coded). We automatically analyzed the decompiled bytecode of the remaining 711 (920 - 209) contracts (step 1), finding that 698 out of the 711 analyzed contracts (98.2\%) satisfy Property \#2 and are indeed proxy contracts.

    \hspace{2em}To resolve the 13 disagreements between our automatically verified labels and the ground truth labels, the first author of this study conducted an additional manual review examining the verified source code and UML class diagrams for each of the 13 identified instances (step 3). Through this process, we observed that in four cases, our automatic sanity check of Property \#2 was incorrect, and Property \#2 was actually met; thus, the ground truth labels were correct. These four missed instances are all LendingPool\footnote{\href{https://etherscan.io/address/0x2b6eaf43499b3db50df84b22c9451fe04599317c\#code}{https://etherscan.io/address/0x2b6eaf43499b3db50df84b22c9451fe04599317c\#code}} contracts where the proxy functionality is implemented within the liquidationCall\footnote{\href{https://etherscan.io/address/0x2b6eaf43499b3db50df84b22c9451fe04599317c\#writeContract\#F5}{https://etherscan.io/address/0x2b6eaf43499b3db50df84b22c9451fe04599317c\#writeContract\#F5}} function. While we can visually confirm the existence of proxy functions at the source code level, detecting this specific case requires more advanced technique than our three lexical patterns at the decompiled bytecode level.

    \begin{figure}[!t]
    \centering
      \includegraphics[width=1.0\columnwidth]{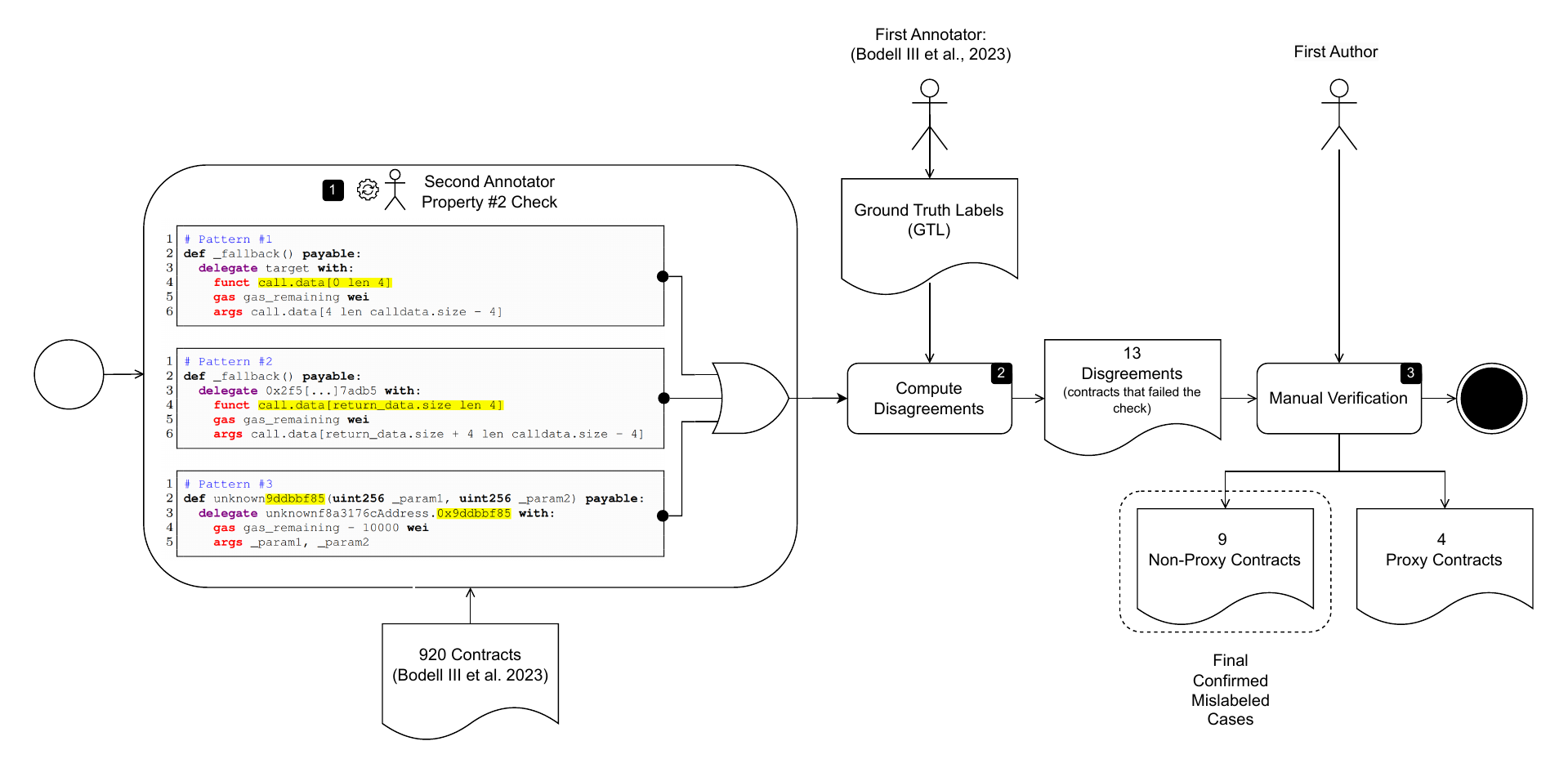}
      \caption{The protocol for examining the ground truth contracts against Property \#2 (proxy must maintain the interface of the object it represents). The three patterns illustrating this property are highlighted in yellow.}
      \label{fig:A1-examine-contracts-against-the-second-property}
    \end{figure}
        
    \hspace{2em} In the other nine cases, Property \#2 was indeed not met. Of these, eight cases involved a delegatecall related to a library call rather than a proxy, and in one case, the delegatecall was used to implement an adapter contract rather than a proxy. An interesting example involving a delegatecall corresponding to a library call is EthVaultImpl\footnote{\href{https://etherscan.io/address/0xa6f8b3ce684e37144745daccb3ef3596e8ee745a\#code}{https://etherscan.io/address/0xa6f8b3ce684e37144745daccb3ef3596e8ee745a\#code}}. EthVaultImpl inherits from EthVault, which includes a fallback function implementing proxy functionality. However, the EthVaultImpl’s fallback completely overrides its parent EthVault’s fallback, effectively nullifying the proxy functionality. As a result, EthVaultImpl cannot function as a proxy. The delegatecall we identified in the bytecode was related to one of EthVaultImpl's functions, specifically the withdraw\footnote{\href{https://etherscan.io/address/0xa6f8b3ce684e37144745daccb3ef3596e8ee745a\#writeContract\#F16}{https://etherscan.io/address/0xa6f8b3ce684e37144745daccb3ef3596e8ee745a\#writeContract\#F16}} function, which invoked the LibTokenManager library. In addition, Listing \ref{listing:adapter-contract-example} highlights the SiuCoin\footnote{\href{https://etherscan.io/address/0x1ff93b2d47d974644bddc62bf66940ebf18703e8\#code}{https://etherscan.io/address/0x1ff93b2d47d974644bddc62bf66940ebf18703e8\#code}} contract, which functions as an adapter contract rather than a proxy. Specifically, an adapter modifies the interface of the object it represents to ensure compatibility with the client. This is evident from the developers' self-admission in lines 2 to 5.

    \begin{quote}
    \textit{``This code allows us to redirect pre-signed calls with different function selectors to our own.''}
    \end{quote}
    
    \hspace{2em} Finally, line 9 demonstrates how the original function selector (msg.sig which is first four bytes of the calldata) is translated into a different selector before delegation. The details of these nine cases in our replication package, which is available online\footnote{\href{https://github.com/SAILResearch/replication-24-amir-upc_sentinel/blob/7cb99786b6de89c725f7e3113be565316d544748/quality\%20checks\%20on\%20ground\%20truths/4.2.4.A2)\%20Verify\%20Against\%20Property\%20\%232.xlsx}{A.2) Verify Against Property \#2.xlsx}}.

\noindent
\begin{minipage}{\linewidth}
  \begin{lstlisting}[frame=single,language=Solidity, caption= The SiuCoin’s fallback function that implements an adapter contract., label={listing:adapter-contract-example}]
[...]
/**
 * @dev This code allows us to redirect pre-signed calls with different 
 * function selectors to our own.
**/
function () public
{
    bytes memory calldata = msg.data;
    bytes4 new_selector = standardSigs[msg.sig];
    require(new_selector != 0);
    assembly {
       mstore(add(0x20, calldata), new_selector)
    }
    require(address(this).delegatecall(calldata));
    [...]
    }
}
[...]
\end{lstlisting}
\end{minipage}

\end{itemize}

After completing the above two verification steps (A.1 \& A.2), we identified that a total of 218 out of the 920 studied cases were not actually proxy contracts as initially labeled in the ground truth. Consequently, we changed the ground truth labels for these 218 cases to Non-Proxy, resulting in a refined ground truth dataset comprising 702 Proxy contracts and 218 Non-Proxy contracts.

\paragraph{B) Sanity Checks of Correct Upgradeability Labeling.}\label{subsec:mislables-gb-upc}In the previous sanity check (A), we identified 218 misclassified contracts that were originally labeled as proxy contracts by~\citet{Bodell23}. However, after thorough verification, these contracts were determined not to be proxy contracts. Since being a proxy is a prerequisite for being classified as an upgradeability proxy, none of these 218 instances could be UPCs in practice. Therefore, we reclassified these instances accordingly as Non-UPC. Out of these 218 instances, 157 were originally labeled as UPC, and 61 were labeled as Non-UPC by~\citet{Bodell23}. Hence, we switched the labels of these 157 misclassified UPC instances to Non-UPC and left the remaining 61 cases intact as they were already labeled correctly. After this adjustment, the refined ground truth consists of 621 UPC and 299 Non-UPC contracts.

\begin{figure}[!t]
\centering
  \includegraphics[width=1.0\columnwidth]{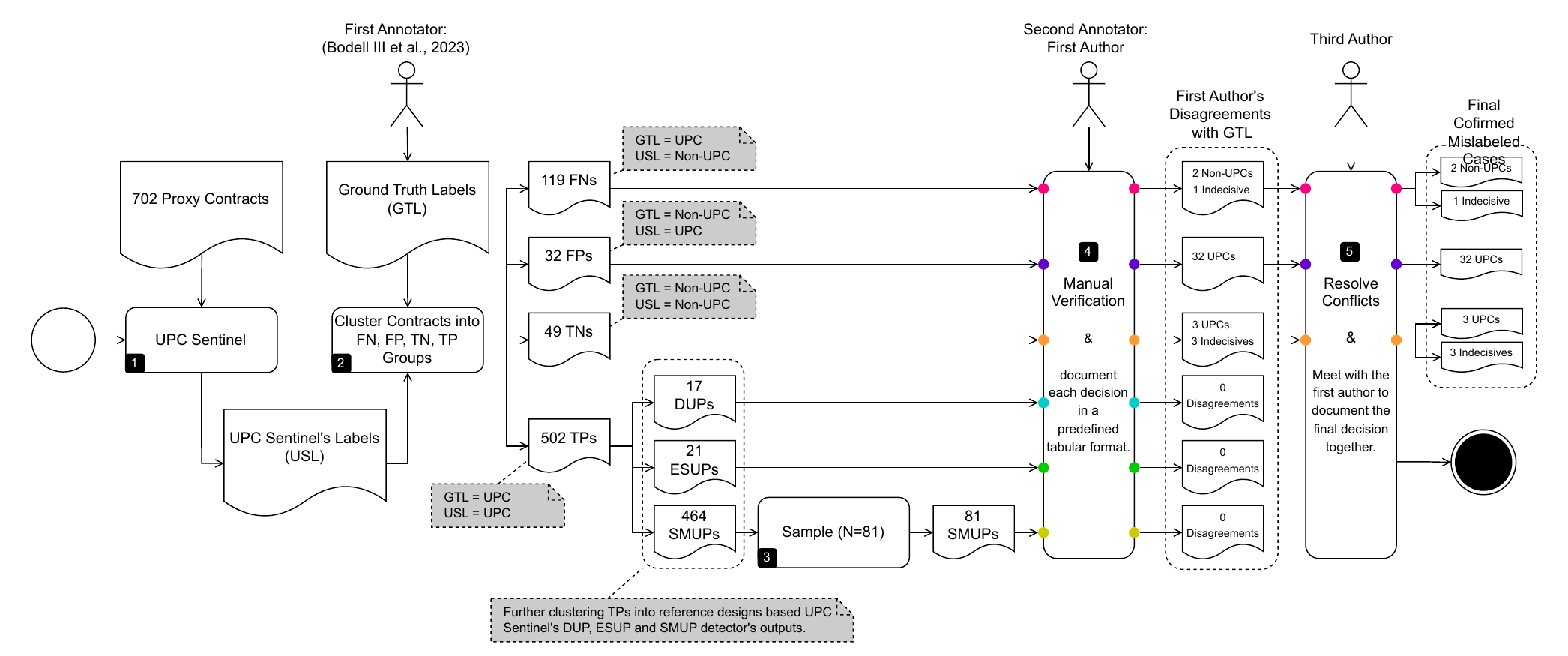}
  \caption{The protocol for examining the ground truth contracts against Property \#2 (proxy must maintain the interface of the object it represents). The three patterns illustrating this property are highlighted in yellow.}
  \label{fig:B-sanity-check-for-upgradeability-labels}
\end{figure}

Given that we are sure that 218 of the contracts are not UPCs, we now focus on manually verifying the upgradeability label of the remaining 702 (920 - 218) contracts, using the protocol shown in Figure~\ref{fig:B-sanity-check-for-upgradeability-labels}. Since verifying ground truth labels for all 702 contracts is a time-consuming and resource-intensive process, we applied UPC Sentinel to these contracts (step1) and clustered them into four groups based on UPC Sentinel’s False Positives (FPs), False Negatives (FNs), True Negatives (TNs) and True Positives (TPs). This clustering allowed for a targeted approach to validation and analysis. Upon applying UPC Sentinel to all 702 contracts, we observed the following distribution: 119 FNs, 32 FPs, 49 TNs, and 502 TPs (step2). 

For the FNs, FPs, TNs groups, we set to analyze all instances, while for the TPs group we opted to verify a sample due to the resource-intensive nature of verifying the entire set of 502 TP instances. To mitigate potential sampling bias, we first group TPs into DUP, ESUP and SMUP strata based on which UPC Sentinel’s detector categorized the contracts as UPC. Out of 502 TPs, 17, 21, and 464 instances adhere to the DUP, ESUP, and SMUP reference designs, respectively. Subsequently, we verified the ground truth labels for all 17 DUP and 21 ESUP instances, as well as a statistically representative random sample of 81 instances from the SMUP class (step 3), totaling 119 TP instances. The sample size for the SMUP class was determined using Cochran's formula~\citep{Nanjundeswaraswamy21}, aiming for a 95\% confidence level and a 10\% confidence interval. These standard parameters are widely accepted in empirical software engineering research to ensure the representativeness of the sample~\citep{Safwat2018,Dayi2019a,Dayi2019b,Zhou20, Zhou2020,ChenQiuyuan21,Yang22}.

For each contract, its associated metadata (source code, implementation contract addresses, transactions, and decompiled bytecode) are the input of the verification process. Then, based on knowledge from upgradeability reference designs and patterns (Section~\ref{subsec-deriving-urds} and Appendix~\ref{subsec:upgradeability-proxy-patterns}), the first author manually analyzed all 119 FNs, 32 FPs, 49 FNs and 119 sample TPs to examine whether they are UPCs or Non-UPC. Since some instances are inactive, if the first author is unable to comprehensively examine the status of the contract due to missing information (e.g., an uninitialized implementation contract or external contract address), the contract is labeled as \textit{indecisive} (step 4). If the label determined by the first author (i.e., second annotator) deviates (41 instances in total) from the ground truth label established by~\citet{Bodell23} (i.e., first annotator), the third author analyzes the contracts, makes the final judgment and labels the contract accordingly (step 5). 

In cases where we refined a ground truth label, both the first and third authors together document the findings in our replication package. Specifically, if a contract is confirmed as a UPC (35 out of 41 instances), we record the details in a systematic tabular format, which includes relevant information such as the specific contract's name and function where the proxy functionality is implemented, the implementation variable, the confirmed upgradeability reference design (i.e., SMUP, ESUP, or DUP) and the specific upgradeability pattern, the contract's name where the upgrade function is implemented, the line of code where the upgrade function begins, and, in the case of DUP or ESUP designs, the address of the implementation or external contract where the upgrade function is actually implemented. This ensures that our analyses are fully reproducible, allowing others to verify, replicate, and apply the same mechanism to identify the upgrade mechanisms for newly discovered UPCs. For indecisive cases (4 out of 41 instances) and those where the ground truth label was UPC but we found the contract to be Non-UPC (2 out of 41 instances), we have documented our detailed reasoning. 

The following outlines the specific details of the analysis conducted for each group, supplemented with concrete examples. For each group, we report the percentage of agreement~\citep{McHugh2012} between the first author’s assigned labels and the ground truth labels. We will further discuss why Cohen’s Kappa was not applicable in this specific context in the threats to validity section (Section~\ref{subsec:threat-construct-validity}).

\begin{itemize}[label=\textbullet, itemsep = 3pt, topsep = 3pt] 
    \item \textbf{B.1) Verifying False Negatives.}\label{bullet:sanitcheck.b.1} FNs are instances whose original ground truth label is UPC, but they are identified as Non-UPC by UPC Sentinel. In particular, out of all 119 FNs, in three instances the labels assigned by the first author deviate from the ground truth ``UPC'' label. Among these, two were labeled as Non-UPC and one as indecisive by the first author, with both classifications later confirmed by the third author. Thus, the observed proportion of agreement between the ground truth label and the first author is 97.5\%. The latter indecisive contract\footnote{\href{ https://etherscan.io/address/0x239a9de61b2d6b1dc9695562a336343e819571e4\#code}{ https://etherscan.io/address/0x239a9de61b2d6b1dc9695562a336343e819571e4\#code}} is an inactive proxy, and its implementation contract address\footnote{\href{ https://etherscan.io/address/0x239a9de61b2d6b1dc9695562a336343e819571e4\#readContract\#F1}{ https://etherscan.io/address/0x239a9de61b2d6b1dc9695562a336343e819571e4\#readContract\#F1}} was never initialized by developers. Consequently, we could not fully verify its upgradeability status to confirm whether the upgradeability logic is implemented inside the implementation contract. We revised the labels for the two instances from UPC to Non-UPC, and excluded the one indecisive instance from the ground truth. Following this adjustment, the refined ground truth now comprises 618 UPC contracts and 301 Non-UPC contracts. Detailed documentation supporting why these three cases are classified as such is included in our replication package online\footnote{\href{https://github.com/SAILResearch/replication-24-amir-upc_sentinel/blob/7cb99786b6de89c725f7e3113be565316d544748/quality\%20checks\%20on\%20ground\%20truths/4.2.4.B.1)\%20Verify\%20False\%20Negatives.xlsx}{B.1) Verifying False Negatives.xlsx}}.

    \item \textbf{B.2) Verifying False Positives.}\label{bullet:sanitcheck.b.2} FPs are instances whose original ground truth label is Non-UPC but are identified as UPC by UPC Sentinel. Upon analysis, the first author determined that all 32 FP instances are indeed UPCs. Thus, the observed proportion of agreement between the ground truth label and the first author is 0\%. Following our verification protocol, these 32 cases were subsequently analyzed by the third author and confirmed as UPCs. Thus, we revised the labels for these 32 instances from Non-UPC to UPC. 
    
    \hspace{2em} Following these adjustments, the refined ground truth now comprises 650 UPC contracts and 269 Non-UPC contracts. Detailed documentation supporting why these 32 cases are classified as UPCs is included in our replication package online\footnote{\href{https://github.com/SAILResearch/replication-24-amir-upc_sentinel/blob/7cb99786b6de89c725f7e3113be565316d544748/quality\%20checks\%20on\%20ground\%20truths/4.2.4.B.2)\%20Verify\%20False\%20Positives.xlsx}{B.2) Verify False Positives.xlsx}}. Specifically, among these, 27 conform to the DUP upgradeability reference design where the upgradeability mechanism was implemented within the proxy’s implementation contract, and five to the SMUP upgradeability reference design where upgradeability mechanism was indeed inside the proxy contracts. For example, the Proxy\footnote{\href{https://etherscan.io/address/0x5f6d994e6ba25a0a23efae15e46a6e79527bdf3f\#code}{https://etherscan.io/address/0x5f6d994e6ba25a0a23efae15e46a6e79527bdf3f\#code}} contract is an instance of the ERC-1822 Universal Upgradeability Proxy Pattern, with its implementation contract, TokenSoftTokenV2\footnote{\href{https://etherscan.io/address/0x7aac67b8cb7f39e080672ca6a32f5a6a964c26a4\#code}{https://etherscan.io/address/0x7aac67b8cb7f39e080672ca6a32f5a6a964c26a4\#code}}, containing an updateCodeAddress\footnote{\href{https://etherscan.io/address/0x7aac67b8cb7f39e080672ca6a32f5a6a964c26a4\#writeContract\#F34}{https://etherscan.io/address/0x7aac67b8cb7f39e080672ca6a32f5a6a964c26a4\#writeContract\#F34}} function that facilitates upgrades according to the ERC-1822 standard. As another example, the NFTFactoryProxy\footnote{\href{https://etherscan.io/address/0x8f9749054390b96cca7b5a41c7215cb6745daaee\#code}{https://etherscan.io/address/0x8f9749054390b96cca7b5a41c7215cb6745daaee\#code}} follows the Unstructured Storage Upgradeability Proxy Pattern, with its upgrade function, setLogicContract\footnote{\href{https://etherscan.io/address/0x8f9749054390b96cca7b5a41c7215cb6745daaee\#writeContract\#F1}{https://etherscan.io/address/0x8f9749054390b96cca7b5a41c7215cb6745daaee\#writeContract\#F1}}, implemented directly within the proxy contract. Consequently, we have updated the ground truth by reclassifying these 32 instances from Non-UPC to UPC. 
    
    \item \textbf{B.3) Verifying True Negatives.}\label{bullet:sanitcheck.b.3} TNs are instances whose original ground truth label is Non-UPC and are identified as Non-UPC by UPC Sentinel as well. In particular, out of all 49 FNs, in six instances the labels assigned by the first author deviate from the ground truth label. Thus, the observed proportion of agreement between the ground truth label and the first author is 87.7\%. As per our protocol, the third author then analyzed and resolved the disagreements. Out of these 6 instances, three were originally labeled as Non-UPC while they are indeed UPCs. For instance, BZxProxy\footnote{\href{https://etherscan.io/address/0x86343be63c60ce182d8b5ac6a84f0722d8d61ae5\#code}{https://etherscan.io/address/0x86343be63c60ce182d8b5ac6a84f0722d8d61ae5\#code}} is a SMUP-based UPC contract with an upgrade function, namely setTarget\footnote{\href{https://etherscan.io/address/0x86343be63c60ce182d8b5ac6a84f0722d8d61ae5\#writeContract\#F8}{https://etherscan.io/address/0x86343be63c60ce182d8b5ac6a84f0722d8d61ae5\#writeContract\#F8}} that enables replacing its implementation contracts. As another example, DutchExchangeProxy\footnote{\href{https://etherscan.io/address/0xe938e574afef9a75d8d6b6b5bd7dc44f3df1add9\#code}{https://etherscan.io/address/0xe938e574afef9a75d8d6b6b5bd7dc44f3df1add9\#code}} is a DUP-based UPC with its upgrade function, namely startMasterCopyCountdown\footnote{\href{https://etherscan.io/address/0x902B4b1882621e02893A1330aa42221939B27447\#writeContract\#F21}{https://etherscan.io/address/0x902B4b1882621e02893A1330aa42221939B27447\#writeContract\#F21}} implemented inside its implementation contract\footnote{\href{https://etherscan.io/address/0x902B4b1882621e02893A1330aa42221939B27447\#code}{https://etherscan.io/address/0x902B4b1882621e02893A1330aa42221939B27447\#code}}. We refined the ground truth reclassifying these three affected instances as UPC. The remaining three instances were confirmed to be indecisive for reasons similar to those discussed in Section B.1. Following these adjustments, the refined ground truth now comprises 653 UPC contracts and 263 Non-UPC contracts. For additional details about this analysis, please refer to our replication package\footnote{\href{https://github.com/SAILResearch/replication-24-amir-upc_sentinel/blob/7cb99786b6de89c725f7e3113be565316d544748/quality\%20checks\%20on\%20ground\%20truths/4.2.4.B.3)\%20Verify\%20True\%20Negatives.xlsx}{B.3) Verify True Negatives.xlsx}}.

    \item \textbf{B.4) Verifying True Positives.}\label{bullet:sanitcheck.b.4} TPs are instances whose original ground truth label is UPC and are also identified as UPC by UPC Sentinel.  Using our established verification protocol (Figure~\ref{fig:B-sanity-check-for-upgradeability-labels}), the first author manually reviewed all 119 contracts to confirm whether they are indeed UPCs (as suggested by the ground truth label). Upon analysis, the first author fully concurred with the ground truth, identifying all 119 cases as UPCs, resulting in an observed agreement rate of 100\%. Detailed documentation justifying the classification of these cases as UPC is included in our online replication package\footnote{\href{https://github.com/SAILResearch/replication-24-amir-upc_sentinel/blob/7cb99786b6de89c725f7e3113be565316d544748/quality\%20checks\%20on\%20ground\%20truths/4.2.4.B.4)\%20Verify\%20True\%20Positives.xlsx}{B.4) Verify True Positives.xlsx}}.

\end{itemize}

After completing the verification steps (B.[1-4]), the ground truth now comprises 653 confirmed UPCs and 263 confirmed Non-UPCs, totaling 916 contracts. Notably, as outlined in steps B.1 and B.3, four cases were excluded due to insufficient evidence or missing information, which prevented us from confirming their upgradeability status.
    \section{Empirical Evaluation of UPC Sentinel}
\label{sec:eval}

In this section, we present a comprehensive evaluation of UPC Sentinel. Research Question 1 (see Section \ref{subsec:eval-on-ge}) assesses the accuracy of UPC Sentinel on the GE ground truth dataset (see Section \ref{subsec:ge-groundtruth}), which contains 3,177 UPCs, focusing on its ability to accurately identify upgradeability proxy contracts. Research Questions 2 (see Section \ref{subsec:eval-on-gb-proxy}), 3 (see Section \ref{subsec:eval-upc-on-gb}), and 4 (see Section \ref{subsec:eval-on-different-patterns}) evaluate the performance of UPC Sentinel in comparison to the USCHUNT method proposed by~\citet{Bodell23}, using the refined GB ground truth dataset (see Section~\ref{subsec:gb-groundtruth}). The authors of USCHUNT provided a replication package containing the GB ground truth dataset and the predicted labels generated by USCHUNT for each data point, which we use to compute USCHUNT's performance across experiments.

\subsection{RQ1: How Accurate is UPC Sentinel in Detecting UPCs When Evaluated Against a Ground Truth Dataset of Confirmed UPCs?} \label{subsec:eval-on-ge}

\subsubsection{Motivation}
In Section~\ref{subsec:ge-groundtruth}, we introduced a novel compiled ground truth dataset, GE, which contains 3,177 confirmed UPCs. This research question aims to evaluate the accuracy of UPC Sentinel using this dataset.

\subsubsection{Experimental Setup}\label{subsec:rq1-exp-setup}
We evaluate UPC Sentinel on the GE ground truth (see Section \ref{subsec:ge-groundtruth}) and report its performance using the accuracy metric. Accuracy, in this context, is defined as the proportion of correctly identified UPCs out of the total positive cases. Since all 3,177 smart contracts collected as part of our data collection are UPCs, every instance represents a positive case. Consequently, in this specific context (similar to ~\citet{Bodell23}), accuracy is synonymous with the recall metric, as there are no False Positives or True Negatives to consider. 

\subsubsection{Findings \& Error Analysis} 

\observation{UPC Sentinel exhibits an accuracy of 99\% on the GE ground truth.}Out of the total 3,177 UPCs, UPC Sentinel successfully detected 3,147 UPCs. However, it failed to detect 30 UPCs. Analyzing the intermediate outputs of UPC Sentinel reveals that out of the 3,147 detected cases, 2,540 (80.7\%), 328 (10.4\%), and 279 (8.9\%) align with the SMUP, DUP, and ESUP reference designs, respectively. Additionally, 79 cases (2.5\%) out of the 3,147 detected represent Variant \#2 of the ESUP or DUP reference designs, where the upgrade functions were located in transitive dependencies (see Section~\ref{subsec-deriving-urds}). For a detailed example, please see Section~\ref{subsec:mislabels-ge}.

\textit{Error analysis. Out of these 30 undetected cases, UPC Sentinel failed to detect the type of 22 due to decompilation issues. For the remaining eight instances, UPC Sentinel was unable to classify them.}

\subsection{RQ2: How Does the Performance of UPC Sentinel’s Proxy Detector Compare to that of the USCHUNT Method Proposed by \texorpdfstring{\citet{Bodell23}}{Bodell et al. (2023)?}} \label{subsec:eval-on-gb-proxy}

\subsubsection{Motivation} The initial step in identifying a UPC involves determining whether a given contract is a proxy contract. Therefore, the accuracy of this task directly impacts the effectiveness of subsequent tasks, such as detecting UPCs. The aim of this research question is to evaluate and compare the performance of UPC Sentinel's proxy detector against USCHUNT, a state of the art technique in detecting upgradeability proxy contracts.

\subsubsection{Experimental Setup} \label{subsec:rq2-exp-setup}
For this case study, we utilize the refined GB ground truth (see Section~\ref{subsec:gb-groundtruth}) originally released by~\citet{Bodell23}, covering data from July 2017 to November 2021. The original GB ground truth included 994 labeled proxy contracts. However, after conducting systematic quality controls, including checks for missing values (see Section~\ref{subsec:mising-values-gb}), schema violations (see Section~\ref{subsec:schema-violation-gb}), duplicates (see Section~\ref{subsec:duplicate-gb}), and mislabels (see Section \hyperref[subsec:mislables-gb-proxy]{4.2.4.A}), the refined GB ground truth now comprises 702 Proxy contracts and 218 Non-Proxy contracts. To provide a fair evaluation, we compare UPC Sentinel’s proxy detector method with USCHUNT~\citep{Bodell23} in two different settings: 

\begin{enumerate}
[label=\roman*., itemsep = 3pt, topsep = 0pt]
    \item \textbf{Mixed Setting}. This setting includes both active and inactive proxy contracts to comprehensively assess the performance of both approaches, providing insights into how UPC Sentinel and USCHUNT handle the full spectrum of proxy contracts.
    \item \textbf{Active-Only Setting}. This setting focuses solely on active proxy contracts whose proxy functionality is at least used once following its creation, helping us measure UPC Sentinel's effectiveness under our defined scope (see Section \ref{subsec:proxy-detector-layer}) where proxies are functioning normally as intended.
\end{enumerate}

 When checking all proxy contracts (702 instances) to see if they have been active as of Sep. 2022, we found 122 instances of inactive proxy contracts. Table \ref{tab:proxy-label-distribution} shows the distribution of labels in the ground truth under each setting. To evaluate UPC Sentinel, we run UPC Sentinel’s proxy detector in each setting. Following this, we compute the Precision, Recall and F1-measure performance metrics, and compare our method with USCHUNT. Also, to assess whether the observed differences in performance metrics are statistically significant, we use McNemar's test (p-value=0.05), which is a non-parametric statistical test designed to analyze paired nominal data, such as binary classification outputs~\citep{Mcnemar1947NoteOT}. Finally, we conduct an error analysis of False Positives and False Negatives to gain further insights.

\begin{table}[!t]
\centering
\caption{The distribution of labels under Mixed and Active-Only settings.}
\label{tab:proxy-label-distribution}
\begin{tabular}{lcccc}
\toprule
\textbf{Setting} & \multicolumn{2}{c}{\textbf{Label}} & \textbf{\#Total} \\
\cmidrule(lr){2-3}
                 & \textit{\#Proxy} & \textit{\#Non-Proxy} &                \\
\midrule
Mixed            & 702            & 218                & 920            \\
Active-Only      & 580            & 218                & 798            \\
\bottomrule
\end{tabular}
\end{table}

\subsubsection{Findings \& Error Analysis} 

\observation{UPC Sentinel’s proxy detector has a statistically higher performance than USCHUNT, across both settings in terms of precision and f1-measures.}Table \ref{tab:comp-bodell-g1-g2} compares the performance of UPC Sentinel’s proxy detector with the USCHUNT method in both Mixed and Active-Only settings. In the Mixed setting, UPC Sentinel outperformed USCHUNT in terms of precision for the Proxy class, achieving 100\% precision, although its recall was lower at 82.6\%, leading to an F1 score of 90.5\%. In contrast, USCHUNT exhibited higher recall for Proxy data at 100\%, but had lower precision at 76.3\%, resulting in an F1 score of 86.6\%. For Non-Proxy data, UPC Sentinel demonstrated strong performance with 100\% recall, though with a reduced precision of 64.1\% and an F1 score of 78.1\%. USCHUNT, however, delivered no positive results for Non-Proxy data, with 0\% precision, recall, and F1 score

In the Active-Only setting, UPC Sentinel maintained consistent, perfect scores across both Proxy and Non-Proxy labels, securing 100\% for precision, recall, and F1. Meanwhile, USCHUNT retained high recall (100\%) but lagged in precision for Proxy data (72.8\%), leading to an F1 score of 84.2\%. For Non-Proxy data, it failed to deliver positive results, with zero precision, recall, and F1 scores. Finally, McNemar's test indicates a statistically significant difference between the performance of UPC Sentinel and USCHUNT.

\begin{table}[!t]
\centering
\caption{The performance comparison between UPC Sentinel’s proxy detector and the USCHUNT method in the Mixed and Active-Only settings.}
\label{tab:comp-bodell-g1-g2}
\begin{tabular}{llcccccc}
\toprule
\textbf{Setting} & \textbf{Method} & \multicolumn{1}{r}{\textbf{Precision}} & \multicolumn{1}{r}{\textbf{Recall}} & \multicolumn{1}{r}{\textbf{F1}} & \multicolumn{1}{r}{\textbf{Precision}} & \multicolumn{1}{r}{\textbf{Recall}} & \multicolumn{1}{r}{\textbf{F1}} \\ 
\midrule
 &  & \multicolumn{3}{c}{\textit{Proxy (702)}} & \multicolumn{3}{c}{\textit{Non-Proxy (218)}} \\ \cmidrule(lr){3-8}  
Mixed & UPC Sentinel & \textbf{100} & 82.6 & \textbf{90.5} & \textbf{64.1} & \textbf{100} & \textbf{78.1} \\
 & USCHUNT & 76.3 & \textbf{100} & 86.6 & 0 & 0 & 0 \\ 
 \midrule
 &  & \multicolumn{3}{c}{\textit{Proxy (580)}} & \multicolumn{3}{c}{\textit{Non-Proxy (218)}} \\ \cmidrule(lr){3-8} 
Active-Only & UPC Sentinel & \textbf{100} & \textbf{100} & \textbf{100} & \textbf{100} & \textbf{100} & \textbf{100} \\
 & USCHUNT & 72.7 & \textbf{100} & 84.2 & 0 & 0 & 0 \\
\bottomrule
\end{tabular}
\end{table}

\begin{figure}[!t]
    \centering
    \begin{minipage}[b]{0.24\textwidth}
        \centering
        \includegraphics[width=\textwidth]{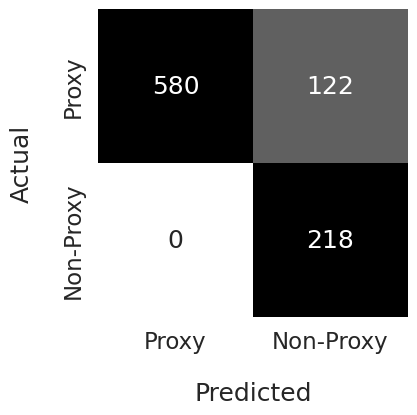}
        \raggedright
        \caption*{\small a. Mixed setting UPC Sentinel}
    \end{minipage}
    \hfill
    \begin{minipage}[b]{0.24\textwidth}
        \centering
        \includegraphics[width=\textwidth]{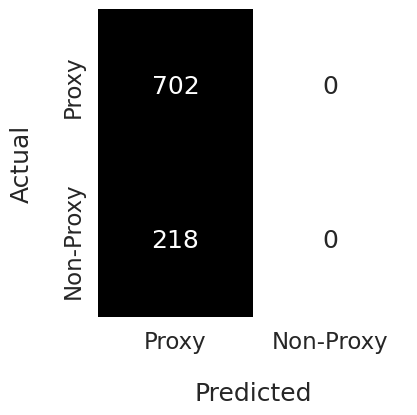}
        \raggedright
        \caption*{\small b. Mixed setting USCHUNT}
    \end{minipage}
    \hfill
    \begin{minipage}[b]{0.24\textwidth}
        \centering
        \includegraphics[width=\textwidth]{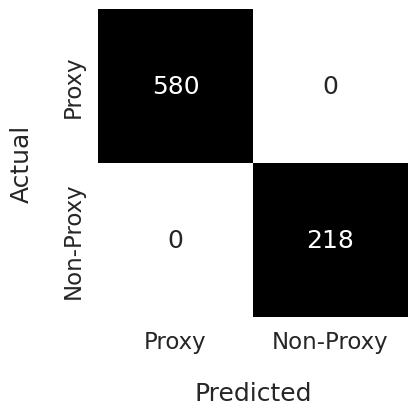}
        \raggedright
        \caption*{\small c. Active-Only setting UPC Sentinel}
    \end{minipage}
    \hfill
    \begin{minipage}[b]{0.24\textwidth}
        \centering
        \includegraphics[width=\textwidth]{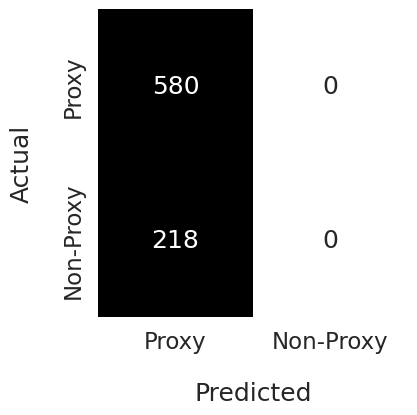}
        \raggedright
        \caption*{\small d. Active-Only setting USCHUNT}
    \end{minipage}
    \caption{Proxy detection confusion matrices for UPC Sentinel and USCHUNT across the Mixed and Active-Only settings.}
    \label{fig:proxy_detector_confusion}
\end{figure}

\textit{Error analysis. Figure \ref{fig:proxy_detector_confusion} shows the confusion matrices for both UPC Sentinel and USCHUNT under the Mixed and Active-Only settings. In the Mixed setting, UPC Sentinel achieved an 82.6\% recall because it cannot detect the 122 inactive proxy contracts (see Section \ref{subsec:rq2-exp-setup}). Since UPC Sentinel categorizes inactive instances as Non-Proxy, this results in a low precision (64.1\%) for the Non-Proxy class. On the other hand, USCHUNT's precision drops to 76.3\% in the Proxy category due to the 218 misclassified instances whose labels were corrected in Section~\hyperref[subsec:mislables-gb-proxy]{4.2.4.A}. In the Active-Only setting, UPC Sentinel achieved perfect precision, recall, and F1 scores across both classes, whereas USCHUNT's precision declined due to the 218 misclassified instances. Finally, because USCHUNT cannot accurately detect those 218 Non-Proxy instances, its performance metrics for the Non-Proxy class are reported as zero across both settings.}

% \subsection{RQ3: How Does the Performance of UPC Sentinel’s Upgradeability Proxy Detector Compare to that of the USCHUNT Method Proposed by~\citet{Bodell23}?} \label{subsec:eval-upc-on-gb}

\subsection{RQ3: How Does the Performance of UPC Sentinel’s Upgradeability Proxy Detector Compare to that of the USCHUNT Method Proposed by~\texorpdfstring{\citet{Bodell23}}{Bodell et al. (2023)?}} \label{subsec:eval-upc-on-gb}

\subsubsection{Motivation}
In this question, we aim to evaluate the performance of UPC Sentinel’s upgradeability detector in identifying UPCs and compare it with USCHUNT, a state-of-the-art method designed to detect UPCs at the source code level.

\subsubsection{Experimental Setup} 
For this experiment, we use the refined GB ground truth (see Section~\ref{subsec:gb-groundtruth}). Following systematic quality control checks—including evaluations for missing values (see Section~\ref{subsec:mising-values-gb}), schema violations (see Section~\ref{subsec:schema-violation-gb}), duplicates (see Section~\ref{subsec:duplicate-gb}), and mislabels (see Sections~\hyperref[subsec:mislables-gb-proxy]{4.2.4.A} and~\hyperref[subsec:mislables-gb-upc]{4.2.4.B})—the refined GB ground truth now consists of 653 UPC contracts and 263 Non-UPC contracts. 

Similar to RQ2 (see Section~\ref{subsec:eval-on-gb-proxy}), we compare the performance of UPC Sentinel with USCHUNT in both the Mixed and Active-Only settings (see Section~\ref{subsec:rq2-exp-setup}). The comparison focuses on the Precision, Recall, and F1-measure performance metrics for detecting upgradeability proxy contracts. Table~\ref{tab:upc-label-distribution} presents the label distribution for the Mixed and Active-Only settings. Additionally, we apply McNemar's test to determine whether the observed differences in performance metrics are statistically significant (p-value = 0.05)~\citep{Mcnemar1947NoteOT}. Finally, we conduct an error analysis of False Positives and False Negatives to gain further insights.

\begin{table}[!t]
\centering
\caption{The distribution of UPC and Non-UPC for the Mixed and Active-Only settings.}
\label{tab:upc-label-distribution}
\begin{tabular}{lcccc}
\toprule
\textbf{Setting} & \multicolumn{2}{c}{\textbf{Label}} & \textbf{\#Total} \\
\cmidrule(lr){2-3}
                 & \textit{\#UPC} & \textit{\#Non-UPC} &                \\
\midrule
Mixed            & 653            & 263                & 916            \\
Active-Only      & 538            & 260                & 798            \\
\bottomrule
\end{tabular}
\end{table}

\subsubsection{Findings \& Error Analysis}

\observation{In the Mixed setting, UPC Sentinel excels in precision and overall F1 scores, while USCHUNT offers higher recall for UPCs.}\label{obs-rq3-01}Table \ref{tab:comp-bodell-g3-and-g4} compares the performance of UPC Sentinel’s upgradeability detector with the USCHUNT method in both Mixed and Active-Only settings. Under the Mixed setting, UPC Sentinel outperforms USCHUNT in terms of precision, achieving a perfect score of 100\% for UPCs, compared to USCHUNT's 79.5\%. This suggests that UPC Sentinel is highly accurate in detecting UPCs without generating false positives. However, in terms of recall, USCHUNT shows superior results with a recall of 94.2\% against UPC Sentinel's 81.8\%. This indicates that USCHUNT is more effective in identifying a larger proportion of UPCs, albeit at the cost of higher false positives, as evidenced by its lower precision. The F1 score, which balances precision and recall, leans in favor of UPC Sentinel at 90\% versus 86.2\% for USCHUNT. In the Non-UPC label, both methods show substantial performance differences. UPC Sentinel achieves a precision of 68.8\% and a perfect recall of 100\%, indicating it identifies all Non-UPC contracts, albeit with some false positives. Conversely, USCHUNT records a higher precision at 73.2\% but a much lower recall of 39.5\%, leading to a significantly lower F1 score of 51.4\%. Finally, McNemar's test indicates a statistically significant difference between the performance of UPC Sentinel and USCHUNT in this setting.

\begin{figure}[!t]
    \centering
    \begin{minipage}[b]{0.24\textwidth}
        \centering
        \includegraphics[width=\textwidth]{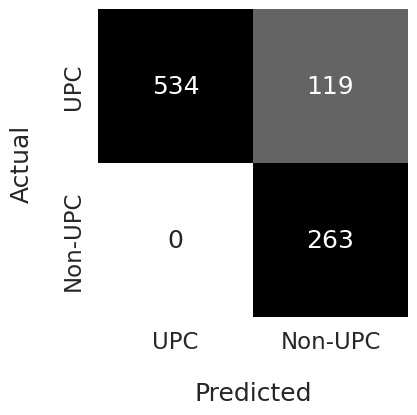}
        \raggedright
        \caption*{\small a. Mixed setting UPC Sentinel}
    \end{minipage}
    \hfill
    \begin{minipage}[b]{0.24\textwidth}
        \centering
        \includegraphics[width=\textwidth]{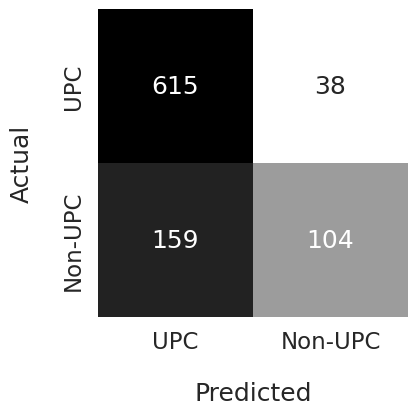}
        \raggedright
        \caption*{\small b. Mixed setting USCHUNT}
    \end{minipage}
    \hfill
    \begin{minipage}[b]{0.24\textwidth}
        \centering
        \includegraphics[width=\textwidth]{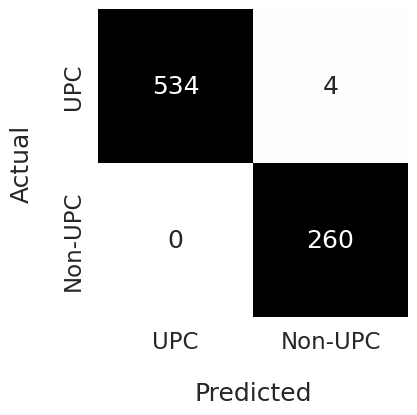}
        \raggedright
        \caption*{\small c. Active-Only setting UPC Sentinel}
    \end{minipage}
    \hfill
    \begin{minipage}[b]{0.24\textwidth}
        \centering
        \includegraphics[width=\textwidth]{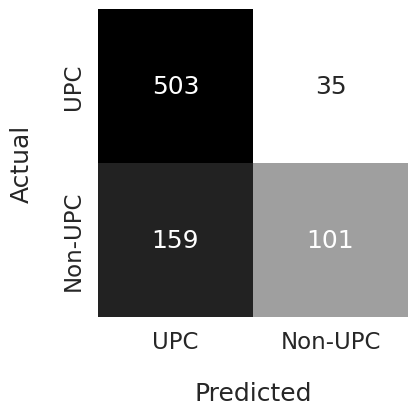}
        \raggedright
        \caption*{\small d. Active-Only setting USCHUNT}
    \end{minipage}
    \caption{Upgradeability proxy detection confusion matrices for UPC Sentinel and USCHUNT across the Mixed and Active-Only settings.}
    \label{fig:upc-detection-confusions}
\end{figure}

\textit{Error analysis. Figure \ref{fig:upc-detection-confusions}.a and \ref{fig:upc-detection-confusions}.b show the confusion matrices for UPC Sentinel and USCHUNT, respectively, under the Mixed setting. UPC Sentinel failed to detect 119 UPCs, with 96.6\% (115) being inactive (i.e., outside the scope of our approach) and 3.4\% (4) active proxy contracts. The detection of inactive cases was limited because both UPC Sentinel require transaction data from proxy contracts to function effectively (see Section \ref{subsec:proxy-detector-layer}). Detailed analysis of the four misclassified active instances showed that three were due to the static analyzer's inability to identify a relevant delegatecall statement (see Section \ref{subsec:upgradeability-detector-layer}), and one due to an unconventional contract storage design. USCHUNT, on the other hand, classified 159 Non-UPCs as UPCs. These 159 are the misclassified instances whose labels we corrected in Section~\hyperref[subsec:mislables-gb-upc]{4.2.4.B}. Additionally, there are 38 UPCs that USCHUNT detected as Non-UPCs, out of which 35 are misclassified instances whose labels were refined in Section~\hyperref[subsec:mislables-gb-upc]{4.2.4.B} from Non-UPC to UPC, and 3 are originally not identified by USCHUNT.}

\smallskip\observation{In the Active-Only setting, UPC Sentinel demonstrates higher performance compared to USCHUNT across both UPC and Non-UPC categories.}Specifically, UPC Sentinel achieves 100\% precision and 99.3\% recall for UPCs, resulting in an F1 score of 99.6\%. In contrast, USCHUNT records a precision of 76\% and a recall of 93.5\% for UPCs, leading to a lower F1 score of 83.8\%. For non-UPC categories, UPC Sentinel also excels with a precision of 98.5\% and a perfect recall of 100\%, translating into an F1 score of 99.2\%. USCHUNT, however, struggles with a precision of 74.3\% and a significantly lower recall of 38.9\%, which results in a much lower F1 score of 51\%. Finally, McNemar's test reveals a statistically significant difference in performance between UPC Sentinel and USCHUNT in this setting.

\textit{Follow-up error analysis. Figure \ref{fig:upc-detection-confusions}.c and \ref{fig:upc-detection-confusions}.d show the confusion matrices for UPC Sentinel and USCHUNT, respectively, under the Active-only setting. In the Active-Only setting, where the focus is on active proxy contracts, UPC Sentinel failed to detect only 4 UPCs, as detailed in Observation \ref{obs-rq3-01}. Nonetheless, USCHUNT those 159 Non-UPCs as UPC, and 35 UPCs as Non-UPCs (i.e., the instances whose ground truth upgradeability labels were refined in Section~\hyperref[subsec:mislables-gb-upc]{4.2.4.B}.)}

\begin{table}[!t]
\caption{The performance comparison between UPC Sentinel’s upgradeability detector and the USCHUNT method in the Mixed and Active-Only settings.}
\label{tab:comp-bodell-g3-and-g4}
\begin{tabular}{llcccccc}
\toprule
\textbf{Dataset} & \textbf{Method} & \textbf{Precision} & \textbf{Recall} & \textbf{F1} & \textbf{Precision} & \textbf{Recall} & \textbf{F1} \\ 
\midrule
 &  & \multicolumn{3}{c}{\textit{UPC (653)}} & \multicolumn{3}{c}{\textit{Non-UPC (263)}} \\ \cmidrule(lr){3-8}  
Mixed & UPC Sentinel & \textbf{100} & 81.8 & \textbf{90} & 68.8 & \textbf{100} & \textbf{81.5} \\
 &  USCHUNT & 79.5 & \textbf{94.2} & 86.2 & \textbf{73.2} & 39.5 & 51.4 \\
 \midrule
 &  & \multicolumn{3}{c}{\textit{UPC (538)}} & \multicolumn{3}{c}{\textit{Non-UPC (260)}} \\ \cmidrule(lr){3-8} 
Activ-Only & UPC Sentinel & \textbf{100} & \textbf{99.3} & \textbf{99.6} & \textbf{98.5} & \textbf{100} & \textbf{99.2} \\
 & USCHUNT & 76 & 93.5 & 83.8 & 74.3 & 38.9 & 51 \\
 \bottomrule
\end{tabular}
\end{table}

% \subsection{RQ4: How Does the Performance of UPC Sentinel’s Upgradeability Pattern Classifier Compare to that of the USCHUNT Method Proposed by~\citet{Bodell23}?}

\subsection{RQ4: How Does the Performance of UPC Sentinel’s Upgradeability Pattern Classifier Compare to that of the USCHUNT Method Proposed by~\texorpdfstring{\citet{Bodell23}}{Bodell et al. (2023)?}}

\label{subsec:eval-on-different-patterns}
% \subsection{\RaggedRight RQ4) How Consistent is UPC Sentinel in Identifying Upgradeability Proxy Contracts Across Different Upgradeability Design Patterns?}
% \label{subsec:eval-on-different-patterns}

\subsubsection{Motivation} 

Motivation While RQ3 has established the effectiveness of UPC Sentinel on the presence of upgradeability features in general (i.e, whether a contract is UPC or Non-UPC), RQ4 empirically validates UPC Sentinel's capabilities of identifying the specific upgradeability pattern a UPC adheres to, essentially evaluating UPC Sentinel’s upgradeability pattern classifier layer (see Section \ref{subsec:upgradeability-pattern-classifier-layer}). This deeper evaluation not only highlights the strengths and weaknesses of UPC Sentinel’s classification capabilities but also provides insights into where the UPC Sentinel can be enhanced to achieve more precise and comprehensive results.

% While RQ2 and RQ3 have established the effectiveness of UPC Sentinel on the presence of upgradeability features in general (i.e, whether a contract is UPC or Non-UPC), our previous findings do not reflect on UPC Sentinel's capabilities in identifying UPCs across various upgradeability patterns, as detailed in Section \ref{fig:back-proxy-pattern-design}. Essentially, our goal in RQ4 is to compare UPC Sentinel and USCHUNT \citep{Bodell23} and determine if their detection mechanism, which determines whether a contract is UPC or not at a more abstract level, is robust across different implementation strategies. For instance, if our truth dataset indicates that Contract X follows pattern Y, we need to confirm that UPC Sentinel correctly identifies Contract X as UPC, regardless of its specific pattern. Understanding this will help us pinpoint the strength of each method and specifically where UPC Sentinel could be enhanced to provide more granular insights or to improve its overall detection accuracy. 

% \subsubsection{Data Collection \& Preprocessing}

\subsubsection{Experimental Setup} 

Given the goal of this research question in evaluating the upgradeability patterns classifier layer, we included all 538 active UPCs from GB ground truth in this analysis (see Table~\ref{tab:upc-label-distribution}). For each UPC entry, the GB ground truth specifies the corresponding patterns, and, in some cases, a single contract is associated with more than one pattern. Overall, the 538 UPCs can be categorized into 19 groups based the set of patterns implemented by a UPC, with the second column of Table ~\ref{tab:upc-sentinel-across-patterns} (labeled "\#Active UPCs") displaying the distribution of active UPCs across each group. For example, the Transparent+ERC1967 category, which constitutes the largest share (43.7\%), represents UPCs that utilize the ERC-1967 Storage Slot alongside the Transparent Upgradeability Proxy Pattern. 

Note that a small subset of 11 data points combines ERC-1967, Unstructured, or Inherited Storage layout principles with patterns from other reference designs, such as ERC-1822, Registry, or Beacon, reflecting design considerations aimed at mitigating storage collisions as discussed in Section 3.5. Even though such combinations of patterns across reference designs seem counter-intuitive, no further details about these patterns were provided by \citeauthor{Bodell23}.'s study. Since these cases are rare and do not significantly impact the overall analysis, we retained the ground truth labels as provided.

In addition, the replication package includes a column indicating whether USCHUNT can detect the specific pattern of the UPC, which we used to compute USCHUNT’s performance. This column for USCHUNT’s prediction has three possible values: “y”, “n”, or “just says it is upgradable”. The value “y” represents cases where USCHUNT successfully detected their patterns, “n” indicates cases where USCHUNT missed the UPC and consequently its pattern, while “just says it is upgradable” refers to cases where USCHUNT identified the UPC but did not determine its specific pattern, either because the UPC does not conform to any known patterns or because USCHUNT could not identify the pattern. As shown in Table \ref{tab:upc-sentinel-across-patterns}, the ground truth also has a “Non-pattern” category, for which we consider USCHUNT to have a match if its predicted label is either “y”, or “just says it is upgradable” to avoid underestimating its performance for this category. For all other patterns, a positive match is counted only if USCHUNT explicitly labels the instance as “y”.

In Section 4.2.4.B (sanity checks on correct upgradeability labeling), we systematically identified 35 mislabeled instances that were reclassified from Non-UPC to UPC (refer to steps 4.2.4.B.2 and 4.2.4.B.3 for details). This set comprises 32 active cases and 3 inactive cases. Since our evaluation focuses exclusively on active cases, only the 32 active instances are categorized as false negatives (“n”) for USCHUNT in this RQ.

As stated in our methodology, since our method always identifies the most specific upgradeability pattern (see Section~\ref{subsec:upgradeability-pattern-classifier-layer}) within the reference design a UPC adheres to, we consider it a positive match if the pattern identified by our method is a subset of the ground truth label. Similarly, for the 11 cases where the label combines patterns across reference designs, we consider UPC Sentinel's output a positive match if the pattern identified by our method is a subset of the contract's ground truth label.

Finally, similar to RQ1 (see Section~\ref{subsec:eval-on-ge}), we report the performance for each group using the accuracy metric, highlighting the percentage of patterns successfully identified by each method. Additionally, we perform McNemar's exact parameter test (suitable for small sample sizes) at a significance level of p=0.05 within each group to evaluate whether the observed differences in performance metrics between UPC Sentinel and USCHunt are statistically significant.

\subsubsection{Findings} 
\observation{UPC Sentinel and USCHUNT perform similarly in terms of pattern detection, each outperforming the other in only one pattern.} Table \ref{tab:upc-sentinel-across-patterns} presents the accuracy of UPC Sentinel and USCHUNT across 19 groups of patterns. Specifically, USCHunt outperforms UPC Sentinel in the Unstructured pattern group, achieving an accuracy of 98.55\% compared to 85.51\%. Conversely, UPC Sentinel outperforms USCHunt in the Non-pattern group, with an accuracy of 85.19\% compared to 55.56\%. These differences are statistically significant based on McNemar's test.

For the remaining groups, UPC Sentinel and USCHunt demonstrate comparable accuracy, with no statistically significant differences. In seven categories (ERC1967, Beacon, Transparent+Unstructured, Diamond, Eternal, Beacon/Inherited, and MasterCopy/Singleton), both systems achieve similar accuracy. UPC Sentinel performs slightly better than USCHunt in five categories (Inherited, ERC1822, Registry/Beacon, Transparent, and Registry), while USCHunt slightly exceeds UPC Sentinel in another five categories (Transparent+ERC1967, Inherited+Eternal, Registry+Inherited, Registry+Unstructured, and ERC1822+ERC1967). Overall, UPC Sentinel successfully identifies the upgradeability pattern type for 92.8\% of UPCs, compared to 89.8\% for USCHunt.

\begin{table}[t]
\caption{Accuracy comparison of UPC Sentinel and USCHunt across different groups of upgradeability proxy patterns. An asterisk (*) indicates a statistically significant difference as determined by McNemar's test.}
\label{tab:upc-sentinel-across-patterns}
\begin{tabular}{llll}
\toprule
\multirow{2}{*}{\textbf{Upgradeability Proxy Pattern}} & \multirow{2}{*}{\textbf{\#Active UPCs}} & \multicolumn{2}{c}{\textbf{Accuracy (Active) \%}}     \\ \cmidrule(lr){3-4} 
                                                       &                                         & \textbf{UPC Sentinel}     & \textbf{USCHUNT}          \\ \midrule
Transparent+ERC1967                                    & 235                                     & 232 (98.72\%)             & \textbf{235 (100.0\%)}    \\
Unstructured                                           & 69                                      & 59 (85.51\%)              & \textbf{68 (98.55\%)$^*$} \\
Non-pattern                                            & 54                                      & \textbf{46 (85.19\%)$^*$} & 30 (55.56\%)              \\
ERC1967                                                & 49                                      & 47 (95.92\%)              & 47 (95.92\%)              \\
Inherited                                              & 48                                      & \textbf{41 (85.42\%)}     & 31 (64.58\%)              \\
Inherited+Eternal                                      & 33                                      & 30 (90.91\%)              & \textbf{33 (100.0\%)}     \\
ERC1822                                                & 12                                      & \textbf{12 (100.0\%)}     & 9 (75.0\%)                \\
Beacon                                                 & 7                                       & 7 (100.0\%)               & 7 (100.0\%)               \\
Registry/Beacon                                        & 6                                       & \textbf{5 (83.33\%)}      & 2 (33.33\%)               \\
Registry+Inherited                                     & 6                                       & 5 (83.33\%)               & \textbf{6 (100.0\%)}      \\
Transparent                                            & 5                                       & \textbf{5 (100.0\%)}      & 4 (80.0\%)                \\
Registry                                               & 3                                       & \textbf{3 (100.0\%)}      & 1 (33.33\%)               \\
Transparent+Unstructured                               & 3                                       & 3 (100.0\%)               & 3 (100.0\%)               \\
Registry+Unstructured                                  & 2                                       & 0 (0.0\%)                 & \textbf{2 (100.0\%)}      \\
Diamond                                                & 2                                       & 2 (100.0\%)               & 2 (100.0\%)               \\
Eternal                                                & 1                                       & 1 (100.0\%)               & 1 (100.0\%)               \\
Beacon/Inherited                                       & 1                                       & 1 (100.0\%)               & 1 (100.0\%)               \\
ERC1822+ERC1967                                        & 1                                       & 0 (0.0\%)                 & \textbf{1 (100.0\%)}      \\
MasterCopy/Singleton                                   & 1                                       & 0 (0.0\%)                 & 0 (0.0\%)                 \\ \midrule
\textbf{Total}                                         & \textbf{538}                            & \textbf{499 (92.8\%)}     & \textbf{483 (89.8\%)}     \\ \bottomrule
\end{tabular}
\end{table}

\smallskip

\begin{tcolorbox}[colframe=black!75!white, colback=white, title=Summary of Findings, parbox=false, breakable, left=1mm, right=1mm ]
        \begin{itemize}[label=\textbullet, itemsep = 3pt, topsep = 0pt]
            \item Our empirical evaluation on our newly compiled ground truth of UPCs demonstrates that UPC Sentinel yields an accuracy of 99\%, demonstrating a high rate of successful identification.
            
            \item In an evaluation using~\citeauthor{Bodell23}'s refined ground truth, UPC Sentinel achieved flawless Precision under both Mixed and Active-Only settings and a Recall of 81.8\% and 99.3\% for the former and lattar settings in turn. Notably, UPC Sentinel surpassed the performance of this preceding study in detecting UPCs.

            \item UPC Sentinel and USCHUNT perform similarly in pattern detection, with each outperforming the other in only one significant pattern. USCHUNT excels in the Unstructured pattern (98.55\% vs. 85.51\%), while UPC Sentinel leads in the Non-pattern group (85.19\% vs. 55.56\%), both differences being statistically significant. For the remaining 17 pattern groups, both tools demonstrate comparable accuracy, with UPC Sentinel achieving 92.8\% overall accuracy compared to 89.8\% for USCHUNT.
            
            \item UPC Sentinel is currently limited to detecting active UPCs due to its behavior-based detection algorithm, highlighting a potential area for future research.

        \end{itemize}
\end{tcolorbox}

% \begin{footnotesize}
%     \begin{mybox}{Summary}
%         \begin{itemize}[label=\textbullet, itemsep = 3pt, topsep = 0pt]
%             \item Our evaluation of our compiled ground truth of UPCs demonstrates that UPC Sentinel yields an accuracy of 99\%, demonstrating a high rate of successful identification.
            
%             \item In an evaluation using~\citeauthor{Bodell23}'s refined ground truth, UPC Sentinel achieved flawless Precision under both Mixed and Active-Only settings and a Recall of 82.6\% and 100\% for the former and lattar settings in turn. Notably, UPC Sentinel surpassed the performance of this preceding study in detecting both proxy contracts and UPCs.

%             \item UPC Sentinel demonstrates equal or higher performance in most (seven out of nine) upgradeability design patterns compared to USCHUNT when UPCs are active, particularly excelling with the Universal Upgradeability Proxy, where it significantly surpasses USCHUNT. Although it shows a marginally lower efficacy in the ERC-1967 Upgradeability Proxy Pattern (Unstructured Storage) and the Beacon Upgradeability Proxy Pattern, its overall prowess is underscored by achieving 100\% recall rate in six of the eight patterns studied, underscoring its robustness and effectiveness across a wide range implementation strategies.

%             \item UPC Sentinel is currently limited to detecting active UPCs due to its behavior-based detection algorithm, highlighting a potential area for future research.
            
%         \end{itemize}
%     \end{mybox}
% \end{footnotesize}

    \section{Discussion}
\label{sec:discussion}

\subsection{Lessons Learned}\label{subsec:lessons-learned}
    In Section~\hyperref[subsec:mislables-gb-proxy]{4.2.4.A}, we identified 209 contracts where no delegatecall statement could be found at the decompiled bytecode level. A detailed examination of the source code for these cases revealed that, in all instances, the proxy functionality was not accessible through the contract's exposed public or external interfaces. This can result in erroneous detection of proxy contracts at the source code level. In the following, we describe the perils of source-code level proxy detection in terms of three patterns of unreachable proxy functionality observed during the analysis of these 209 cases.
    
\begin{figure}[t]
    \centering
      \includegraphics[scale=0.6]{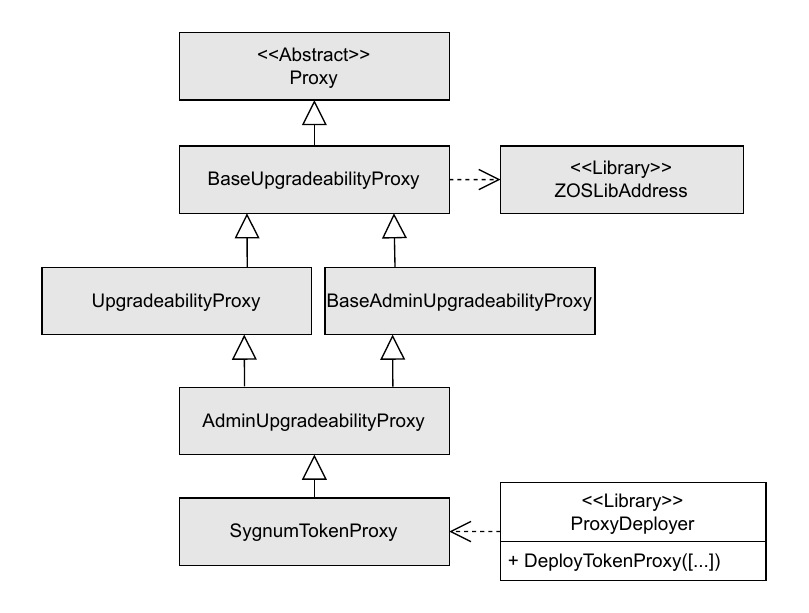}
      \caption{Simplified class diagram of the ProxyDeployer contract. The gray background contracts are the upgradeability proxy contracts.}
      \label{fig:case-1}
    \end{figure}

    \smallskip\noindent \textbf{Case \#1) Lack of proxy exposure}. Figure \ref{fig:case-1} shows a simplified\footnote{Visit \href{https://etherscan.io/viewsvg?t=1&a=0xa8305b9cfd9590d328e1ac5b13bdd647daec5652}{https://etherscan.io/viewsvg?t=1\&a=0xa8305b9cfd9590d328e1ac5b13bdd647daec5652} for a detailed class diagram.} UML class diagram of the eight contracts that are shown under the ProxyDeployer\footnote{\href{https://etherscan.io/address/0xa8305b9cfd9590d328e1ac5b13bdd647daec5652\#code}{https://etherscan.io/address/0xa8305b9cfd9590d328e1ac5b13bdd647daec5652\#code}} contract in Etherscan. The ProxyDeployer library simplifies the deployment of UPCs by creating a SygnumTokenProxy, an AdminUpgradeabilityProxy variant, to enable seamless upgrades for Sygnum's tokens. The base Proxy and BaseUpgradeabilityProxy provide delegation of calls and dynamic implementation updates, with ZOSLibAddress confirming valid smart contract addresses. BaseAdminUpgradeabilityProxy adds an admin-only upgrade mechanism, while AdminUpgradeabilityProxy and UpgradeabilityProxy establish initialization procedures and admin controls.

    While the upgradeability proxy mechanism does exist here in this example, the ProxyDeployer is not a proxy itself. ProxyDeployer (i.e., main contract\footnote{An Ethereum Contract Account (CA) at a specific address can consist of multiple contracts, with one serving as the main contract and the others as its dependencies. The main contract is the one that integrates and interacts with other contracts. It orchestrates the logic, permissions, and data flow between smaller, modular contracts which might handle more specific tasks like managing users, handling transactions, or implementing business logic. The main contract acts as the front-facing interface of the CA, effectively encapsulates the core functionality and serves as the interface for users and other external entities interacting with a decentralized application.}) does not expose any public or external function through which one can directly use the proxy functionality (i.e., violation of Property \#1). Instead, the ProxyDeployer operates as a factory, following the Factory Design Pattern. It depends on (rather than inherits from) the upgradeability infrastructure to deploy instances of a UPC to Ethereum. More specifically, when one calls the deployTokenProxy function of the ProxyDeployer, it deploys a new SygnumTokenProxy contract to Ethereum and returns its address. Thus, due to this dependency, one needs to upload the entire set of dependencies (i.e., all the gray contracts in Figure \ref{fig:case-1}) to be able to compile and verify the source code of the ProxyDeployer. 
    
    This example shows that sequential scanning (i.e., one contract at a time) the source code of verified contracts under an address, without considering their semantic relationships between the main contract and its dependencies, can result in erroneous detection of proxy contracts at the source code level. On the other hand, the decompiled bytecode generated by Panoramix does not suffer from this issue. More specifically, the decompiled bytecode includes all the public and external functions that the main contract exposes. For every public or external function, Panoramix linearizes its cross-contract calls; for instance, if public Function F1 of Contract C1 calls Function F2 of Contract C2, Panoramix copies F2’s logic under F1 during the decompilation process. Therefore, if F2 implements a proxy, its logic will be found under F1 after decompilation. This simplification allows us to search the decompiled logic of the public and external functions for the two inherent properties of a proxy contract (see Section \ref{subsec:proxy-detector-layer}). 
    
    As such, analyzing the decompiled bytecode provides a more reliable view of the contract's behavior and interaction patterns than analyzing the official, verified source code. Finally, out of 209 instances, 171 suffer from this issue (i.e., verified source code not being reliable).
    
    \begin{tcolorbox}[colframe=white!75!white, colback=gray!10, boxrule=0.5mm, sharp corners]
    \textbf{Takeaway:} An Ethereum Contract Account (CA) is only classified as a proxy if the proxy functionality is served by one of the contract’s exposed public or external functions. While detecting this at the source code level requires advanced cross-contracts techniques to analyze the relationship between the main contract and its dependencies, Panoramix's decompiled bytecode captures such relationships in simplified linear format for reliable proxy detection.
    \end{tcolorbox}
    
\begin{figure}[t]
    \centering
      \includegraphics[scale=0.75]{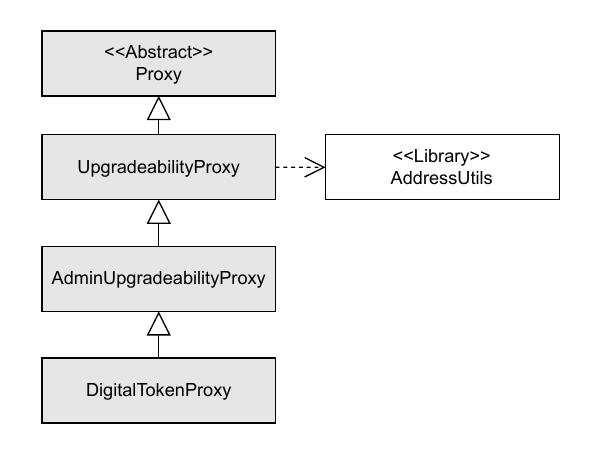}
      \caption{Simplified class diagram of the AddressUtils contract. The gray background contracts are dead upgradeability proxy contracts.}
      \label{fig:case-2}
    \end{figure}

    \smallskip\noindent\textbf{Case \#2) Dead proxy code}. A dead proxy contract is a contract that neither the main contract nor its dependencies are dependent upon. Figure \ref{fig:case-2} shows a simplified\footnote{Visit \href{https://etherscan.io/viewsvg?t=1&a=0xf2262f092063cd19050f7e037cbf6401b58e9035}{https://etherscan.io/viewsvg?t=1\&a=0xf2262f092063cd19050f7e037cbf6401b58e9035} for a detailed class diagram.} UML class diagram of the five contracts shown under the AddressUtils\footnote{\href{https://etherscan.io/address/0xf2262f092063cd19050f7e037cbf6401b58e9035\#code}{https://etherscan.io/address/0xf2262f092063cd19050f7e037cbf6401b58e9035\#code}} contract in Etherscan. The Proxy contract facilitates delegation to a designated implementation contract. The UpgradeabilityProxy extends this by allowing the implementation address to be upgraded. The AdminUpgradeabilityProxy adds administrative control, allowing only authorized users to perform upgrades. The DigitalTokenProxy leverages these mechanisms specifically for digital tokens, using AdminUpgradeabilityProxy to enable secure and seamless updates to the token's contract, minimizing disruption for users. AddressUtils is a library that provides a function to check if a given address is a contract.
    
    As shown in Figure \ref{fig:case-2}, the main contract (i.e., AddressUtils) does not depend on the other contracts. While it is likely that all five contracts belong to one particular software system, one can verify the AddressUtils’ source code on Etherscan by only submitting its source during the verification process. Although not reachable, the source codes of the other four contracts can also be seen under the AddressUntil verified source code in Etherscan. We conjecture that this issue is related to Etherscan’s contract verification process, which does not check for such dead contracts. In other words, we conjecture that one can upload dead contracts along with relevant ones during the contract verification process, and Etherscan will show the sources of all of them after the process is successfully completed under the CA address. 
    
    To validate our hypothesis, we deployed an exact AddressUtils\footnote{\href{https://sepolia.etherscan.io/address/0xe09c93edad08dffc684e8b51da0cd30b9dd50653\#code}{https://sepolia.etherscan.io/address/0xe09c93edad08dffc684e8b51da0cd30b9dd50653\#code}} contract to Sepolia Ethereum testnet powered by Etherscan. Then, we verified the deployed contract on Etherscan using just the AddressUtil’s source code, and we could successfully verify it. Then, we deployed another AddressUtils\footnote{\href{https://sepolia.etherscan.io/address/0x0f741460ae5f5b8cdbf1ae9181adbf5927fcfc66\#code}{https://sepolia.etherscan.io/address/0x0f741460ae5f5b8cdbf1ae9181adbf5927fcfc66\#code}} contract using a similar configuration. This time, we uploaded the sources of all five contracts (i.e., Proxy, UpgradeabilityProxy, AdminUpgradeabilityProxy, DigitalTokenProxy, and AddressUtils) during the verification. The verification again went successful. When we checked the verified source code, we observed all five contracts’ source codes, even though four of them are dead code. This confirms our earlier hypothesis that the Etherscan verification process does not exclude the dead contracts out of the verified code.
    
    This example shows that the verified source code under a CA account can have noise, negatively affecting the proxy detection at the source code level. On the other hand, the contract bytecode does not suffer from this issue, as the Solidity compiler (solc) includes only the main contract and its dependencies during the compilation process, excluding any non-reachable or dead contracts. Since the Panoramix input is the bytecode, its output is not affected by the presence of dead contracts. Out of 209 instances, 34 contain dead proxy codes at source code level. 

    \begin{tcolorbox}[colframe=white!75!white, colback=gray!10, boxrule=0.5mm, sharp corners]
    \textbf{Takeaway:} The presence of dead contracts, due to Etherscan's lack of checks during source code verification, can cause erroneous proxy detection at the source code level, but this issue is avoided at the decompiled bytecode level as compilers ignore dead contracts while generating bytecode, providing a clean input to the Panoramix decompiler.
    \end{tcolorbox}

\begin{figure}[t]
    \centering
      \includegraphics[scale=0.75]{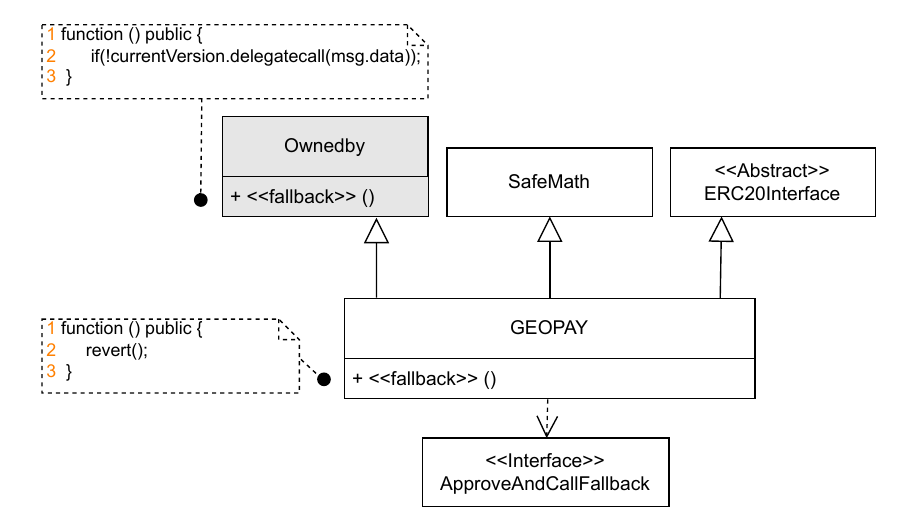}
      \caption{Simplified class diagram of the GEOPAY contract. The gray background contracts are the upgradeability proxy contracts.}
      \label{fig:case-3}
    \end{figure}

    \smallskip\noindent\textbf{Case \#3) Overriding proxy functionality}. Figure \ref{fig:case-3} shows a simplified\footnote{Visit \href{https://etherscan.io/viewsvg?t=1&a=0xfa82f0a05b732deaf9ae17a945c65921c28b16dd}{https://etherscan.io/viewsvg?t=1\&a=0xfa82f0a05b732deaf9ae17a945c65921c28b16dd} for a detailed class diagram.} UML class diagram of the five contracts that are shown under the GEOPAY\footnote{\href{https://etherscan.io/address/0xfa82f0a05b732deaf9ae17a945c65921c28b16dd\#code}{https://etherscan.io/address/0xfa82f0a05b732deaf9ae17a945c65921c28b16dd\#code}} contract in Etherscan. The GEOPAY contract is an ERC-20 token named "GEO Utility Token". It integrates safe arithmetic operations through the SafeMath contract and supports standard ERC20 functions like transferring tokens, approving spenders, and querying balances. The Ownedby proxy contract ensures ownership management while facilitating upgradeability by relaying calls to an up-to-date implementation via a fallback function. Additionally, GEOPAY includes an ApproveAndCall mechanism for executing functions in approved contracts and allows the owner to transfer out accidentally received ERC-20 tokens or ether. 
    
    The GEOPAY contract inherits from the Ownedby contract. Although the Ownedby’s fallback implements a proxy, the GEOPAY’s fallback overrides it, effectively nullifying the proxy functionality. Therefore, GEOPAY is not a proxy. Similar to the previous two cases, this case shows another instance of code no longer being reachable, in this case due to overriding the proxy functionality by the main contract. Hence, overlooking the impact of OOP features like method overriding can lead to inaccurate identification of UPCs at the source code level. Conversely, the Panoramix decompiled bytecode captures and encodes such inheritance effects, providing a more precise detection solution. Out of 209 instances, four suffer from this issue.

    \begin{tcolorbox}[colframe=white!75!white, colback=gray!10, boxrule=0.5mm, sharp corners]
    \textbf{Takeaway:} Overriding inherited methods can change a smart contract's behavior and disable key functions like proxy mechanisms. Researchers should note that inherited contracts may lose their original functionalities if proxy functions are overridden.
    \end{tcolorbox}

\subsection{On the Applications of Upgradeability Proxy Detection}\label{subsec:disc-on-the-application} 

Detecting UPCs serves as a gateway to a multitude of subsequent research avenues. We envision another layer on top of our proposed architecture (see Section~\ref{subsec:solution-domain}) that leverages the knowledge we obtained in the first three layers towards applications that enable different monitoring tasks on upgradeability proxy contracts. While the applications themselves are not exclusive to our approach, UPC Sentinel offers a more accurate method for identifying active UPCs, enabling it to cover a more complete slice of all upgradeable proxies.

Prior literature has started exploring a variety of applications for analyzing the evolution, impact, and security issues of upgradeable proxy contracts. For instance, recovering the contract release history addresses the challenge of tracking version changes in smart contracts, as Etherscan lacks a direct linkage to source repositories and integrated version history tools. \citet{qasse2023} introduced an initial method to detect new versions through emitted upgrade events, though this approach depends on developers consistently implementing and triggering upgrade events. \citeauthor{Bodell23} further analyzed upgrade frequency by counting the invocations of upgrade functions. Another example application is the evolutionary analysis of smart contracts, which examines how contracts evolve over time to address security, efficiency, functionality, and regulatory requirements. \citeauthor{qasse2023} have set out to investigate reasons for changes between releases, such as bug fixes and feature additions. Another application is to assess how updates influence user trust, engagement, and asset liquidity, as trust in blockchain relies on perceived immutability and transparent governance, and frequent upgrades without clear communication can erode user confidence, raising concerns about potential risks or centralization ~\citet{qasse2023}. This includes studying changes in transaction volume, user retention, user diversity, and the liquidity of assets associated with the contract~\citep{DappRadar_2021,FasterCapital_2024}. 

A notable limitation in the aforementioned application areas is that the domain's current focus is primarily restricted to open-sourced (a.k.a., verified) contracts. This constraint arises due to the reliance of existing UPC detectors on source code availability. Yet, recent studies indicate that the majority of smart contracts remain unverified~\citep{Junzhou2020,Huang2021,Wenkai2024,Ting2019,Chen21_Maintenance, Jiachi_DefectChecker_2022}, suggesting that current findings may not fully generalize across the entire spectrum of deployed contracts. 

This is where UPC Sentinel can make an impact by detecting UPCs at the bytecode level. Our empirical evaluation on two different ground truth datasets - encompassing more than 4,000 data points - demonstrate that UPC Sentinel is i) effective, ii) reliable against source code pitfalls (see Section \ref{subsec:lessons-learned}), and iii) capable of precisely detecting active UPCs across a wide range of known upgradeability proxy patterns, regardless of source code availability. This broader applicability allows future studies to conduct deeper analyses that are not constrained by the availability of source code, enabling more comprehensive assessments of the overall upgradeability landscape across the entire Ethereum ecosystem. As such, UPC Sentinel can even be a potential candidate to serve as an autonomous instrument or seamlessly integrate with Etherscan, vigilantly monitoring UPCs for any alterations in their implementation contracts.

\subsection{Time-Cost of UPC Sentinel}
Our experiments were conducted on a high-performance computing system equipped with 60 CPU cores, 500 GB of RAM, and a network bandwidth of 90 Mbps, ensuring efficient handling of large-scale data processing tasks and swift data transfer. For data retrieval, we used the free tier of Google BigQuery to query the \texttt{crypto\textunderscore ethereum} dataset. In the following sections, we discuss the time-cost of running UPC Sentinel and the various measures implemented to improve its time complexity.

\smallskip\noindent\textbf{Time-cost of initializing the proxy detector.} Given an input contract address, this component analyzes the contract’s transactions to determine its proxy status and implementation contract addresses~\citep{Ebrahimi23}. The proxy detector algorithm, developed in our previous study~\citep{Ebrahimi23}, is implemented as a BigQuery SQL query executed via the Google BigQuery API (see Section \ref{subsec:proxy-detector-layer}). Since this query involves scanning, filtering, and joining the `trace' table, which contains billions of transactions, executing it for every individual contract address would be cost-prohibitive. Instead, the query is designed to retrieve all active proxy contracts and their corresponding implementation contracts within a specified time range that marks the start and end of the study period. The resulting table is then automatically downloaded locally and hashed (i.e., \texttt{<proxy address: string, implementation addresses: list>}). This hashing mechanism enables \(\mathcal{O}(1)\) verification of whether a given contract is a proxy. The process took 20.2 minutes to identify, download and load all active proxy contracts as of Sep. 2022

\smallskip\noindent\textbf{Time-cost of initializing the bytecode decompiler.} UPC Sentinel’s upgradeability detector layer operates on the decompiled bytecodes generated by the Panoramix decompiler for input proxy contracts, their implementation contracts, and their dependency contracts. Given the high prevalence of type-1 clones (i.e., contracts with identical bytecodes) in the Ethereum ecosystem, this component includes a memory mechanism to avoid redundant decompilation of contracts with identical bytecodes. Additionally, it is equipped with multi-core processing (set to 60 cores in this study) and a decompilation timeout parameter (set to 60 minutes per input contract), allowing for parallel decompilation of multiple contracts. We used our multi-core decompilation across other UPC Sentinel’s components to speed up the process. For instance, if a proxy contract has several implementation contracts, instead of decompiling them one at a time, we decompile all of them in parallel. Finally, to accelerate and reduce the cost of on-the-fly retrieval of contract bytecodes, this component computes a bytecode hash for all Ethereum contracts using a BigQuery SQL query. The resulting table is automatically downloaded locally and hashed into a mapping of \texttt{<contract address: string, bytecode hash: string>}. Additionally, another mapping is created between \textit{distinct} bytecodes and their corresponding bytecode hashes (\texttt{<bytecode hash: string, bytecode: string>}). Once initialized, using these two mappings, any contract's bytecode can be fetched in \(\mathcal{O}(1)\) time. The initialization process including retrieving, downloading and loading datasets took 13.6 minutes.

\smallskip\noindent\textbf{Time-cost of running UPC Sentinel on the GE \& GB datasets.} We conducted a time-cost analysis of running UPC Sentinel on the GE (see Section \ref{subsec:eval-on-ge}) and GB (see Section \ref{subsec:eval-upc-on-gb}) ground truth dataset with 3,177 and 920 contracts, respectively. More specifically, we break down UPC Sentinel's total execution time across each dataset into four aspects: the proxy detector initialization time, the bytecode decompiler initialization time, the total decompilation time, and the time taken to run UPC Sentinel's detectors, all shown in Table 1. This breakdown allows us to understand the impact of each component on the overall performance. Finally, we report the average computation time per contract. Since initialization occurs only once, we also report the average computation time per contract, excluding the initialization time for the proxy detector and bytecode decompiler.

Table 1 shows the time cost of UPC Sentinel across the two datasets. For the GE ground truth dataset with 3,177 contracts, the entire process took 4.29 hours to complete, resulting in an average processing time of 4.9 seconds per contract. Similarly, for the GB ground truth dataset with 920 contracts, the process took 2.06 hours, yielding an average processing time of 8 seconds per contract. Having excluded the initialization time for the proxy detector and bytecode decompiler, this resulted in an adjusted average of 4.2 seconds for the GE dataset and 5.9 seconds for the GB dataset.

Compared to USCHUNT, which has an average processing time of 0.76 seconds per contract, UPC Sentinel’s processing time is significantly higher, with an average time cost that is 7.8 times greater \(\left(\frac{5.9}{0.76}\right)\). We identified that decompilation is the most time-consuming aspect in both datasets. To optimize this, we plan to pre-decompile all Ethereum contracts as a future improvement. This approach is feasible because, despite the deployment of over 50 million contracts on Ethereum as of Sep. 2022, only 1.33\% have \textit{distinct} bytecodes. By pre-decompiling these \textit{distinct} bytecodes, we can eliminate the need for on-the-fly decompilation and retrieve the decompiled version of any contract in \(\mathcal{O}(1)\) time. Under this scenario, the total processing time would be reduced to 13.78 minutes seconds for the GB dataset, averaging 0.89 seconds per contract, which closely aligns with USCHUNT’s performance. Alternatively, a faster decompiler is another viable solution to enhance the time cost.

\begin{table}[t]
\centering
\caption{The breakdown of UPC Sentinel time-cost across GE and GB datasets.}
\resizebox{\textwidth}{!}{
\begin{tabular}{lcccccccc} 
\toprule
\textbf{Dataset} & \textbf{\#Contracts} & \rotatebox{90}{\makecell[l]{\textbf{Proxy Detector Init Time (min)}}} & \rotatebox{90}{\makecell[l]{\textbf{Decompiler Init Time (min)}}} & \rotatebox{90}{\makecell[l]{\textbf{Decompilation Time (hr)}}} & \rotatebox{90}{\makecell[l]{\textbf{Detectors Time (min)}}} & \rotatebox{90}{\makecell[l]{\textbf{Total Execution Time (hr)}}} & \rotatebox{90}{\makecell[l]{\textbf{Avg Time per Contract (sec)}}} & \rotatebox{90}{\makecell[l]{\textbf{Avg Time per Contract} \\ \textbf{Excluding Init. (sec)}}}  \\ 
\midrule
GE       & 3,177                    & 20.16                                  & 13.6                                 & 3.51                            & 12.85                           & 4.29                              & 4.9                                  & 4.2                                                   \\
GB       & 920                      & 20.16                                   & 13.6                                 & 1.27                            & 13.78                           & 2.06                               & 8                                    & 5.9                                                   \\
\bottomrule
\end{tabular}
}
\label{tab:execution-time-comparison}
\end{table}

    \section{Related Work}\label{sec:related-work}
In our prior study, we analyzed the prevalence of proxy contracts within a dataset of 50 million Ethereum smart contracts, revealing an increasing deployment of proxy contracts over time. We also investigated how these contracts are integrated into applications and the prevalence of various proxy contract types~\citep{Ebrahimi23}.~\citet{Meisami23} conducted a comprehensive survey of upgradeable smart contract patterns using proxies, compiling and categorizing various design patterns. Building on this foundation, ~\citet{Bodell23} advanced the study by characterizing proxy-based upgradeable smart contracts (USCs) in blockchains, emphasizing their critical role in managing billions of USD in digital assets. This follow-up work introduced six sets of syntactic features and nine semantic features to identify eleven USC design patterns, and reported 2,546 security and safety issues across six categories, thereby deepening the understanding of USC vulnerabilities. Additionally,~\citet{qasse2023} examined smart contract upgrading patterns, their prevalence, and the reasons for upgrades, such as introducing new features or fixing vulnerabilities. 

Further expanding on upgradeability,~\citet{Salehi22} evaluated six upgradeability patterns and developed a framework to measure the number of upgradeable contracts on Ethereum, revealing significant control by single Externally Owned Addresses (EOAs) and multi-signature wallets.~\citet{Frowis2022} assessed the impact and adoption of CREATE2, focusing on its security implications. Meanwhile,~\citet{Bui21} proposed the "Comprehensive-Data-Proxy pattern" to address security resilience and scalability issues in smart contract upgradeability.~\citet{Chen20-2} conducted a survey on the use of self-destruct, identifying key reasons for contract destruction and providing recommendations for its improved use. Lastly,~\citet{Feist19} introduced the Slither framework, automating the detection of vulnerabilities and optimizing code, including the detection of upgradeable proxy contracts (UPCs). This body of work collectively highlights the advancements and ongoing challenges in managing the upgradeability of Ethereum smart contracts.

In the following two subsections, we compare prior studies on both proxy detection and upgradeability proxy detection with UPC Sentinel to clearly present and make accessible the nuanced differences and implications of various studies to the reader.

\subsection{Comparative Analysis of Proxy Detection Techniques}

Five studies reviewed earlier proposed techniques for detecting proxy contracts as a component of their upgradeability proxy detection method. In the following, we carefully compare these five studies based on the characteristics of their proxy detectors. A summary is shown in Table \ref{tab:proxy-detection-benchmark}.

\smallskip\noindent\textbf{1. Detection approach}. This characteristic refers to how a study detects proxy contracts, either through a run-time dynamic analysis (via transaction analysis) or by analyzing the contract's source code (via static analysis). Dynamic analysis investigates contracts' behavior during execution, while static analysis examines a given source code without running it. UPC Sentinel~\citep{Ebrahimi23} and~\citet{Salehi22} focus on a dynamic approach, analyzing contracts during execution using transactional data. On the other hand,~\citet{Bodell23},~\citet{qasse2023}, and~\citet{Feist19} opt for static analysis, inspecting the source code. Unlike static analysis, dynamic analysis does not require access to the source code. This is particularly beneficial for contracts where source code is not available (see Section \ref{subsec:problem-domain}). Dynamic analysis relies on the availability of transaction data. If a contract has a limited or non-representative transaction history (e.g., due to being new or rarely used), the analysis may not be comprehensive.

\smallskip\noindent\textbf{2. Proxy signatures}. This criterion defines when a contract qualifies as a proxy contract. A contract is considered a proxy contract if it satisfies the two inherent properties of proxy contracts, Property \#1 and Property \#2, as outlined in Sections \ref{sec:background} and \ref{subsec:proxy-detector-layer}. UPC Sentinel’s proxy detector (see Section \ref{subsec:proxy-detector-layer}) and the proxy detector by \citet{Salehi22} adopt a more nuanced approach by checking for both properties, resulting in a more accurate detection method. Conversely, ~\citet{Bodell23} check for the presence of the delegatecall statement (i.e., Property \#1) within the fallback function's control flow graph.~\citet{qasse2023} employs a regular expression (``delegatecall'') to discern proxy contracts, whereas ~\citet{Feist19} focuses solely on the presence of a delegatecall keyword. While this approach holds merit, it is prone to false positives. The rationale behind this criticism lies in the fact that techniques based on searching keywords (e.g., delegatecall keyword) might not always convey the full context, especially since smart contracts often have a higher comment rate than other languages, and language keywords such as delegatecall can sometimes be found in comments. Furthermore, not every delegatecall operation is indicative of a proxy contract. For example, a contract that has a delegatecall operation but the signature of the function where the delegatecall is used is not similar to the function to which calls are delegated is indeed an adapter contract and not a proxy contract. In other words, a proxy must preserve the interface of the object it represents (Property \#2). In Listing \ref{listing:adapter-contract-example} we provided an example demonstrating that, despite the use of a delegatecall in a fallback function, the contract behaves as an adapter than a proxy.

\begin{table}[t]
\centering
\caption{Comparison of studies on proxy detection, detailing detection approaches, signature checks, analysis level, implementation location, and usage status. The blank cells indicate the items for which we could not find any information in the studies.}
\resizebox{\textwidth}{!}{%
\begin{tabular}{lcccccccccccccccccc}
\toprule
\textbf{Study}& \multicolumn{10}{c}{\textbf{Proxy detection characteristics}} \\
\cmidrule(l){2-11}
 & \multicolumn{2}{c}{\begin{tabular}[c]{@{}c@{}}Detection\\ approach\end{tabular}} & \multicolumn{2}{c}{\begin{tabular}[c]{@{}c@{}}Proxy\\ signatures\end{tabular}} & \multicolumn{2}{c}{\begin{tabular}[c]{@{}c@{}}Analysis\\  level\end{tabular}} & \multicolumn{2}{c}{\begin{tabular}[c]{@{}c@{}}Proxy\\ function\end{tabular}} & \multicolumn{2}{c}{\begin{tabular}[c]{@{}c@{}}Usage\\ status\end{tabular}} \\
\cmidrule(lr){2-3} \cmidrule(lr){4-5} \cmidrule(lr){6-7} \cmidrule(lr){8-9} \cmidrule(lr){10-11}
 & \rotatebox{90}{Dynamic} & \rotatebox{90}{Static} & \rotatebox{90}{Property \#1} & \rotatebox{90}{Property \#2} & \rotatebox{90}{Source Code} & \rotatebox{90}{Bytecode} & \rotatebox{90}{Fallback Function} & \rotatebox{90}{Any Function} & \rotatebox{90}{Active} & \rotatebox{90}{Inactive} \\
\midrule
UPC Sentinel & * &  & * & * &  &  &  & * & * &  \\
\citep{Bodell23} &  & * & * &  & * &  & * &  & * & * \\
\citep{qasse2023} &  & * & * &  & * &  &  & * & * & * \\
\citep{Salehi22} & * &  & * & * &  &  & * &  & * &  \\
\citep{Feist19} &  & * & * &  & * &  & * &  & * & * \\
\bottomrule
\end{tabular}%
}
\label{tab:proxy-detection-benchmark}
\end{table}

\smallskip\noindent\textbf{3. Analysis level}. This criterion demonstrates whether the detection approach operates at the source code or bytecode level. This is specific to studies that use a static analysis approach. Source code refers to the original, human-readable version of a contract, while bytecode is the machine-readable version used for execution on the Ethereum Virtual Machine (EVM). In this case,~\citet{Bodell23},~\citet{qasse2023} and~\citet{Feist19} conduct their analysis at the source code level. As highlighted earlier in Section~\ref{sec:introduction}, the majority of smart contracts remain unverified~\citep{Chen21_Maintenance}. Thus, studies that rely on the source code would suffer if the smart contract is not verified on Etherscan. In Section \ref{subsec:lessons-learned}, we discussed that source code can contain noise (e.g., dead or unreachable contracts), which can lead to inaccuracies in tools that operate on smart contract source code. It is noteworthy that none of the studies reviewed address the detection of proxy contracts at the bytecode level. Detecting proxy contracts at the bytecode level eliminates the need for access to the contract's source code but introduces the added complexity of decompiling the bytecode. As part of our methodology for identifying upgradeable proxy contracts, we derived three equivalent patterns corresponding to the two inherent properties of proxy contracts at the decompiled bytecode level. These patterns can serve as effective static measures for proxy contract detection. Further exploration of this approach is left for future work.

\smallskip\noindent\textbf{4. Proxy function}. This criterion reveals where the proxy functionality is implemented within a contract. Typically, a proxy is implemented in the fallback function. Yet, in practice, the proxy functionality can be in any function, provided the function's signature where the delegatecall is placed resembles the one receiving the delegated calls (i.e., Property \#2). Studies by~\citet{Salehi22},~\citet{Bodell23}, and~\citet{Feist19} search for proxies in the fallback function, traditionally associated with proxy implementation. However, UPC Sentinel~\citep{Ebrahimi23} and~\citet{qasse2023} extend their search to any function, potentially enabling them to detect non-traditional proxy contracts. 

\smallskip\noindent\textbf{5. Usage status}. This criterion indicates the types of proxy contracts a method can detect, distinguishing between active and inactive proxies. An active proxy contract has used its proxy functionality at least once following its creation transaction. An inactive proxy, however, has never utilized its proxy functionality post-creation. UPC Sentinel~\citep{Ebrahimi23} and~\citet{Salehi22} can only detect active proxies, limiting their detection scope to contracts that have used their proxy functionality after creation. Meanwhile,~\citet{Bodell23},~\citet{qasse2023}, and~\citet{Feist19} can detect both active and inactive proxies because they statically detect proxy contracts. We further discuss the significance of inactive proxy contracts in our discussion of threats to construct validity (Section~\ref{subsec:threat-construct-validity}) and explain how UPC Sentinel can be extended to detect such cases. 

Given the above-mentioned comparison, we picked the proxy detection method proposed by~\citet{Ebrahimi23} to detect proxy contracts (see Section \ref{subsec:proxy-detector-layer}). 

\subsection{Comparative Analysis of Upgradeability Proxy Detection Techniques}\label{subsec:comparive-analysis-upc}
Similarly, we compare UPC Sentinel with the last four recent studies in the field~\citep{Salehi22, Bodell23, qasse2023, Feist19} based on four different criteria. In the following, we detail our comparison across each criterion.

\smallskip\noindent\textbf{1. Detection approach.} The first criterion refers to the methodology utilized for identifying upgradeability proxy smart contracts. This can be achieved either through static, dynamic, or a hybrid method. The dynamic approach involves analyzing smart contract transactions, while the static method involves running an analyzer on the smart contract source or bytecode. In the given studies,~\citet{Salehi22},~\citet{Bodell23}, ~\citet{qasse2023}, and UPC Sentinel employ a hybrid approach, utilizing both dynamic and static methods. On the other hand,~\citet{Feist19} rely solely on static methods. 

More specifically,~\citet{Salehi22} used a dynamic method to extract the external dependencies, then applied a static analyzer to further validate if the given contract implements an upgradeability function. ~\citet{Bodell23} undertook the task of running a static analyzer on the proxy's source code, applying advanced techniques such as context-sensitive, flow-insensitive, and cross-contract data-flow analysis. Similarly, this study developed an "explorer querying" component to retrieve a contract's dependencies, embed related contracts into the same compilation unit, before running their cross-contract analysis. However, this approach is limited in scenarios where the proxy has multiple implementation contracts, one of which handles the upgrade mechanism, necessitating more advanced cross-contract analysis to account for multiple implementations~\citep{Bodell23}. 

UPC Sentinel addresses this gap, because its proxy detector conducts behavioral analysis, seamlessly detects all proxy's implementation contracts, and subsequently feeds them to the upgradeability detector layer to be analyzed one by one, ensuring detection of such nuanced cases. Additionally, UPC Sentinel integrates an additional dynamic analysis as part of its DUP detector, interacting directly with deployed contracts using a blockchain node to verify compliance with the ERC-2535 Diamonds Upgradeability Proxy interface. This enhancement addresses scenarios where the DiamondCutFacet contract has never been invoked, making it undetectable through transaction analysis alone.
   
~\citet{qasse2023} employs an innovative approach, leveraging the definitions of upgradeability patterns to detect a proxy as an upgradeability proxy if its source code exhibits certain features (e.g., Proxiable interface or IDiamond interface) relevant to these patterns. If the detected proxy does not comply with the defined features, they classify the proxy as a Transparent Upgradeability Proxy pattern. For instance, concerning the ERC-1822 Universal Upgradeable Proxy pattern, the author extracted the proxy's delegatecall transactions, and subsequently, the implementation contract (i.e., \texttt{to\textunderscore address} field) using a dynamic method and then checked if the implementation contract implements the Proxiable interface. This methodology is thoughtful and reflects a deep understanding of the underlying concepts. However, while relying on the unique features of upgradeability patterns is indeed a promising approach, it may not guarantee that the upgradeability function is necessarily implemented. In contrast, UPC Sentinel, and the methods proposed by~\citet{Salehi22} and ~\citet{Bodell23} only classify a proxy as an upgradeability one if they find at least one public or external function (e.g. an upgrade function) that substitutes the old implementation contract with a new one, potentially leading to higher precision.

~\citet{Feist19} introduced Slither, a tool that identifies security vulnerabilities and coding issues in smart contracts. As highlighted by~\citet{Bodell23}, Slither comes with native detection for proxy upgradeability. Yet, there are some limitations to its capabilities. Slither's method of upgradeability detection is mainly focused on searching for the terms ``upgradeable'' and ``proxy'', and on the presence of delegatecall in the fallback. In addition, This approach cannot analyze cross-contract interactions, which restricts its overall efficacy.

Finally, we acknowledge that when a proxy contract is inactive (i.e., its proxy functionality has never been utilized since creation), there is no transaction history linking the proxy to its implementation contract or external contracts. Consequently, such dependencies cannot be identified through transaction analysis alone. This limitation impedes the detection of inactive DUP-based UPCs (e.g., ERC-1822 Universal Upgradeable Proxy) and ESUP-based UPCs (e.g., Beacon Upgradeability Proxy), where the upgrade function is implemented outside the proxy. Although none of the existing UPC detection methods, including UPC Sentinel, currently address this issue, future research should prioritize developing mechanisms to directly read the state variables of the proxy contract—typically initialized during proxy creation—to extract the implementation or beacon contract addresses. We further discuss this in Section~\ref{subsec:threat-construct-validity}.

\smallskip\noindent\textbf{2. Analysis level}. The second criterion is pertinent to static detection methods and differentiates between the analysis conducted on the smart contract's source code and its bytecode. In this comparison,~\citet{Bodell23},~\citet{qasse2023}, and~\citet{Feist19} operate at the source code level, while UPC Sentinel and~\citet{Salehi22}'s one analyze at the decompiled bytecode level. Similar to proxy detection methods, upgradeability detector methods that operate at the source code level are limited to verified smart contracts and can be error-prone, as the source code may include unreachable (upgradeability) proxy code (Section~\ref{subsec:lessons-learned}). In contrast, UPC Sentinel remains unaffected by this limitation, as it operates directly on the bytecode. This approach inherently excludes dead functionality, given that the Solidity compiler automatically omits any unreachable or non-executable code during the compilation process.

\begin{table}[t]
\centering
\caption{Comparison of studies on upgradeability proxy detection, detailing detection approaches, analysis level, upgradeability reference design, and upgradeability proxy pattern.}
\resizebox{\textwidth}{!}{%
\begin{tabular}{lcccccccccccccccccc}
\toprule
\textbf{Study} & \multicolumn{17}{c}{\textbf{Upgradeability Proxy Detection Characteristics}} \\
\cmidrule(lr){2-18}
 & \multicolumn{2}{c}{\begin{tabular}[c]{@{}c@{}}Detection\\ Approach\end{tabular}} & \multicolumn{2}{c}{\begin{tabular}[c]{@{}c@{}}Analysis\\ Level\end{tabular}} & \multicolumn{5}{c}{\begin{tabular}[c]{@{}c@{}}Upgradeability\\ Reference Design\end{tabular}} & \multicolumn{8}{c}{\begin{tabular}[c]{@{}c@{}}Upgradeability\\ Proxy Pattern\end{tabular}} \\
\cmidrule(lr){2-3} \cmidrule(lr){4-5} \cmidrule(lr){6-10} \cmidrule(lr){11-18}
 & \rotatebox{90}{Dynamic} & \rotatebox{90}{Static} & \rotatebox{90}{Source Code} & \rotatebox{90}{Bytecode} & \rotatebox{90}{SMUP} & \rotatebox{90}{ESUP V\#1} & \rotatebox{90}{ESUP V\#2} & \rotatebox{90}{DUP V\#1} & \rotatebox{90}{DUP V\#2} & \rotatebox{90}{Beacon Upgradeability Proxy} & \rotatebox{90}{Registry Upgradeability Proxy} & \rotatebox{90}{ERC-1822 Universal Upgradeable Proxy } & \rotatebox{90}{ERC-2535 Diamonds Upgradeability Proxy } & \rotatebox{90}{Transparent Upgradeability Proxy} & \rotatebox{90}{ERC-1967 Upgradeability Proxy} & \rotatebox{90}{Inherited Storage Upgradeability Proxy} & \rotatebox{90}{Eternal Storage Upgradeability Proxy} \\
\midrule
UPC Sentinel & * & * &  & * & * & * & * & * & * & * & * & * & * & * & * & * & * \\
\citep{Bodell23}  & * & * & * &  & * & * &  & * &  & * & * & * & * & * & * & * & * \\
\citep{qasse2023} & * & * & * &  & * & * &  & * &  &  &  & * & * & * &  &  &  \\
\citep{Salehi22}  & * & * &  & * & * & * &  & * &  & * &  & * &  & * &  &  &  \\
\citep{Feist19}   &  & * & * &  & * & * &  & * &  &  &  &  &  &  &  &  &  \\
\bottomrule
\end{tabular}%
}
\label{tab:my_label}
\end{table}

\smallskip\noindent\textbf{3. Upgradeability reference design.} The third criterion is the type of upgradeability reference design that the detection method can identify. We identified three URDs in Section \ref{subsec:upgradeability-detector-layer}: SMUP, ESUP (with two variants v\#1 and v\#2), and DUP (with three variants v\#1 and v\#2 and variant v\#3). UPC Sentinel can detect UPCs from all reference designs and their variants.~\citet{Bodell23},~\citet{Salehi22}, and~\citet{Feist19} have demonstrated the ability to detect upgradeability proxies across all three reference designs. However, we could not find detailed information concerning their effectiveness in detecting the second and third variants of the externally supervised and delegated reference designs. It is worth acknowledging that the proposed methods show promising results, but it is important to consider that their recall in each category could be intricately tied to the specific implementation of the approach. Thus, a more comprehensive evaluation might be needed to fully understand their capabilities and limitations, especially in the context of varied upgradeability proxy patterns.

\smallskip\noindent\textbf{4. Upgradeability pattern}. The fourth criterion illustrates the specific upgradeability patterns that can be detected by the studies, including the eight patterns we introduced in Section \ref{subsec:upgradeability-proxy-patterns}. We flag a given study to detect a specific pattern if we find any information that shows the detection of the pattern. Most of the studies detect a wide array of these patterns, with~\citet{Bodell23} and UPC Sentinel covering all of them.~\citet{Salehi22} and~\citet{qasse2023} miss certain patterns, and~\citet{Feist19} seems to only be able to detect the Transparent proxy and lack detection for the remaining patterns.

Based on the aforementioned comparison, it is evident that the UPC Sentinel's design is comprehensive, encompassing both static and dynamic analyses, ensuring applicability to all smart contracts regardless of source code availability, and supporting a broad range of upgradeability proxy patterns.

    \section{Threats to Validity}
\label{sec:threats}

\subsection{Construct Validity} \label{subsec:threat-construct-validity}
% We rigorously evaluated UPC Sentinel on two different independent ground truth datasets, and meticulously compared our findings with previous studies whenever possible. Our evaluation suggests that UPC Sentinel shows promising improvements in performance relative to the methodologies used in earlier studies. Additionally, we conducted a comprehensive comparative analysis to shed light on the differences between UPC Sentinel with the recent studies in the field. 
% concerning evaluation, we carefully applied UPC Sentinel to two different ground truth datasets. The first dataset from Etherscan (GE ground truth) was automatically collected and then manually verified for 94 instances. Additionally, we evaluated UPC Sentinel against another dataset (GB ground truth) from a prior study~\citep{Bodell23}. An extensive comparative analysis was conducted, and disagreements on labeling were transparently addressed.

In terms of our limitations, UPC Sentinel currently cannot detect inactive (upgradeability) proxy contracts.~\citet{Ebrahimi23} analyzed a statistically representative random sample of 385 contracts drawn from the entire set of Ethereum contracts, including both open-source and closed-source ones. Their study found that 90 (23.4\%) of these contracts were proxy contracts, of which 62 (16.1\%) were active and 28 (7.27\%) were inactive. However, since inactive proxies are not necessarily obsolete (e.g., they may be activated in the future or upgraded in response to evolving needs) one can apply UPC Sentinel periodically to monitor inactive cases over time. Hence, such inactive proxies can be identified right after they are used for the first time. Additionally, we propose a future improvement plan to extend UPC Sentinel for more proactive detection of inactive (upgradeability) proxy contracts. 

\smallskip\noindent\textbf{Proxy detector layer:} Our method in Section \ref{subsec:proxy-detector-layer} currently uses the two conceptual properties of a proxy contract and assesses contracts’ transactions against them to detect *active* proxy contracts. However, due to lack of transactional activities for inactive proxy contracts this method is not effective for the latter type of contracts. In order to support inactive proxy contracts, future studies can leverage insights that we obtained during our study regarding possible detection of inactive proxies based on the decompiled bytecode only. 

In particular, we found information equivalent to the two conceptual properties of proxy contracts (see Section \ref{subsec:proxy-detector-layer}) at the decompiled bytecode level. Concerning Property \#1, one can reliably identify if the decompiled bytecode of a given contract contains a function with a delegatecall statement by searching for a ``delegate '' keyword. Concerning Property \#2, one can assess the decompiled bytecode of the function where the delegate keyword is found with our three identified lexical patterns (see Figure~\ref{fig:pattern-concepual-property}), to verify if the function has a similar interface to that of its implementation contract. In Section \ref{subsec:upgradeability-detector-layer}, these lexical patterns were present in 99.8\% of a representative sample of active proxy contracts. As a result, transaction activity may no be longer important for detecting proxy contracts, given these two equivalent properties in the decompiled bytecode. However, given the focus of this study on active proxy contracts, we leave the investigation of inactive proxy contracts as fruitful future work.

\smallskip\noindent\textbf{UPC detector layer:} For each of our three upgradeability reference designs, we explain how one could adapt them to support inactive proxies. For the SMUP-based UPCs, once an inactive proxy is identified by the improved proxy detection layer discussed above, no further changes are needed to the A: SMUP Detector module (see Figure \ref{fig:A-smup-detector}), as there are no design differences between inactive and active UPCs of this type. 

For the DUP-based UPCs, where the upgradeability logic is implemented inside the proxy's implementation contracts, our algorithm (see Figure \ref{fig:C-dup-detector}) would require the list of the proxy's implementation contract addresses. This list is currently obtained by analyzing the proxy’s historical transactions (i.e., dynamic analysis) for active proxy contracts (see Section \ref{subsec:proxy-detector-layer}). However, this approach proves ineffective for inactive proxy contracts due to unavailability of their transactional data. To address this, one can obtain the latest implementation contract address of an inactive proxy contract by reading the proxy’s storage. Specifically, one could run a static analyzer to extract the storage slot of the implementation variable from the decompiled bytecode of an inactive proxy contract, then use a blockchain node service provider like Infura to read the value of this slot (i.e., the implementation contract address) directly from the proxy’s storage. This approach could be effective as the implementation contract address is typically initialized by developers using proxy’s constructor or after deployment via calling an initializer function.

Concerning the ESUP-based UPCs, our method (see Figure \ref{fig:B-esup-detector}) relies on analyzing historical transactions to obtain the target contract address (i.e., the external contract that the proxy query to fetch the current implementation contract address right before the delegating calls). To handle this for inactive cases, one can locate the storage slot where the external contract address is stored in the proxy's storage and use an Infura node to read its address. The rest of our algorithms for the DUP and ESUP detectors would remain the same for inactive cases.

To ensure that our evaluations are not influenced by the presence of duplicates, we identified and removed repeated instances of the same contract address in the employed ground truth dataset (see Section~\ref{subsec:duplicate-gb}). Assuming two contracts C1 and C2 in the ground truth, there are in fact two interpretations of duplicates: (1) C1 and C2 have identical bytecodes but are deployed at different Ethereum contract addresses; (2) C1 and C2 have identical Ethereum contract addresses. 

We only remove duplicates under interpretation (2), since under interpretation (1) C1 and C2 still have distinct addresses (i.e., different entities) and their own unique runtime metadata and state (e.g., bytecodes, transactions, implementation contracts, storage values). For instance, they may differ in their implementation contracts or dependencies, which makes them distinct input contracts. However, with interpretation (2) both contracts, including their addresses, bytecodes, runtime transactions, are completely identical, basically representing the same deployed contract. In the ground truth dataset provided by~\citet{Bodell23}, we identified 40 instances where the same contract address appeared multiple times (i.e., interpretations (2)). By retaining only one unique entry per contract address, we ensured that each contract is represented only once, thereby preventing any skewing of results due to duplicate data points.

% Furthermore, the presence of duplicate contracts does not affect the effectiveness of UPC Sentinel. There are two types of duplicates: i) Contracts that have an identical address. Each deployed contract in Ethereum has its own unique address that distinguishes it from others. This contract address also represents the input of UPC Sentinel. We included a preprocessing step in Section~\ref{subsec:rq2-preprocessing}, namely “C) Exclude duplicate contracts” where we removed 41 of such duplicates as they all represent a similar contract. Therefore, our findings are affected by the presence of such duplicates. ii) Contracts with identical bytecode (a.k.a., type-1 clones). Type-1 clones are those that have distinct addresses but identical bytecodes. UPC Sentinel analyzes and treats each contract independently since each can have different runtime properties, such as storage state, set of transactions, and dependencies. Therefore, our method remains accurate and reliable regardless of the presence of type-1 clones.

In our study, we opted not to use Cohen’s Kappa to measure inter-rater agreement due to concerns about its reliability with imbalanced datasets. Research by \citeauthor{Deloitte_2023} and \citeauthor{Li2023} underscores Kappa’s sensitivity to category prevalence, which often leads to unstable or misleading results when one category is either very rare or very common, even in the presence of high observed percentage agreement between raters.

This issue arose in RQ1, where the first and third authors (i.e., the raters) independently verified the upgradeability status of a sample of 94 contracts. This sample was drawn from a dataset of 3,177 contracts flagged as UPC by Etherscan, where the positive category (UPC class) was expected to be highly prevalent. As a result, Kappa behaves counterintuitively, yielding a disproportionately low value even when there is substantial agreement between raters (in 92 out of 94 cases (97.87\%)). A similar scenario appeared in RQ3, where we compared \citeauthor{Bodell23}’s ground truth upgradeability labels (i.e., first rater) with our assigned upgradeability labels (i.e., second rater) across four groups (i.e., FN, TN, TP, FP). Since one category (UPC or Non-UPC) was significantly overrepresented in each group, it led to category imbalances that rendered Kappa unreliable. To address these limitations, we instead reported the observed proportion of agreement~\citep{McHugh2012}, which directly reflects the consistency between raters without being skewed by category prevalence.

Furthermore, the Kappa statistic is effective when the positive classification rate is balanced, meaning that raters are equally likely to classify subjects as positive or negative (approximately 50\%). However, when one rater significantly favors one category, the Kappa measure becomes unreliable~\citep{Gwet2002,Byrt1993}. This issue arose in RQ2, where we compared ~\citeauthor{Bodell23}’s ground truth proxy labels (i.e., the first rater) with our verified proxy labels (i.e., the second rater). In this case, all 920 analyzed contracts in the ground truth were labeled as proxy, and our analyses determined 702 contracts as proxy. This extreme imbalance in labeling rendered the Kappa statistic unreliable, resulting in a zero value despite the relatively high percentage agreement between our labels and the ground truth. To address this, we reported the observed proportion of agreement~\citep{McHugh2012}, which directly reflects the consensus between our verified labels and the ground truth ones. 

% Finally, patterns such as Inherited Storage Upgradeability Proxy, Eternal Storage Upgradeability Proxy and ERC-1967 Upgradeability Proxy (see Section \ref{subsec:upgradeability-proxy-patterns}) primarily focus on the storage structure of the proxy and its implementation to minimize the risk of storage collisions. While these storage patterns were originally introduced to design SMUP-based UPCs by~\cite{openzeppelin_2020_proxy_pattern}, their storage ideas have become sufficiently versatile to be integrated in UPCs across all three upgradeability reference designs. For instance, the ERC1967 storage idea can be utilized to design the storage of either Transparent Upgradeability Proxy (i.e., SMUP), ERC-1822 Universal Upgradeable Proxy (i.e., DUP), or Beacon Upgradeability Proxy (i.e., ESUP). While storage patterns ensure storage consistency, adopting them along with other upgradeability proxy patterns does not introduce a new reference design nor an upgradeability pattern. From the detection perspective, contracts’ storage structure does not impact the overall UPC Sentinel’s algorithm. This is because decompilers (e.g., Panoramix in our case) flatten the source code by converting complex structures such as inheritance hierarchies, which is the case for Eternal and Inherited storage, into simpler, more linear formats. This simplification allows us to analyze all contracts equally regardless of how their storage structure is designed. %which satisfies R5. 

%\goliva{Are there alternative explanations for the things that we're observing (e.g., confounding factor)?}
\subsection{Internal Validity} 
While evaluating decompilers is out of the scope of our current research, we acknowledge that Panoramix may not always complete the decompilation process or may produce inaccurate decompiled code, as evidenced by our experiments and corroborated by a recent study \citep{Liu23_2}. However, evidence suggests that the impact of such cases on our study's results is not significant. 

More specifically,~\cite{Liu23_2} performed an empirical study on the top five state-of-the-art smart contract decompilers, including Erays \citep{Zhou18}, Vandal~\citep{Brent18}, Panoramix~\citep{Palkeo_2019}, EthervmDec\footnote{\url{https://ethervm.io/decompile}}, and Gigahorse~\citep{Grech19}. The authors analyzed the decompilation success rates for these five decompilers on a representative random sample of 16,000 smart contracts. Their results indicate that, aside from Erays, the other four decompilers achieve notably high success rates (all above 98\%), with the Panoramix decompiler reaching a success rate of 98.2\%. Additionally, the authors found that among the included decompilers, Panoramix generates contract sources with the highest similarity scores to original sources. which is crucial for accurately detecting upgradeability proxy contracts at the decompiled bytecode level, as it allows for more reliable identification of subtle code patterns. Erays and Gigahorse, for example, generated code with near-zero similarity scores, complicating the UPC detection process at decompiled bytecode level. These findings indicate that Panoramix is of superior quality from a code similarity perspective, which is another reason why we chose Panoramix in this research.

In our experiments, specifically RQ1, we were unable to determine the type of 22 (0.7\%) out of 3177 UPCs due to decompilation failures. Similarly, in RQ3, one (0.1\%) out of 920 contracts featured an inconsistent storage structure which is due to Panoramix not recognizing the storage pattern. Despite this, we observe that the impact of such decompilation problems appears minimal or negligible, as less than one percent of the UPC instances were not detected across two different analyses. Finally, we believe that our methodology and upgradeability reference designs (see Section \ref{subsec:upgradeability-detector-layer}) offer a comprehensive roadmap for detecting UPCs, and we encourage researchers to replicate our study using other decompilers to further validate our findings and enhance the robustness of this research area.

Finally, we conducted a thorough assessment of the two ground truth datasets (GE and GB), including systematic sanity checks to validate their quality. Following established data quality guidelines~\citep{Bhatia2024}, we defined clear and well-justified criteria for any dataset modifications. Each adjustment was documented in detail, including the rationale and supporting evidence, adhering to empirical software engineering principles to promote transparency\citep{Ralph2021}. Furthermore, all results and findings have been made publicly accessible through our replication package to support peer review, reproducibility, and future research.

% Finally, we performed several manual studies to verify the labels of our employed ground truths across the paper. We followed coding best practices and also leveraged our expertise in smart contract analysis~\citep{Oliva20a, Oliva20b, Oliva20c, Oliva21, Oliva23a, Ebrahimi23}

% ~\citep{Oliva20c, Oliva20b, Oliva21, Oliva23a, Oliva23b}.

%\goliva{Do our results generalize?} 
\subsection{External Validity} 
Our collected ground truth dataset (GE) comprises active UPCs deployed from Jan. 2018 through Sep. 2022, as identified by Etherscan. This dataset is particularly valuable because it reflects verified UPCs from nearly every month within this timeframe, as illustrated in Figure \ref{fig:ge-upc-count-per-mont}. The consistency and broad temporal coverage of the dataset suggest that it potentially captures the overall trend and frequency of UPC usage over these years. The steady escalation in UPCs, especially from 2020 onward also aligns with the increasing trend in the ratio of EOAs who deployed proxy contracts~\citep{Ebrahimi23}, That said, we acknowledge that this set is still a subset of UPCs for which at least two versions are reported by Etherscan. Such UPCs may follow specific known storage standards, making it easy for Etherscan to identify and track the changes in their implementation contracts. While this may introduce some bias, especially since not all UPCs may follow these standards, we believe that the insights gained from this dataset are still valuable for evaluating UPC Sentinel.

To further mitigate concerns about potential bias, we also employed another ground truth dataset from a prior study \citep{Bodell23}, and evaluated UPC Sentinel based on that as well. This ground truth includes contracts initially classified as UPCs by another tool, Slither, then refined by the \cite{Bodell23}. One potential limitation of this ground truth is that all its contracts are verified (i.e., their source code is available on the Etherscan blockchain explorer), while the majority of Ethereum smart contracts are not verified. Hence, this ground truth might not fully allow us to measure the recall metric of UPC Sentinel and USCHUNT. That said, it is important to note that preparing a balanced dataset is extremely time-consuming and resource-intensive. Among previous studies, only~\citet{Bodell23} evaluated their methods against a ground truth. Consequently, by utilizing two different ground truths in this study, one of them proposed by us, we ensured that our evaluation approach matches or surpasses the rigour of other studies in the field.

In addition, our employed dataset may not be representative of the true negative cases. However, we believe UPC Sentinel is able to effectively handle true negatives (i.e., contracts that are not (upgradeability) proxy contracts) for two reasons. First, our proxy detection method has 100\% precision. Hence, this step excludes a tremendous number of true negatives with perfect precision. Second, when it comes to upgradeability detection, we designed UPC Sentinel based on the fundamental features of the most prevalent upgradeability proxy patterns (a.k.a, our upgradeability reference designs). This systematic design ensures that UPC Sentinel is deterministic and reliable and decreases the likelihood of false negative cases. That is also confirmed by our perfect precision while evaluating UPC Sentinel in Section \ref{subsec:eval-upc-on-gb}. Finally, our results might not generalize to other programmable blockchain platforms. 

% Finally, we only focus on active proxies. However, inactive proxies may have unique characteristics. Therefore, our results may not generalize to inactive proxy contracts. 

    \section{Conclusion}
\label{sec:conclusion}

% The advent of Ethereum and its introduction of smart contracts has undeniably transformed the blockchain ecosystem, laying the foundation for a decentralized digital economy. As the demand for maintenance and evolution of decentralized applications grows, the challenge of reconciling the principle of immutability with the practical necessity of timely updates has come to the forefront. While several mechanisms have been proposed to bridge this gap, the proxy smart contract has emerged as a prominent solution, offering both flexibility and efficiency. Recognizing the significance of upgradeability proxy patterns (UPCs), we identified the necessity for an effective method to detect and trace UPCs, which is essential for enhancing the transparency, security, and reliability of the Ethereum ecosystem.

The increasing complexity and adoption of decentralized applications have highlighted the critical need to reconcile blockchain immutability with the practical necessity of updates. Proxy contracts, particularly upgradeability proxy contracts (UPCs), have emerged as a leading solution to this challenge, offering a balance between flexibility and efficiency. Despite their growing popularity and critical role, current methods for detecting upgradeability proxy contracts (UPCs) still face several gaps, particularly in handling closed-source contracts, limited pattern coverage, and suboptimal orchestration of static and dynamic analyses. These limitations underscore the need for effective tools to detect and classify UPCs.

In response, we developed UPC Sentinel, a three-tier algorithm that systematically orchestrates both static and dynamic analysis techniques on decompiled smart contract bytecode to identify UPCs. Each layer of UPC Sentinel incorporates well-defined techniques to achieve its objectives with precision. Given a smart contract address: (i) the proxy detector layer accurately determines whether the contract is a proxy by analyzing its behavior (i.e., transactional activities) against the two inherent properties of proxy contracts; (ii) the upgradeability detector layer combines static and dynamic analysis techniques to assess the upgradeability status of identified proxies, based on adherence to three defined abstract upgradeability reference designs; and (iii) the pattern classifier layer provides fine-grained classification, identifying the specific upgradeability pattern the detected UPC conforms to. 

Our comprehensive empirical evaluations, conducted on two distinct ground truth datasets, demonstrate that the UPC Sentinel approach is both reliable and effective within the defined scope of this study (i.e., active proxy contracts). Specifically, for our newly collected GE dataset comprising 3,177 UPCs, UPC Sentinel achieved a near-perfect accuracy of 99\%. Similarly, for the GB dataset, it achieved a perfect precision of 100\% and a near-perfect recall of 99.3\%, outperforming the state-of-the-art methods (see Table~\ref{tab:comp-bodell-g3-and-g4}). Notably, UPC Sentinel overcomes the reliance on source code availability that limits many existing methods, enabling it to detect active UPCs at the bytecode level across a wide range of upgradeability proxy patterns and their variants. This broader applicability allows for a more comprehensive analysis of the Ethereum ecosystem's upgradeability landscape, extending beyond open-sourced contracts.

While our study focused on active proxy contracts, UPC Sentinel may require further adaptation to handle inactive proxy contracts. We provided concrete future plan on how to extend UPC Sentinel to enable detection of inactive UPCs. Finally, we hope that our methodology and replication package—including the publicly available source code of UPC Sentinel, ground truth datasets, and detailed findings—will serve as valuable resources for the research community, facilitating further advancements in the field and enabling researchers to enhance, refine, and build upon our work in future studies.

    \section*{Declarations}

\subsection*{Data Availability Statement}
\label{sec:Data_Availability_Statement}

A supplementary materials package, including the UPC Sentinel source code, two ground truth datasets, and detailed information on our manual studies, is available online: \url{https://github.com/SAILResearch/replication-24-amir-upc_sentinel}. 

\subsection*{Funding and/or Conflicts of Interests/Competing Interests}
\label{sec:Conflict_Of_Interest}
The authors declare that they have no known competing interests or personal relationships that could have (appeared to) influenced the work reported in this article.

   \bibliographystyle{IEEEtranN} %sorted by order of appearance
    \bibliography{11-references} 
    
    % \newpage
    \clearpage 
% 	\pagebreak

    \begin{appendices}

\section{The landscape of smart contract upgradeability methods}\label{subsec:landscape-of-upgradeability-methods}
Given the immutable nature of the Ethereum blockchain, introducing modifications to deployed smart contracts presents a unique challenge. Over time, several methods have emerged, each attempting to bridge the gap between the need for change and the essence of immutability. These methods vary in complexity, effectiveness, and application, but all aim to extend the life cycle of a contract while preserving its core attributes. In this section, we will shortly review the different upgradeability techniques employed in smart contracts, providing insights into their mechanisms.

\begin{itemize}[label=\textbullet, itemsep = 3pt, topsep = 3pt, wide = 0pt]

    \item{\textbf{Proxy contract}}~\citep{OpenZeppelin_2017b}. The Proxy Pattern offers a means to enhance smart contracts post-deployment. In this setup, there is a forward-facing proxy contract that channels interactions to a linked implementation contract. The real operations and rules are housed in this implementation contract. If updates are needed, a new implementation contract is deployed, and the proxy is adjusted to this fresh version (see Section \ref{subsec:back-proxy-contract} for details). Consequently, users keep interacting with the unchanged proxy, oblivious to the changes behind the scenes. This method allows continuous adaptability while retaining the original contract's address and stored assets. The Ethereum community has played a pivotal role in the evolution and standardization of upgradeability mechanisms, with the proxy pattern standing out prominently both in practice and theory~\citep{Salehi22, Ebrahimi23}. Specifically, in recent years, practitioners have proposed several Ethereum proposals and standards. These standards aim to standardize various aspects of the proxy contracts, improve their functionality, and ensure best practices are followed. 
    
    \item{\textbf{Data separation}}~\citep{smartcontract_upgrades}. In this method, the smart contract's data and logic are decoupled and stored in separate contracts. The idea is to keep the state or data in one contract (often referred to as the ``data contract'' or ``storage contract'') and the business logic or functions in another (the ``implementation contract''). In the event an upgrade is needed, only the implementation contract is replaced or modified, while the data contract remains unchanged. Since the data stays consistent and is not moved, it ensures the integrity of the contract's state throughout the upgrade process. Eternal storage is one way to decouple contract storage from its logic.
    
    \item{\textbf{Create2 mechanism}~\cite{Feist19}}. The \texttt{CREATE2} method offers a unique spin on smart contract upgradeability. It allows developers to pre-determine a contract's address even before it is deployed. By using this approach, a contract can be re-deployed at the same address as long as specific conditions, often involving initial states or salts, are met. This means that while the old contract code becomes defunct, the new version takes its place at the familiar address. This technique ensures consistent contract interaction points, even after upgrades. 
    
    \item{\textbf{Data migration}}~\citep{Bandara_2020}. In this method, when a contract needs an upgrade, a new version of the contract is deployed. The existing data from the old contract is then transferred or ``migrated'' to this new contract. Users subsequently interact with the new version. While this method can ensure that upgrades incorporate both logic and data changes, it is essential to manage the migration process meticulously to prevent data loss or inconsistency, and to ensure seamless transition for users.
    
    \item{\textbf{Strategy pattern}}~\citep{smartcontract_upgrades}. The Strategy Pattern is a design pattern familiar to object-oriented programming. This pattern introduces a method for changing an object's behavior at runtime, making it applicable for smart contract upgradeability. Instead of modifying the contract directly, various algorithms (or ``strategies'') are defined and encapsulated separately, allowing them to be swapped as needed. By employing this pattern, smart contracts can adapt and evolve by merely switching out the strategy component, providing a modular approach to introduce new functionalities or rectifications without altering the core contract structure.
    
    \item{\textbf{Self-destruct mechanism}}~\citep{Chen20-2}. The \texttt{self-destruct} functionality in Solidity allows a smart contract to be ``destroyed'', releasing its remaining Ether balance to a specified address. When used as an upgradeability mechanism, a contract can be programmed to self-destruct under certain conditions, essentially removing its code from the blockchain. Subsequently, a new version of the contract with upgraded logic is deployed. While using self-destruct clears the contract's bytecode and reclaims the Ethereum state's occupied space, it has some challenges. First, all data associated with the contract is permanently lost upon self-destruction. Therefore, any needed state or data must be migrated beforehand to a new version or an external storage solution. Second, once executed, the self-destruct action cannot be reversed. Finally, after self-destruct, the contract's address cannot be reused, and users have to be notified to migrate to the new address. For instance, if the users mistakenly send their assets to the destroyed contract, their assets will be lost forever. Given these challenges, using self-destruct for upgradeability is generally not the preferred method due to its risks and complexities associated with data migration and user transition.
\end{itemize}

\section{Upgradeability Proxy Patterns}\label{subsec:upgradeability-proxy-patterns}
We explain the well-established upgradeability patterns and proposals that utilize the proxy pattern at their core.~\citet{Meisami23} compiled a comprehensive list of all recognized proxy patterns. From this list, we selected the eight proxy patterns that are known to be used for upgradeability for our study. An ERC (Ethereum Request for Comments) serves as the foundational framework for proposing new features, processes, or changes within the Ethereum ecosystem, focusing specifically on application-layer standards for Ethereum's smart contracts, such as token standards.

\begin{itemize}[label=\textbullet, itemsep = 3pt, topsep = 3pt, wide = 0pt]
    \item \textbf{Inherited Storage Upgradeability Proxy Pattern}~\citep{openzeppelin_2020_proxy_pattern}. The pattern is designed to facilitate seamless smart contract upgrades while ensuring consistent storage across versions, thereby maximizing flexibility and reliability in contract evolution. Figure \ref{fig:Inherited-Storage-Upgradeability-Proxy-Pattern} depicts a simplified class diagram for this pattern. The UpgradeabilityStorage holds the implementation contract address, ensuring that every new implementation inherits this setup through the Upgradeable contract. This inheritance chain maintains the storage layout defined in previous versions and prevents inadvertent overrides of the implementation contract address by subsequent implementations. The Proxy acts as the front-facing interface for all interactions being responsible for delegating calls to specific implementations. The UpgradeabilityProxy inherits from both the Proxy and UpgradeabilityStorage to manage which version of the implementation is currently operational, allowing for upgrades. 
    
    \begin{figure}[!htbp]
     \centering
     \includegraphics[scale=0.75]{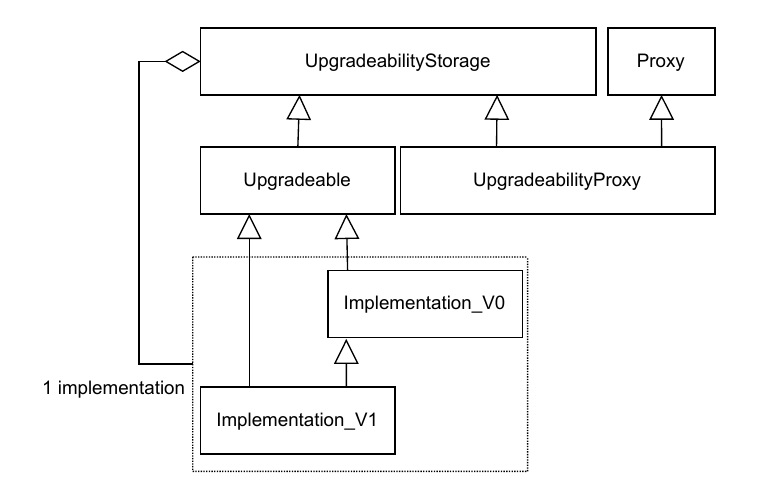}
     \caption{Key components of the Inherited Storage Upgradeability Proxy Pattern.}
      \label{fig:Inherited-Storage-Upgradeability-Proxy-Pattern}
    \end{figure}    
    
    \item \textbf{Eternal Storage Upgradeability Proxy Pattern}~\citep{openzeppelin_2020_proxy_pattern}. This pattern is designed to enable robust and flexible upgrades for smart contracts while maintaining a consistent, immutable storage structure. Figure \ref{fig:Eternal-Storage-Upgradeability-Proxy-Pattern} depicts a simplified class diagram for this pattern. The UpgradeabilityStorage holds essential data for upgrade management  (e.g., the implementation contract address). The Proxy routes interactions to appropriate implementation versions via delegatecall. The UpgradeabilityProxy combines the functionalities of the UpgradeabilityStorage and the Proxy to facilitate the transition between implementations, ensuring smooth updates. The key component in this pattern is the EternalStorage contract, which establishes a universal and immutable storage schema composed of mappings for each variable type inherited by subsequent implementation contracts and the EternalStorageProxy contract. This direct inheritance ensures that all upgrades and behavioral changes remain compatible with the initial storage structure laid out by EternalStorage, preventing discrepancies in data handling across versions. The new implementation contract must adhere to the storage schema of the preceding version to maintain data consistency.
    
    \begin{figure}[!htbp]
     \centering
     \includegraphics[scale=0.75]{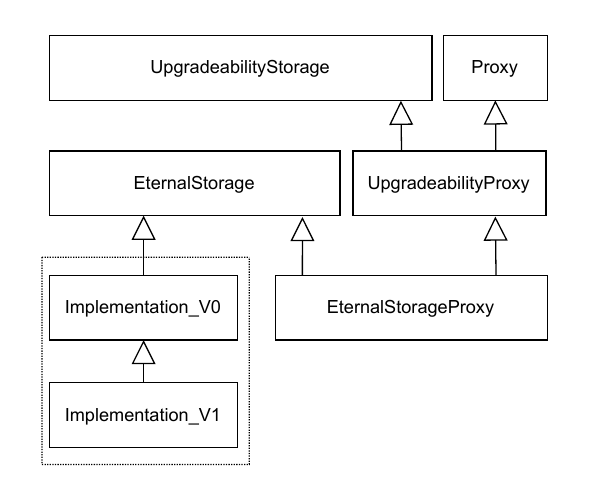}
     \caption{Key components of the Eternal Storage Upgradeability Proxy Pattern.}
      \label{fig:Eternal-Storage-Upgradeability-Proxy-Pattern}
    \end{figure} 

    \item \textbf{Registry Upgradeability Proxy Pattern}~\citep{Peh18}. This pattern provides a centralized, manageable, and secure solution for tracking and updating different versions of contract implementations. Figure \ref{fig:Registry-Upgradeability-Proxy-Pattern} shows a simplified class diagram for this pattern. IRegistry is an interface that defines the structure for a Registry contract. Registry implements the IRegistry interface and is deployed at a different Ethereum address. This contract manages a registry of versions and their corresponding implementation contract addresses, allowing the owner to register new versions or query the implementation address associated with a particular version. Proxy is a contract that serves as a forwarder or a delegate to the current implementation contract. UpgradeabilityProxy is an instance of the Proxy that query the Registry each time before delegating a call to fetch the appropriate implementation contract address. This ensures that the UpgradeabilityProxy always uses the latest version registered in the Registry without requiring manual updates, enabling more dynamic upgrade patterns.
    
    \begin{figure}[!htbp]
     \centering
     \includegraphics[scale=0.75]{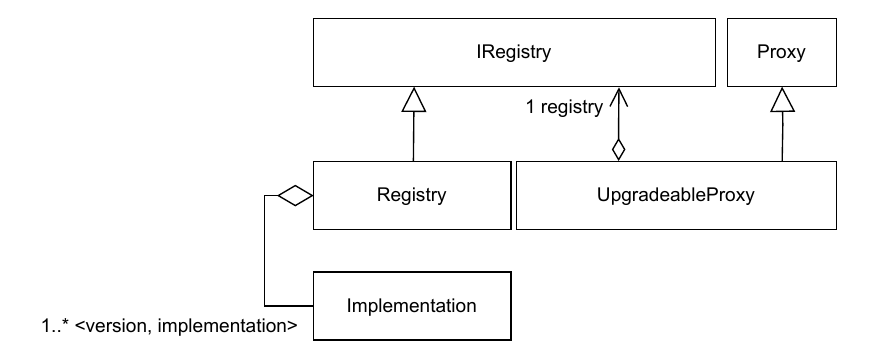}
     \caption{Key components of the Registry Upgradeability Proxy Pattern.}
      \label{fig:Registry-Upgradeability-Proxy-Pattern}
    \end{figure}     
    
    \item \textbf{Beacon Upgradeability Proxy Pattern}~\citep{beacon_proxy}. The pattern is designed to facilitate the dynamic updating of the implementation contract, providing an efficient and centralized way to manage upgrades across multiple proxies. Figure \ref{fig:Beacon-Upgradeability-Proxy-Pattern} illustrates a simplified class diagram for this pattern. The IBeacon interface specifies a minimal requirement for beacon contracts, primarily to return the current implementation address. The UpgradeableBeacon is a specific implementation of the IBeacon interface that allows the implementation address to be dynamically updated via an upgrade function. The Proxy contract acts as a stable intermediary that delegates all function calls to an implementation address. The BeaconProxy extends the Proxy by integrating directly with a beacon (conforming to the IBeacon interface) to fetch the latest implementation address before each call, ensuring that the proxy always delegates calls to the the latest implementation contract, thereby combining dynamic upgradeability with operational stability.
    
    \begin{figure}[!htbp]
     \centering
     \includegraphics[scale=0.75]{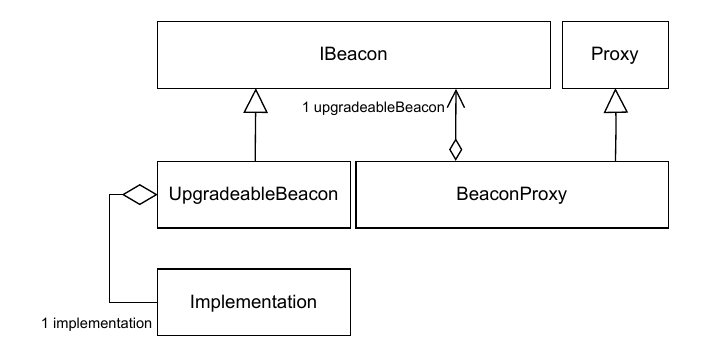}
     \caption{Key components of the Beacon Upgradeability Proxy Pattern.}
      \label{fig:Beacon-Upgradeability-Proxy-Pattern}
    \end{figure} 
    
    \item \textbf{ERC-1967 Upgradeability Proxy Pattern (Unstructured Storage)}~\citep{eip1967}. This pattern incorporates the idea of Unstructured Storage, introduced by OpenZeppelin, where proxy-specific information, such as the implementation contract address, is stored in designated fixed slots not allocated by the compiler. These storage slots are calculated using a hashing method. Specifically, a string identifier (e.g., “eip1961.proxy.implementation”) is hashed using the \texttt{keccak256} hashing function. This approach ensures that these critical addresses do not conflict with the storage layout of the implementation contracts, thereby avoiding inadvertent overwrites. Figure \ref{fig:ERC1967-Storage-Upgradeability-Proxy-Pattern} shows a simplified class diagram for this pattern. The Proxy serves as the core interface, channeling interactions to the correct contract versions via delegatecall, and the ERC1967Upgrade provides a suite of functionalities to manage and execute upgrades of smart contracts. This abstract contract utilizes specific, well-defined storage slots for the administration and control of proxy contracts, such as for the current implementation, the admin, and potentially a beacon. The ERC1967Proxy integrates these functionalities by inheriting from both the Proxy and ERC1967Upgrade, enabling secure upgrades through public interfaces. This standard has been widely embraced across various upgradeability patterns as a best practice for designing proxy storage architecture.
    
    \begin{figure}[!htbp]
     \centering
     \includegraphics[scale=0.75]{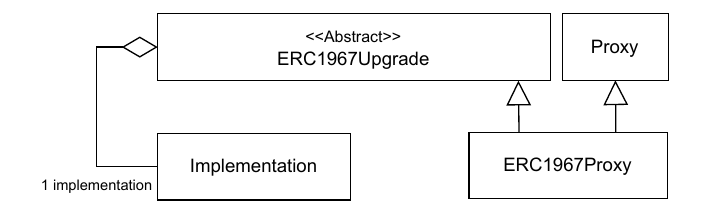}
     \caption{Key components of the ERC-1967 Upgradeability Proxy Pattern.}
      \label{fig:ERC1967-Storage-Upgradeability-Proxy-Pattern}
    \end{figure}

    \item \textbf{ERC-1822 Universal Upgradeable Proxy Standard (UUPS)}~\citep{eip_1822}. This pattern provides a flexible and cost-effective way to upgrade smart contracts while preserving their state and reducing gas costs. Figure \ref{fig:UUPS-Upgradeability-Proxy-Pattern} illustrates a simplified class diagram for this pattern. The ERC1967Utils library defines and manages storage slots in accordance with the ERC-1967 standard (a.k.a., the unstructured storage) for upgradeable proxies, specifically focusing on organizing the storage structure to safely and transparently handle changes in the implementation storage slot. The Proxiable contract utilizes the specific implementation storage slot from ERC1967Utils to securely initiate an upgrade. The UUPSProxy also utilizes the ERC1967Utils library to read from the designated implementation storage slot. To allow for a secure upgrade process, every implementation contract must implement the Proxiable interface. This consistency ensures that the implementation contracts write to the same implementation slot that the UUPSProxy reads from. To initiate the upgrade, the owner uses the UUPSProxy to delegate the upgrade task to the implementation contract’s upgrade function. As the delegatecall operates within the UUPSProxy's context, the upgrade effectively updates the proxy’s storage with the new implementation address.
    
    \begin{figure}[!htbp]
     \centering
     \includegraphics[scale=0.75]{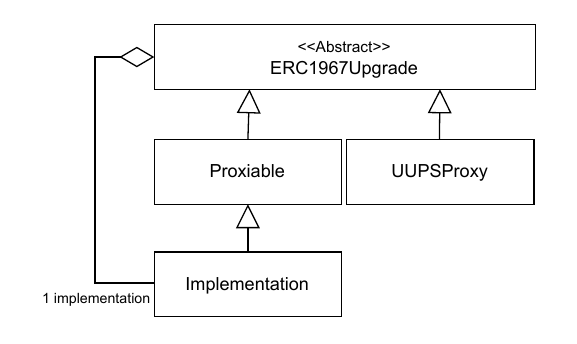}
     \caption{Key components of the ERC-1822 Universal Upgradeable Proxy Pattern (UUPS).}
      \label{fig:UUPS-Upgradeability-Proxy-Pattern}
    \end{figure} 

    \item \textbf{ERC-2535 Diamonds Upgradeability Proxy Pattern}~\cite{EIP-2535}. This pattern is a sophisticated framework for Ethereum smart contracts that allows for modular design and scalable upgradability. This standard uses a DiamondProxy contract that delegates functionality to various Facet contracts (a.k.a., implementation contracts). Each Facet, deployed separately, can add specific functions to the DiamondProxy, enhancing modularity. Figure \ref{fig:Diamond-Upgradeability-Proxy-Pattern} depicts a simplified class diagram for this pattern. DiamondProxy acts as the central contract that delegates calls to specific functions in Facets. LibDiamond facilitates the addition, replacement, or removal of function, handles diamond storage, mapping function selectors to facets, and enforces access controls. IDiamondCut provides an interface for adding, replacing, or removing facets' functions, defining the structure for these operations. IDiamondLoupe provides introspection functions, helping to identify the facets and functions within a diamond. DiamondLoupeFacet and DiamondCutFacet implement these interfaces, respectively, managing the facets' organization and modifications through the LibDiamond. These facets interact by using delegatecall from the DiamondProxy to execute logic defined in various facets. Facets do not store data within their own contract storage. Instead, they can manipulate state variables stored in the DiamondProxy, the central proxy contract. Similar to the ERC-1822 UUPS pattern, for effective upgrades, the owner employs the DiamondProxy to delegate the upgrade task to the DiamondCutFacet’s DiamondCut upgrade function. As the delegatecall operates within the DiamondProxy's context, the upgrade effectively updated the proxy’s storage, adjusting the mapping of function selectors accordingly.

    \begin{figure}[!htbp]
     \centering
     \includegraphics[scale=0.75]{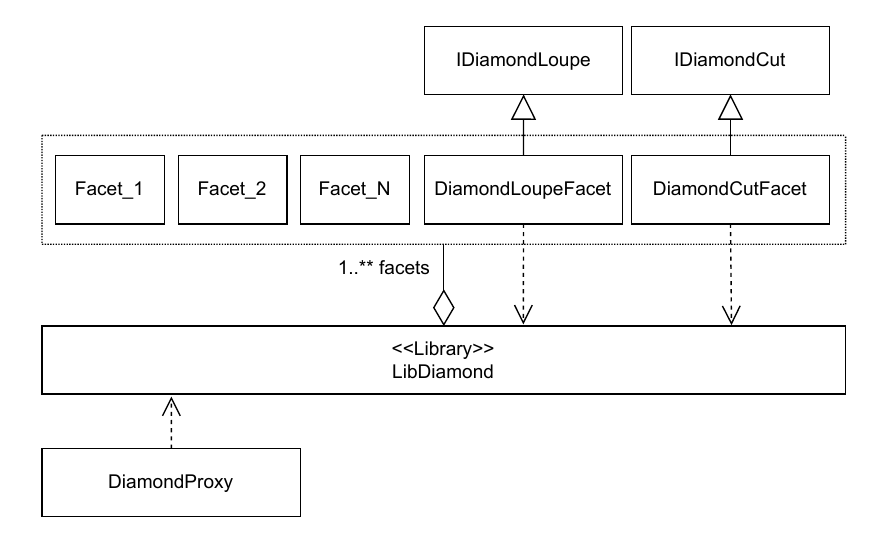}
     \caption{Key components of the ERC-2535: Diamonds Upgradeability Proxy Pattern.}
      \label{fig:Diamond-Upgradeability-Proxy-Pattern}
    \end{figure} 
    
    \item \textbf{Transparent Upgradeability Proxy Pattern}~\citep{palladino_2021}. Upgradeable proxies, which delegate function calls to an implementation contract, can run into issues with function clashes because Solidity uses 4-byte function identifiers. This encoding can lead to different functions from separate contracts (e.g., a proxy and its implementation contracts) accidentally sharing the same identifier, posing a risk of executing incorrect functions when called through the proxy. Transparent Upgradeability Proxy pattern addresses this by differentiating the behavior based on the caller's identity: if the admin (with rights to upgrade the proxy) calls, the proxy handles the function itself; if any other user calls, the function is delegated to the logic contract. This prevents admins from accidentally triggering logic functions when intending to manage the proxy, thus enhancing  control. Figure \ref{fig:Transparent-Upgradeability-Proxy-Pattern} depicts a simplified class diagram for this pattern. As shown, this pattern reuses components from ERC-1967 Upgradeability Proxy Pattern along with a new TransparentUpgradeableProxy component that distinguishes caller’s identity and redirects the calls accordingly.

    \begin{figure}[!htbp]
     \centering
     \includegraphics[scale=0.75]{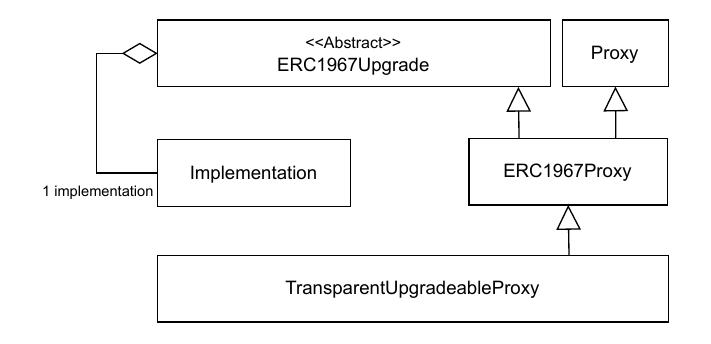}
     \caption{Key components of the Transparent Upgradeability Proxy Pattern.}
      \label{fig:Transparent-Upgradeability-Proxy-Pattern}
    \end{figure} 
   
\end{itemize}

\section{Panoramix decompiler} \label{apx:panoramix-decompiler}
Panoramix is a decompiler designed to reverse-engineer smart contract bytecode~\citep{Palkeo_2019}. This tool is integrated into Etherscan, the most widely used blockchain explorer. Listing \ref{listing:a-decompiled-bytecode} provides an excerpt from the decompiled bytecode of a real-world contract\footnote{\href{https://etherscan.io/address/0xee6a57ec80ea46401049e92587e52f5ec1c24785}{0xee6a57ec80ea46401049e92587e52f5ec1c24785}}. Typically, decompiled bytecode consists of two main segments: i) the storage section (lines 3-8), which defines all variables declared at the contract level, and ii) the business logic (lines 10-42), which specifies the public and external functions/interfaces accessible for interaction. The decompiler translates bytecode back into a high-level representation, flattening the structure by integrating inherited functions and resolving internal contract calls to present a consolidated and linearized view of the code.

More specifically, lines 10 to 14 define two getter functions that return the addresses of the admin and the implementation contract. The fallback function, located between lines 16 and 24, implements the proxy functionality. Line 18 demonstrates that the fallback function delegates the call to the implementation contract address stored in the stor3608 variable (i.e., the implementation variable). Additionally, line 19 verifies that the target function selector is derived from the first four bytes of the calldata, ensuring a consistent interface between the proxy and its implementation (i.e., the secondary property). Lines 26 to 36 define two functions, ``upgradeTo'' and upgradeToAndCall'', which allow the implementation variable to be updated with a new value. Finally, the ``changeAdmin'' function (lines 38-42) permits the modification of the proxy admin address.

\noindent
\begin{figure}[t]
% \begin{minipage}{\linewidth}
\begin{lstlisting}[frame=single,language=Solidity, caption={An example of a decompiled bytecode.},label={listing:a-decompiled-bytecode}]
# Panoramix decompiler. 

def storage:
  stor3608 is uint128 at storage 0x360[...]bbc offset 160
  stor3608 is addr at storage 0x360[...]bbc
  stor3608 is uint256 at storage 0x360[...]bbc
  storB531 is uint128 at storage 0xb531[...]103 offset 160
  storB531 is addr at storage 0xb531[...]103

def admin(): 
  [...]

def implementation(): 
  [...]

def _fallback() payable: 
  [...]
  delegate uint256(stor3608.field_0) with:
     funct call.data[0 len 4]
       gas gas_remaining wei
      args call.data[4 len calldata.size - 4]
  if not delegate.return_code:
      revert with ext_call.return_data[0 len return_data.size]
  return ext_call.return_data[0 len return_data.size]

def upgradeTo(address _implementation):
  require calldata.size - 4 >= 32
  [...]
  addr(stor3608.field_0) = _implementation
  [...]

def upgradeToAndCall(address _implementation, bytes _data) payable: 
  require calldata.size - 4 >= 64
  [...]
  addr(stor3608.field_0) = _implementation
  [...]

def changeAdmin(address _admin):
  require calldata.size - 4 >= 32
  [...]
  addr(storB531.field_0) = _admin_
  [...]


\end{lstlisting}
% \end{minipage}
\end{figure}

    \end{appendices}
    
\end{document}